\documentclass[journal]{IEEEtran}

\usepackage{amsmath}
\usepackage{graphicx}
\usepackage{subfig} 
\usepackage{tabu}  

\usepackage[colorlinks,linkcolor=blue,anchorcolor=blue,citecolor=green]{hyperref}
\usepackage[table]{xcolor} 

\usepackage{multirow}
\usepackage{dsfont}
\usepackage{stfloats}
\usepackage{amsmath}
\usepackage{cite}
\usepackage{booktabs}
\usepackage{algorithm}
\usepackage{algorithmic}
\usepackage{amssymb}
\usepackage{makecell}
\usepackage{bm}
\usepackage{caption}
\usepackage{xcolor}
\usepackage{tikz}
\definecolor{mypink}{RGB}{243, 196, 255} 
\definecolor{myblue}{RGB}{207, 233, 244} 
\definecolor{tp}{RGB}{0,118,121} 
\definecolor{tn}{RGB}{247,238,238} 
\definecolor{fp}{RGB}{255,171,96} 
\definecolor{fn}{RGB}{239,95,112} 
\definecolor{bareland_color}{rgb}{0.502,0.0,0.0}
\definecolor{cropland_color}{rgb}{0.655, 0.733,0.106}
\definecolor{vegetation_color}{rgb}{0.274,0.710,0.474}
\definecolor{water_color}{rgb}{0.110,0.549,0.741}
\definecolor{building_color}{rgb}{0.710, 0.274, 0.274}
\definecolor{road_color}{rgb}{0.870,0.721,0.274}
\definecolor{developed_space_color}{rgb}{0.580,0.580,0.580}
\definecolor{background_color}{rgb}{0.949,0.937,0.914}

\ifCLASSINFOpdf

\else
 
\fi

\begin{document}

\captionsetup{font = {small}}

\title{ChangeMamba: Remote Sensing Change Detection with Spatio-Temporal State Space Model}

\author{
        Hongruixuan~Chen\dag,~\IEEEmembership{Graduate Student Member,~IEEE,}
        Jian~Song\dag,
        Chengxi~Han,~\IEEEmembership{Graduate Student Member,~IEEE,}
        Junshi~Xia,~\IEEEmembership{Senior Member,~IEEE,}
        Naoto~Yokoya,~\IEEEmembership{Member,~IEEE}
        
\thanks{Manuscript accepted in June 20, 2024. This work was supported in part by the Council for Science, Technology and Innovation (CSTI), the Cross-ministerial Strategic Innovation Promotion Program (SIP), Development of a Resilient Smart Network System against Natural Disasters (Funding agency: NIED), the JSPS, KAKENHI under Grant Number 22H03609, JST, FOREST under Grant Number JPMJFR206S, Microsoft Research Asia, and Next Generation AI Research Center, The University of Tokyo. \emph{(\dag: equal contribution; Corresponding author: Naoto Yokoya)}}
\thanks{Hongruixuan Chen, Jian Song, and Naoto Yokoya are with Graduate School of Frontier Sciences, The University of Tokyo, Chiba, Japan (e-mail: qschrx@gmail.com; song@ms.k.u-tokyo.ac.jp; yokoya@k.u-tokyo.ac.jp).}
\thanks{Chengxi Han is with State Key Laboratory of Information Engineering in Surveying, Mapping, and Remote Sensing, Wuhan University, Wuhan, China (e-mail: chengxihan@whu.edu.cn).}
\thanks{Junshi Xia is with RIKEN Center for Advanced Intelligence Project (AIP), RIKEN, Tokyo, 103-0027, Japan (e-mail: junshi.xia@riken.jp).}

}

\markboth{IEEE TRANSACTIONS ON GEOSCIENCE AND REMOTE SENSING, 2024}%
{Shell \MakeLowercase{\textit{\emph{et al.}}}: Bare Demo of IEEEtran.cls for IEEE Journals}

\maketitle

\begin{abstract}
Convolutional neural networks (CNN) and Transformers have made impressive progress in the field of remote sensing change detection (CD). However, both architectures have inherent shortcomings: CNN are constrained by a limited receptive field that may hinder their ability to capture broader spatial contexts, while Transformers are computationally intensive, making them costly to train and deploy on large datasets. Recently, the Mamba architecture, based on state space models, has shown remarkable performance in a series of natural language processing tasks, which can effectively compensate for the shortcomings of the above two architectures. In this paper, we explore for the first time the potential of the Mamba architecture for remote sensing CD tasks. We tailor the corresponding frameworks, called MambaBCD, MambaSCD, and MambaBDA, for binary change detection (BCD), semantic change detection (SCD), and building damage assessment (BDA), respectively. All three frameworks adopt the cutting-edge Visual Mamba architecture as the encoder, which allows full learning of global spatial contextual information from the input images. For the change decoder, which is available in all three architectures, we propose three spatio-temporal relationship modeling mechanisms, which can be naturally combined with the Mamba architecture and fully utilize its attribute to achieve spatio-temporal interaction of multi-temporal features, thereby obtaining accurate change information. On five benchmark datasets, our proposed frameworks outperform current CNN- and Transformer-based approaches without using any complex training strategies or tricks, fully demonstrating the potential of the Mamba architecture in CD tasks. Further experiments show that our architecture is quite robust to degraded data. The source code is available in \url{https://github.com/ChenHongruixuan/MambaCD}.
\end{abstract}

\begin{IEEEkeywords}
  State space model, Mamba, binary change detection, semantic change detection, building damage assessment, spatio-temporal relationship, optical high-resolution images
\end{IEEEkeywords}

\IEEEpeerreviewmaketitle

\section{Introduction}\label{sec:1}
\par \IEEEPARstart{C}{hange} detection (CD) has been a popular field within the remote sensing community since the inception of remote sensing technology. It aims to detect changes in objects on the Earth's surface from multi-temporal remote sensing images acquired at different times. Depending on the desired result from the change detector, CD tasks can be categorized into three types, namely binary CD (BCD), semantic CD (SCD), and building damage assessment (BDA). Now, CD techniques play an important role in many fields, including land cover change analysis, urban sprawl studies, disaster response, geographic information system (GIS) updating and ecological monitoring \cite{Lu2004, Coppin2004, ZHENG2021Building, guo2021deep, chen2022unsupervised, chen2023land, Xiao2024TTST}.

\par Optical high-resolution remote sensing imagery has become one of the most applied and researched data sources in the field of CD, which can provide detailed textural and geometric structural information about surface features, allowing us to achieve more refined changes \cite{ZHENG2021Building}. Due to the increased heterogeneity within the same land-cover feature brought about by the increase in spatial resolution, traditional pixel-based CD methods \cite{Wu2014, Nielsen2007, Sharma2007} are difficult to achieve satisfactory detection results. To cope with this, researchers proposed object-based CD methods \cite{Hussain2013, Sun2021a, Chen2023Fourier}, which take the object consisting of homogeneous pixels as the basic unit for CD. However, these methods rely on hand-designed “shallow” features, which are not representative and robust enough to cope with complex ground conditions in multi-temporal high-resolution images \cite{Wu2022Unsupervised,Xiao2024EDiffSR}.

\par The emergence of deep learning has brought new models and paradigms for CD, greatly improving the efficiency and accuracy of CD. After the early days of developing image patch-based approaches with tiny deep learning models \cite{Gong2017Superpixel, Chen2019a, Chen2019Deep}, the release of a number of large-scale benchmark datasets \cite{Shi2022Deeply, Yang2022Asymmetric, Song_2024_WACV} has allowed us to start designing and introducing larger and deeper models. Convolutional neural network (CNN)-based methods have been dominant within the field of CD since the introduction of fully convolutional networks (FCNs) to the field by Daudt \emph{et al.} \cite{CayeDaudt2018}. By employing models within the computer vision field as well as a priori knowledge of the CD task, many representative methods have been proposed \cite{Zhang2020, Zheng2022, ZHENG2021Building, guo2021deep, CAO2023full, Lv2023Hierarchical, Liu2022Building}. Despite the decent results achieved by these methods, the inherent shortcomings of the CNN model structure, \emph{i.e.}, the inability of the limited receptive field to capture long-range dependencies between pixels, make these methods still fall short when dealing with complex and diverse multi-temporal scenes in images with different spatial-temporal resolutions \cite{Chen2022Remote}. The emergence of the visual Transformer \cite{dosovitskiy2020image} provides an effective way to solve the above problems. By means of the stacked self-attention modules, the Transformer architecture can fully model the relationship between all the pixels of the whole image. Currently, more and more CD architectures adopt Transformer architectures as the encoder to extract representative and robust features \cite{Bandara2022Transformer, Chen2022Dual, Zhang2022SwinSUNet, Zhang2023Relation} and utilize it in the decoder to capture the spatio-temporal relationships between multi-temporal features \cite{Chen2022Remote, LiTransUNetCD2022}, exhibiting superior performance. 

\par Nonetheless, the self-attention operation demands quadratic complexity in terms of image sizes, resulting in expensive computational overhead, which is detrimental in dense prediction tasks, e.g., land-cover mapping, object detection, and CD, in large-scale remote sensing datasets. Some available solutions, such as limiting the size of the computational window or the step size to improve attention efficiency, come at the cost of imposing limits on the size of the reception field \cite{Liu_2021_ICCV, Xie2021SegFormer}. As a viable alternative to the Transformer architecture, state space models (SSMs), especially structured state space sequence models (S4) \cite{gu2021efficiently}, have shown cutting-edge performance in continuous long-sequence data analysis, with the promising attribute of linear scaling in sequence length. The Mamba architecture \cite{gu2023mamba} further improves on the S4 model by employing a selection mechanism that allows the model to select relevant information in an input-dependent manner. By combining it with hardware-aware implementations, Mamba outperforms Transformers on a number of downstream tasks with high efficiency. Very recently, Mamba architecture has been extended to image data and shown promising results on some visual tasks \cite{liu2024vmamba, zhu2024vision}. However, its potential for high-resolution optical images and different CD subtasks is still under-explored. How to design a suitable architecture based on a priori knowledge of CD tasks is undoubtedly valuable for subsequent research in the community.

\par In this paper, we present the first attempt at introducing Mamba architecture for CD tasks and propose several spatio-temporal state space models (collectively called ChangeMamba) to realize effective and robust CD. ChangeMamba can effectively model the global spatial context and spatio-temporal relationships in multi-temporal images, thereby showing very promising results on the three subtasks of CD, namely BCD, SCD, and BDA. Specifically, ChangeMamba is based on the recently proposed VMamba architecture \cite{liu2024vmamba}, which adopts a Cross-Scan Module (CSM) to unfold image patches in different spatial directions to achieve effective modeling of global contextual information from images. Since CD tasks require the detector to adequately learn spatio-temporal features from multi-temporal images, we design three spatio-tepmoral relationship modeling mechanisms and naturally combine them with Mamba architecture. The proposed three mechanisms can adequately model the spatio-temporal relationships between multi-temporal features by scanning them in different ways, and thus effectively detect the different categories of changes, including binary changes, semantic changes, and building damage levels.

\par In summary, the main contributions of this paper are as follows:
\begin{enumerate}
    \item For the first time, we explore the application of the Mamba architecture in the field of remote sensing change detection, thereby achieving high accuracy, efficient, and robust CD. 
    \item Based on the Mamba architecture and incorporating the unique characteristics of CD tasks, we design the corresponding network frameworks for each of the three CD tasks, \emph{i.e.}, BCD, SCD, and BDA.
    \item Three spatio-temporal relationship modeling mechanisms are proposed, tailored for CD tasks, with the aid of the Mamba architecture to fully learn spatio-temporal features. 
    \item The proposed three frameworks show very competitive and even SOTA performance on five benchmark datasets, demonstrating their superiority. The source code for this work is publicly available for contributions to subsequent possible research.
\end{enumerate}

\par The remainder of this paper is organized as follows. Section \ref{sec:2} reviews related work. Section \ref{sec:3} details the three ChangeMamba architectures proposed for BCD, SCD, and BDA, respectively. The experimental results and discussion are provided in Sections \ref{sec:4}. Section \ref{sec:5} draws the conclusion.

\section{Related Works}\label{sec:2}
\par In this section, we focus on reviewing deep learning-based CD methods using optical high-resolution remote sensing images as data sources. 

\subsection{CNN-Based Method}
\par The rapid evolution of CNN and its prowess in automatically extracting hierarchical features have led to wide applications in CD tasks. After the early stage of the patch-based framework \cite{Gong2017Superpixel, Chen2019Deep}, Daudt \emph{et al.} \cite{CayeDaudt2018} presented the first attempt to introduce FCNs into the BCD field and three basic FCNs architectures were developed. After that, many representative network architectures were proposed. Zhang \emph{et al.} \cite{Zhang2020} sought to tailor networks for semantic segmentation tasks more suitable for the BCD task by integrating feature change maps as a deep supervised signal, thus refining the final prediction. To make the framework suitable for handling multi-source images, Chen \emph{et al.} \cite{Chen2019a} proposed a framework by combining long- and short-term memory (LSTM) networks with CNNs. Furthermore, Fang \emph{et al.} \cite{Fang2022SNUNet} designed a Siamese architecture with denser connections to facilitate more effective shallow information exchange for BCD. To combat the challenge of uneven foreground-background distributions within BCD, Han \emph{et al.} \cite{Han2023HANet} employed a novel sampling approach and an attention mechanism to integrate global information more effectively. After that, they introduced an innovative self-attention module that not only enhances long-distance dependency, but also specifically targets the refinement of BCD predictions \cite{Han2023CGNet}. Zhao \emph{et al.} \cite{Zhao2023Exchanging} explored the effect of different fusion strategies on BCD performance and designed a dual encoder-decoder architecture to improve detection performance. To solve the problem of insufficient samples, the transfer learning technique was also introduced. Guo \emph{et al.} \cite{guo2021deep} developed an attention-based network for semi-supervised building updating. Cao \emph{et al.} \cite{CAO2023full} designed a full-level fused cross-task transfer learning architecture for building change detection. Recently, Chen \emph{et al.} \cite{Chen2023Exchange} has developed a sample generation method by simply exchanging image patches, which can train change detectors on single-temporal images in an unsupervised manner.

\par Compared to BCD, which only aims to identify “where” changes have occurred, SCD further seeks to determine “what” the changes are. Although this poses a greater challenge, it is more important for practical application analysis \cite{Rodrigo2019Multitask, Zheng2022}. 
Early explorations based on CNNs for SCD is the study \cite{Rodrigo2019Multitask}, which adopts a multitask learning framework to simultaneously predict binary change maps and land cover maps, using the latter to assist in generating the final semantic change prediction. Mou \emph{et al.} \cite{Mou2019} proposed to combine recurrent neural network (RNN) and CNN for SCD. Yang \emph{et al.} \cite{Yang2022Asymmetric} introduced a benchmark dataset for SCD along with a new evaluation metric. Ding \emph{et al.} \cite{Ding2022Semantic} proposed a more powerful architecture to address the insufficient exchange of information between the semantic and change decoders. A new loss function was introduced to force the network to correlate two decoders. Zheng \emph{et al.} \cite{Zheng2022} continued to strengthen the interactions between multi-temporal features, designing a temporal-symmetric transformer (TST) module to model the spatio-temporal relationships. In \cite{peng2021scdnet}, channel attention was used to embed change information into temporal features. In \cite{tian2023temporal}, temporal salient features were fused to extract and improve the representation of changes. In \cite{Saha2019} and \cite{Wu2022Unsupervised}, unsupervised SCD methods were also studied, using pre-trained CNNs or unsupervised learning techniques to extract hierarchical features, then comparing these features to determine changes. 

\par Due to the lack of suitable benchmark datasets caused by the scarcity of disaster events and difficulty in annotation, the methodology progression of BDA has been relatively slow. The establishment of the xBD benchmark dataset in the xView2 Challenge \cite{Gupta_2019_CVPR_Workshops}, in turn, provides a large-scale benchmark for the field. The procedural approach to BDA typically encompasses two primary components: identifying the location of the building and assessing the level of damage incurred. In the xBD dataset, there are four damage levels, i.e., “no damage”, “minor damage”, “major damage”, and “destroyed”. This bifurcation essentially positions BDA as a nuanced subset of SCD, with a focus on quantifying damage levels. Among established approaches, the xView2 challenge \cite{Gupta_2019_CVPR_Workshops} introduced a baseline method that employs a residual UNet for building localization and damage classification, trained with pre and post-disaster imagery. The Siamese-UNet, the winning approach of the xView2 challenge \cite{Gupta_2019_CVPR_Workshops}, utilized a UNet framework for both pixel-level building localization and damage-level classification. It initialized the UNet for the damage classification task with the parameters from the UNet for the building localization task. A further advancement \cite{weber2020building} utilized Mask-RCNN architecture \cite{He2017Mask} to address BDA, incorporating Siamese encoders, multitask heads, and instance-level predictions. Zheng \emph{et al.} \cite{ZHENG2021Building} introduced a deep object-based architecture that encompasses shared and task-specific encoders, alongside an FCN-based multitask decoder. This development enhances task integration with an object-based post-processing strategy, resulting in more refined outcomes.

\par The development of CNN-based CD methods has matured significantly over time. However, despite their excellent ability to extract local features, CNN architectures inherently struggle to capture long-distance dependencies \cite{dosovitskiy2020image}. This limitation becomes particularly critical in CD tasks, where changed regions, compared to background areas, are often sparse and scattered. The network's ability to model relationships beyond local areas is essential for accurately identifying these changed regions \cite{Chen2022Remote, Bandara2022Transformer}. 


\subsection{Transformer-Based Method}
\par Recently, the promising results have been obtained by Transformers in the computer vision field \cite{dosovitskiy2020image, Xie2021SegFormer}. Their superior long-distance modeling capabilities have made them effectively overcome the challenges faced by the CNNs. As a result, many studies have begun to address CD tasks using Transformer architectures. In the field of BCD, Chen \emph{et al.} \cite{Chen2022Remote} emerged as a pioneering approach, transforming multi-temporal images into semantic tokens to model spatial-temporal relationships within a token-based framework, enhancing the CD process. Bandara \emph{et al.} \cite{Bandara2022Transformer} proposed a pure Transformer-based Siamese network, which has a Siamese Transformer encoder coupled with a multilayer perceptron (MLP) decoder. This architecture eliminates the need for a CNN-based feature extractor. Similarly, Zhang \emph{et al.} \cite{Zhang2022SwinSUNet} proposed a pure Transformer-driven model, adopting a Siamese U-shaped structure to fully recover detailed change information. Li \emph{et al.} \cite{LiTransUNetCD2022} introduced a hybrid model that blends the global contextual modeling capability of Transformers with the CNN block and UNet architecture. It eliminates UNet's redundant information and enhances feature quality through a difference enhancement module, leading to more precise detection results.

\par For Transformer-based SCD, Niu \emph{et al.} \cite{Niu2023SMNet} combined preactivated residual and Transformer blocks to extract semantic features from multi-temporal image pairs, which is crucial for detecting subtle changes. The model's efficacy is further amplified by multitask prediction branches and a custom loss function. In addition, Ding \emph{et al.} \cite{Ding2024Joint} introduced the semantic change Transformer, a CSWin Transformer adaptation, to explicitly learn semantic transitions. The model has a triple encoder-decoder architecture to enhance spatial features and incorporates spatio-temporal dynamic and task-specific prior information, leading to a decrease in learning disparities. This approach significantly boosts accuracy on benchmarks, adeptly managing scenarios with sparse labels.

\par Transformer-based BDA methods are currently understudied. Chen \emph{et al.} \cite{Chen2022Dual} proposed the first pure Transformer-based BDA architecture called DamFormer. It has a Siamese Mix Transformer (MiT) \cite{Xie2021SegFormer} encoder and a lightweight all-MLP decoder to efficiently process multi-temporal image pairs. DamFormer employs a multi-temporal fusion module for improved information integration, and a dual-task decoder for precise building localization and damage classification. By adopting strategies like weight sharing and innovative fusion mechanisms, it efficiently handles high-resolution imagery with lower computational costs, making it effective and streamlined for BDA.

\par Although Transformer-based methods have significantly enhanced CD accuracy through their powerful nonlocal modeling capabilities, they come with substantial computational resource demands. Specifically, as the number of tokens increases, the consumption of computational resources escalates quadratically, posing challenges for processing large-scale input, such as optical high-resolution imagery. Recently, the emerging Mamba architecture \cite{gu2023mamba} not only rivals Transformers in non-local modeling capabilities but also maintains computational overhead at a linear growth rate, which opens up the possibility of efficient and accurate CD. 

\begin{figure*}[!t]
  \centering
\includegraphics[width=6.95in]{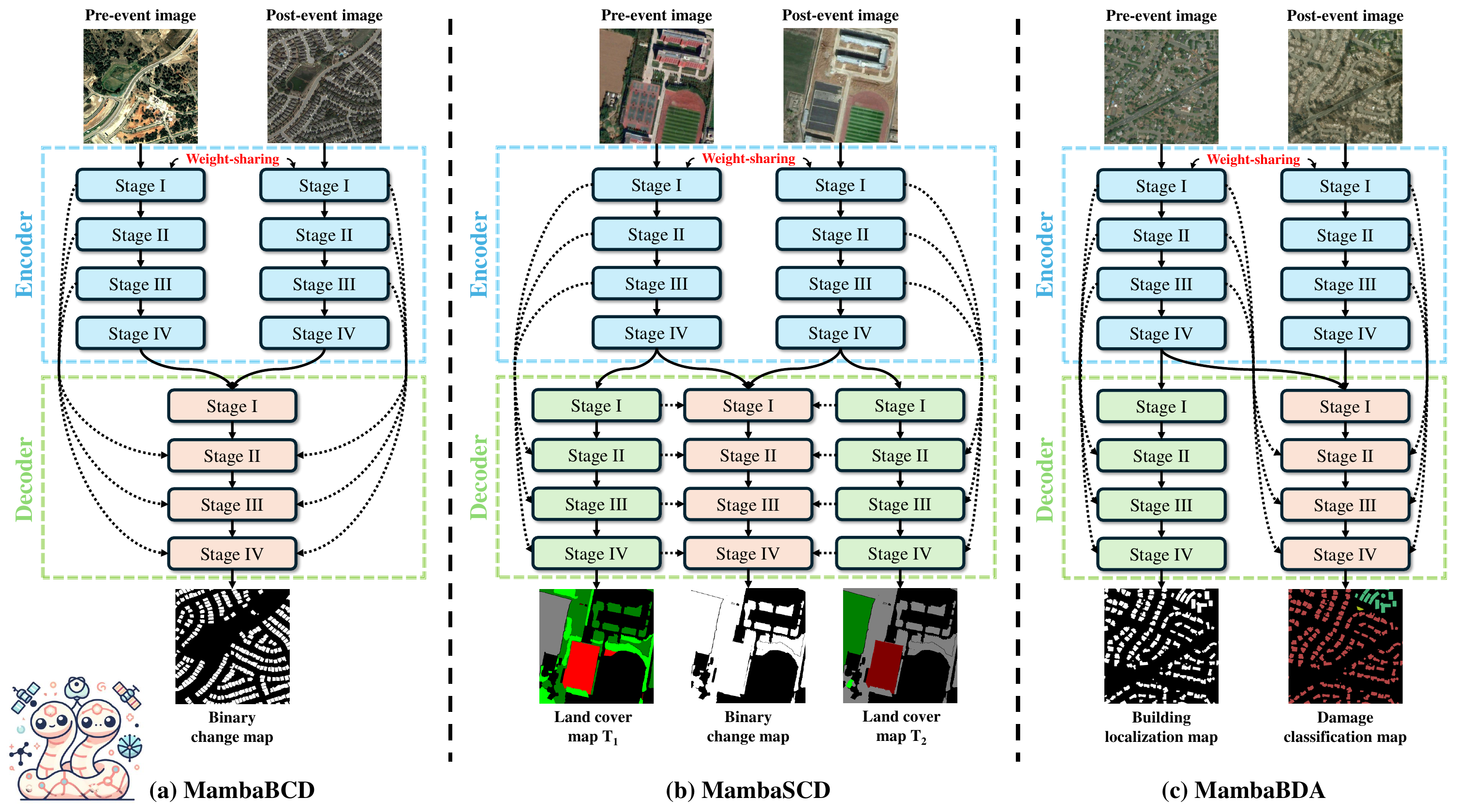}
  \caption{The network framework of the proposed (a) MambaBCD for binary change detection, (b) MambaSCD for semantic change detection, and (c) MamabaBDA for building damage assessment. The specific structure of the encoder and decoder is shown in Figs. \ref{fig:Mamba_encoder}, \ref{fig:Change_decoder} and \ref{fig:Semantic_decoder}.}
  \label{fig:Mamba_framework}
\end{figure*}

\subsection{State Space Model}
\par The SSM concept first popped up with the S4 model \cite{gu2021efficiently}, offering a new way to handle contextual information globally, which caught attention due to its attractive property of scaling linearly in sequence length. Based on the S4 model, Smith \emph{et al.} \cite{smith2022simplified} proposed a new S5 model by introducing MIMO SSM and efficient parallel scan into the S4 model. After that, the H3 model \cite{fu2022hungry} further advanced these concepts, achieving a performance on par with that of Transformers in language modeling tasks. Recently, Gu \emph{et al.} \cite{gu2023mamba} proposed a data-dependent SSM layer and built a generic language model backbone called Mamba, which outperforms Transformers at different scales on large-scale real data and scales linearly in sequence length. Very recently, Visual Mamba \cite{liu2024vmamba} and Vision Mamba \cite{zhu2024vision} have extended the Mamba architecture to 2D image data, showing superior performance on many computer vision tasks. Inspired by this progress, some pioneering work has introduced the Mamba architecture to the field of remote sensing image processing \cite{he2024pan, chen2024rsmamba}. Their studies show that the Mamba architecture can yield performance comparable to the advanced CNN and Transformer architectures on scene classification \cite{chen2024rsmamba} and pansharpening \cite{he2024pan} tasks. However, these works still focus on low-level tasks or classification tasks, and all of these tasks are single-temporal tasks \cite{Chen2023Exchange}. The potential of the Mamba architecture for multi-temporal remote sensing image-related scenarios and dense prediction tasks remains to be explored.

\section{Methodology}\label{sec:3}

\subsection{Preliminaries}
\par The SSM-based models and Mamba are inspired by linear time-invariant systems, which map a 1-D function or sequence
$x\left(t\right) \in \mathcal{R}$ to the response $y\left(t\right) \in \mathcal{R}$ through a hidden state $h\left(t\right) \in \mathcal{R}^{N}$. These systems are usually formulated as linear ordinary differential equations (ODEs) as
\begin{equation}
\left\{
    \begin{aligned}
      &h^{\prime}\left(t\right) =\mathbf{A} h(t)+\mathbf{B} x(t)  \\
      &y(t) =\mathbf{C} h(t)
    \end{aligned}
\right.
\label{eq:ODEs}
\end{equation}
where $\mathrm{\mathbf{A}}\in \mathcal{R}^{N \times N}$ is the evolution parameter and
$\mathrm{\mathbf{B}} \in \mathcal{R}^{N\times 1}, \mathrm{\mathbf{C}} \in \mathcal{R}^{1\times N}$ are the projection parameters.

\par The S4 model \cite{gu2021efficiently} is a discrete counterpart of the continuous system, as continuous systems face significant challenges when integrated into deep learning algorithms. It contains a time scale parameter $\Delta$ that is used to convert the continuous parameters $\mathrm{\mathbf{A}}$ and $\mathrm{\mathbf{B}}$ into their discrete counterparts $\overline{\mathrm{\mathbf{A}}}$ and $\overline{\mathrm{\mathbf{B}}}$. A prevalent way to achieve this transformation is to use a zero-order hold (ZOH) approach:
\begin{equation}
\left\{
    \begin{aligned}
      &\overline{\mathbf{A}}=\exp (\Delta \mathbf{A})  \\
      &\overline{\mathbf{B}}=\left(\Delta \mathbf{A}\right)^{-1}\left(\exp \left(\Delta \mathbf{A}\right)-\mathbf{I}\right)\Delta \mathbf{B}
    \end{aligned}
\right.
\label{eq:ZOH}
\end{equation}

\par After that, the discretization of Eq. (\ref{eq:ODEs}) can be formulated as 
\begin{equation}
\left\{
    \begin{aligned}
      &h^{\prime}_{t} =\overline{\mathbf{A}} h_{t-1}+&\overline{\mathbf{B}} x_{t}  \\
      &y_{t} =\mathbf{C} h_{t}
    \end{aligned}
\right.
\label{eq:discretation}
\end{equation}

\par The output of $x$ with sequence length $\mathbf{L}_{x}$ can be calculated directly by the following global convolution operation
\begin{equation}
\left\{
    \begin{aligned}
      &\overline{\mathbf{K}}=\left(\mathbf{C} \overline{\mathbf{B}}, \mathbf{C} \overline{\mathbf{A}} \overline{\mathbf{B}}, \ldots, \mathbf{C}\overline{\mathbf{A}}^{\mathbf{L}_{x}-1} \overline{\mathbf{B}} \right)  \\
      &\mathbf{y}=\mathbf{x} * \overline{\mathbf{K}}
    \end{aligned}
\right.
\label{eq:global_conv}
\end{equation}
where $\overline{\mathbf{K}} \in \mathcal{R}^{\mathbf{L}_{x}}$ is a structured convolutional kernel and $*$ denotes a convolutional operation. 

\par Mamba architecture \cite{gu2023mamba} builds on the S4 model by proposing a selection mechanism that allows the model to filter out irrelevant information and recall relevant information\footnote{Mamba architectures sometimes are abbreviated as the S6 model since they are S4 models with a selection mechanism and computed with a scan \cite{gu2023mamba}.}. In addition, a hardware-aware algorithm has been developed for recursive computation of the model but will not be implemented in the extended state, optimizing the GPU memory layout. These characteristics allow the Mamba architecture not only to obtain promising results on languages and other data with long sequences, but also to achieve fast training and inference with memory and computational overhead linearly proportional to the sequence length. 

\subsection{Problem Statement}
\par In this paper, we focus on three sub-tasks within the CD field. They are binary change detection (BCD), semantic change detection (SCD), and building damage assessment (BDA). The definitions of the three tasks are as follows.

\subsubsection{Binary Change Detection}
\par BCD is the basic and the most intensively studied task in the CD field. It focuses on “where” change occurs. According to the category of interest, BCD can also be divided into category-agnostic CD focusing on general land-cover change information, and single-category CD, e.g., building CD and forest CD. Given a training set represented as $\mathcal{D}^{bcd}_{train} = \left\{\left(\mathrm{\mathbf{X}}^{\mathrm{T}_{1}}_{i}, \mathrm{\mathbf{X}}^{\mathrm{T}_{2}}_{i}, \mathrm{\mathbf{Y}}^{bcd}_{i}\right)\right\}_{i=1}^{N^{bcd}_{train}}$, where $\mathrm{\mathbf{X}}^{\mathrm{T}_{1}}_{i}, \mathrm{\mathbf{X}}^{\mathrm{T}_{2}}_{i}\in \mathcal{R}^{ \mathrm{H}\times  \mathrm{W}\times C}$ is the $i$-th multi-temporal image pair acquired in $\mathrm{T}_{1}$ and $\mathrm{T}_{2}$, respectively, and $\mathrm{\mathbf{Y}}^{bcd}_{i} \in \left\{0,1\right\}^{\mathrm{H}\times  \mathrm{W}}$ is the corresponding label, the goal of BCD is to train a change detector $\mathcal{F}_{\theta}^{bcd}$ on $\mathcal{D}^{bcd}_{train}$ that can predict change maps reflecting accurate “change / non-change” binary information on new sets. 

\subsubsection{Semantic Change Detection}
\par Compared to BCD, SCD focuses not only on “where” the change occurs, but also on the “what” of the change, i.e. “from-to” semantic change information \cite{Ding2022Semantic}. The training set in the SCD task can be represented as $\mathcal{D}^{scd}_{train} = \left\{\left(\mathrm{\mathbf{X}}^{\mathrm{T}_{1}}_{i}, \mathrm{\mathbf{X}}^{\mathrm{T}_{2}}_{i}, \mathrm{\mathbf{Y}}^{\mathrm{T}_{1}}_{i}, \mathrm{\mathbf{Y}}^{\mathrm{T}_{2}}_{i}, \mathrm{\mathbf{Y}}^{bcd}_{i}\right)\right\}_{i=1}^{N_{train}}$. Compared to BCD, the corresponding land-cover labels $\mathrm{\mathbf{Y}}^{\mathrm{T}_{1}}_{i}, \mathrm{\mathbf{Y}}^{\mathrm{T}_{2}}_{i}\in \left\{0,1,\cdots, \mathbf{C}^{lcm} \right\}^{\mathrm{H}\times  \mathrm{W}}$ with $\mathbf{C}^{lcm}$ categories of $\mathrm{\mathbf{X}}^{\mathrm{T}_{1}}_{i}$ and $\mathrm{\mathbf{X}}^{\mathrm{T}_{2}}_{i}$ are additionally required. The goal of SCD is to train a semantic change detector $\mathcal{F}_{\theta}^{scd}$ on $\mathcal{D}^{scd}_{train}$ that can predict land-cover maps of multi-temporal images and binary change maps between them on new sets as accurately as possible. By combining the information in predicted land-cover maps and binary change maps, the from-to" semantic change information can be derived.

\subsubsection{Building Damage Assessment}
\par BDA is a special “one-to-many” SCD task \cite{ZHENG2021Building}. BDA requires recognizing not only the “where” of the change (damage) occurs but also post-event states of the object of interest (building) given the pre-event state of the object of interest. The training set in the BDA task can be denoted as $\mathcal{D}^{bda}_{train} = \left\{\left(\mathrm{\mathbf{X}}^{\mathrm{T}_{1}}_{i}, \mathrm{\mathbf{X}}^{\mathrm{T}_{2}}_{i}, \mathrm{\mathbf{Y}}^{loc}_{i}, \mathrm{\mathbf{Y}}^{clf}_{i}\right)\right\}_{i=1}^{N_{train}}$, where $\mathrm{\mathbf{Y}}^{loc}_{i} \in \left\{0,1\right\}^{\mathrm{H}\times  \mathrm{W}}$ is the label of the object-of-interest (building) at $\mathrm{T}_{1}$, and $\mathrm{\mathbf{Y}}^{clf}_{i} \in \left\{0,1,\cdots, \mathbf{C}^{dam}\right\}^{\mathrm{H}\times  \mathrm{W}}$ is the post-event state (damage level) of the object of interest at $\mathrm{T}_{2}$.

\begin{figure*}[!t]
  \centering
\includegraphics[width=6.0in]{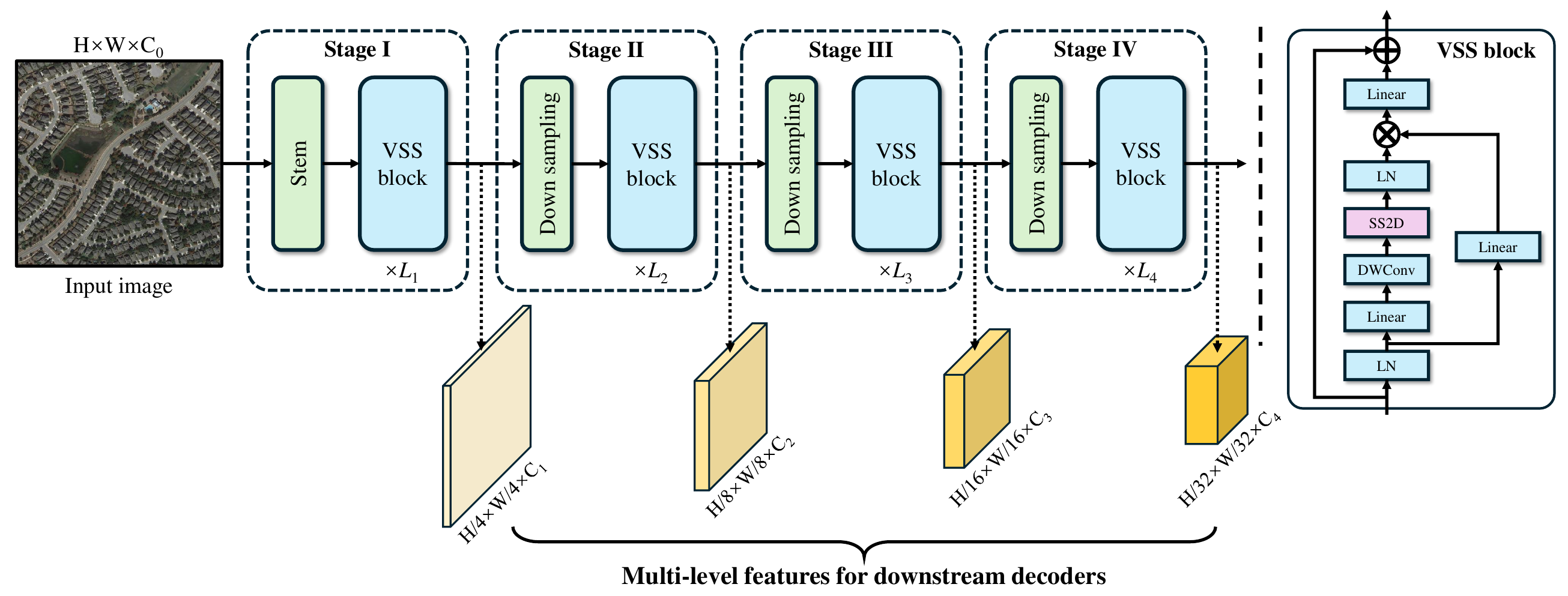}
  \caption{The specific structure of the encoder network is based on the visual state space model used in our three architectures. }
  \label{fig:Mamba_encoder}
\end{figure*}

\begin{figure}[!t]
  \centering
  \subfloat[]{
    \includegraphics[width=2.2in]{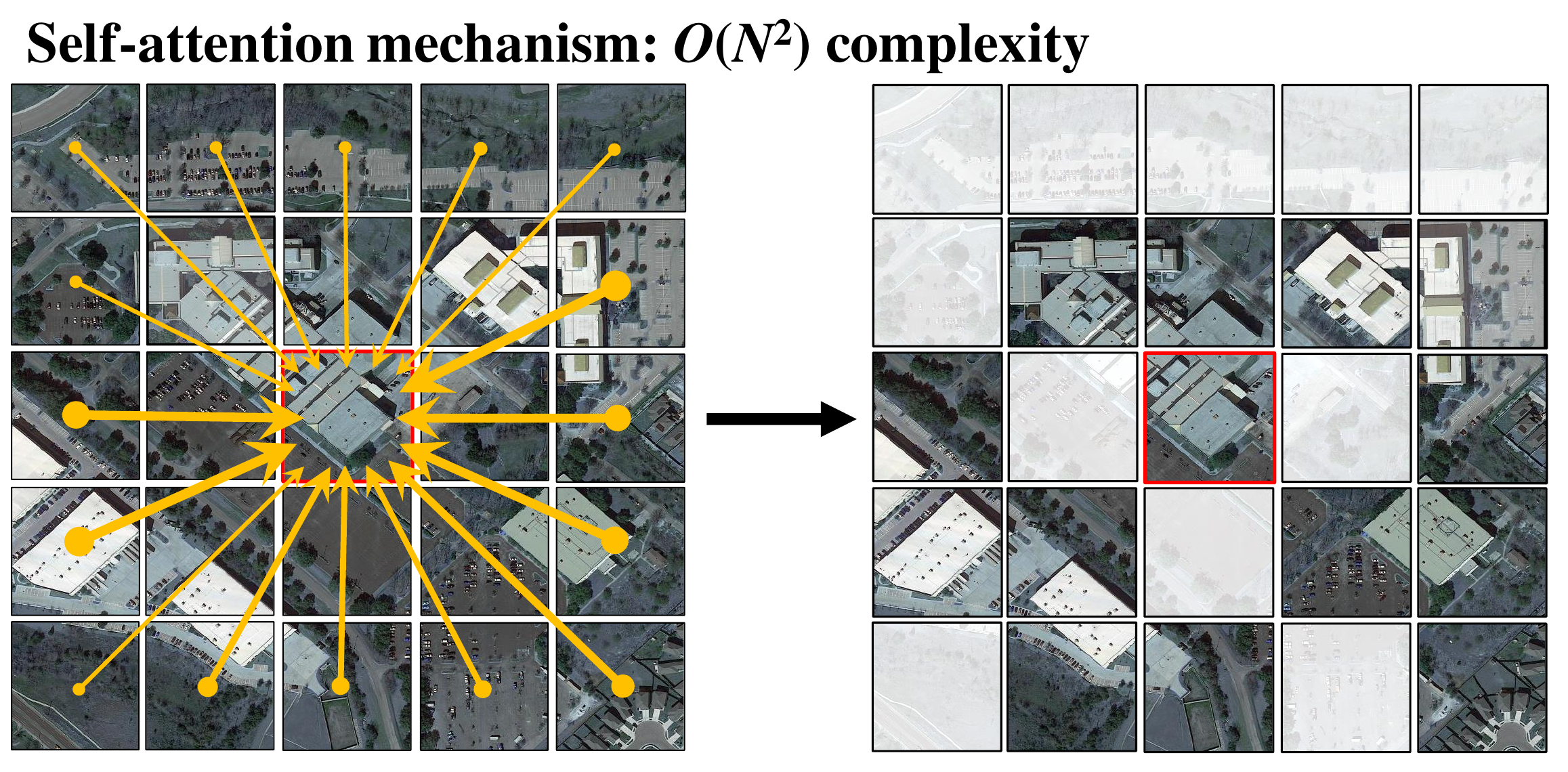}
  \label{fig_second_case}}  

  \subfloat[]{
    \includegraphics[width=3.3in]{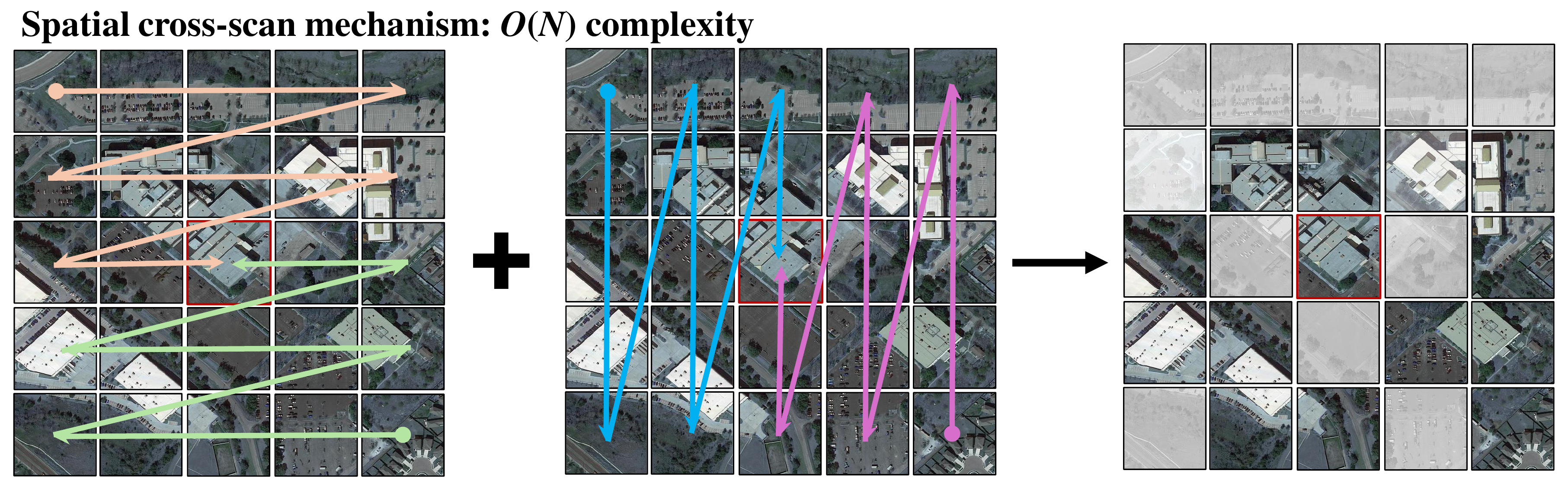}
  \label{fig_first_case}}
  \caption{Illustration of the (a) self-attention mechanism \cite{Vaswani2017Attention, dosovitskiy2020image} and (b) cross-scan mechanism \cite{liu2024vmamba} to capture global contextual information. }
  \label{fig:attention_compare}
\end{figure}

\subsection{Network Architecture}
\par According to the commonalities and properties of the three CD subtasks, we design the corresponding network frameworks based on the Mamba architecture, called MambaBCD, MambaSCD, and MamabBDA, respectively. Fig. \ref{fig:Mamba_framework} shows the network framework of these three architectures. Among them, the encoders of the three networks are weight-sharing siamese networks based on the Visual State Space Model (VMamba) architecture \cite{liu2024vmamba}. VMamba can adequately extract the robust and representative features of the input images that benefit from the Mamba architecture and an efficient 2D cross-scan mechanism (as shown in Fig. \ref{fig:attention_compare}) for downstream tasks. All three architectures have a change detector of the same structure for learning spatio-temporal relationships from the features extracted by the encoder. In addition, MambaSCD and MambaBDA additionally possess semantic decoders for other tasks, \emph{i.e.}, land-cover classification and building localization. The specific structure of the encoders and decoders in our architectures will be elaborated in Sections \ref{sec:encoder} and \ref{sec:decoder}. In this subsection, we focus on the input, output, and internal information flow of these three network architectures.

\subsubsection{MambaBCD}
\par MambaBCD is the architecture designed for the BCD task. First, the siamese encoder network denoted as $\mathcal{F}^{enc}_{\theta}$ extracts multi-level features from the input multi-temporal images as $\left\{\mathbf{F}^{\mathrm{T}_{1}}_{i,j}\right\}^{4}_{j=1}=\mathcal{F}^{enc}_{\theta}\left(\mathrm{\mathbf{X}}_{i}^{\mathrm{T}_{1}}\right)$ and $\left\{\mathbf{F}^{\mathrm{T}_{2}}_{i,j}\right\}^{4}_{j=1}=\mathcal{F}^{enc}_{\theta}\left(\mathrm{\mathbf{X}}_{i}^{\mathrm{T}_{2}}\right)$. Next, these multi-level features are fed into a tailored change decoder $\mathcal{F}^{cdec}_{\theta}$. Based on the Mamba architecture, the change decoder can fully learn the spatio-temporal relationship from multi-level features through three different mechanisms, and gradually obtains an accurate BCD result, formulated as $\mathrm{\mathbf{P}}_{i}^{bcd}=\mathcal{F}^{cdec}_{\theta}\left(\left\{\mathbf{F}^{\mathrm{T}_{1}}_{i,j}\right\}^{4}_{j=1}, \left\{\mathbf{F}^{\mathrm{T}_{2}}_{i,j}\right\}^{4}_{j=1}\right)$. The binary change map can be obtained as $\hat{\mathbf{Y}}^{bcd}_{i} = \text{argmax}_{c}\mathrm{\mathbf{P}}_{i}^{bcd}$.

\subsubsection{MambaSCD}
\par MambaSCD is the architecture designed for the SCD task. As shown in Fig. \ref{fig:Mamba_framework}-(b), MambaSCD adds two additional semantic decoders for land cover mapping tasks based on MambaBCD, denoted as $\mathcal{F}^{\mathrm{T}_{1}}_{\theta}$ and $\mathcal{F}^{\mathrm{T}_{2}}_{\theta}$. In addition to being treated as input as $\mathcal{F}^{cdec}$ to learn spatio-temporal relationships to predict BCD results, the multi-level features extracted by the encoder are also fed into $\mathcal{F}^{\mathrm{T}_{1}}_{\theta}$ and $\mathcal{F}^{\mathrm{T}_{2}}_{\theta}$ to predict the land-cover map of the corresponding temporal image, formulated as $\mathrm{\mathbf{P}}_{i}^{\mathrm{T}_{1}}=\mathcal{F}^{\mathrm{T}_{1}}_{\theta}\left(\left\{\mathbf{F}^{\mathrm{T}_{1}}_{i,j}\right\}^{4}_{j=1}\right)$ and $\mathrm{\mathbf{P}}_{i}^{\mathrm{T}_{2}}=\mathcal{F}^{\mathrm{T}_{2}}_{\theta}\left(\left\{\mathbf{F}^{\mathrm{T}_{2}}_{i,j}\right\}^{4}_{j=1}\right)$. After obtaining the land-cover maps $\hat{\mathbf{Y}}^{\mathrm{T}_{1}}_{i} = \text{argmax}_{c}\mathrm{\mathbf{P}}_{i}^{\mathrm{T}_{1}}$ and $\hat{\mathbf{Y}}^{\mathrm{T}_{2}}_{i} = \text{argmax}_{c}\mathrm{\mathbf{P}}_{i}^{\mathrm{T}_{2}}$ and the binary change map $\hat{\mathbf{Y}}^{bcd}_{i}$, the semantic change information of $\mathrm{T}_{1}\rightarrow \mathrm{T}_{2}$ can be obtained by performing the mask operation on $\hat{\mathbf{Y}}^{\mathrm{T}_{1}}_{i}$ and $\hat{\mathbf{Y}}^{\mathrm{T}_{2}}_{i}$ using $\hat{\mathbf{Y}}^{bcd}_{i}$ \cite{Rodrigo2019Multitask, Zheng2022, chen2023land}.

\subsubsection{MambaBDA}
\par MambaBDA is the architecture presented for the BDA task. Unlike the typical SCD task, the BDA task only needs to predict land-cover maps (building location maps) for pre-event images. Therefore, in the MambaBDA architecture shown in Fig. \ref{fig:Mamba_framework}-(c), there is only one semantic decoder $\mathcal{F}^{loc}_{\theta}$ to predict the building localization map as $\mathrm{\mathbf{P}}_{i}^{loc}=\mathcal{F}^{loc}_{\theta}\left(\left\{\mathbf{F}^{\mathrm{T}_{1}}_{i,j}\right\}^{4}_{j=1}\right)$. Like MambaBCD and MambaSCD, a change decoder denoted as $\mathcal{F}^{clf}_{\theta}$ then learns spatio-temporal relationships from multi-temporal features to classify the damage level of buildings as $\mathrm{\mathbf{P}}_{i}^{clf}=\mathcal{F}^{clf}_{\theta}\left(\left\{\mathbf{F}^{\mathrm{T}_{1}}_{i,j}\right\}^{4}_{j=1}, \left\{\mathbf{F}^{\mathrm{T}_{2}}_{i,j}\right\}^{4}_{j=1}\right)$. The obtained building location map $\hat{\mathbf{Y}}^{loc}_{i} = \text{argmax}_{c}\mathrm{\mathbf{P}}_{i}^{loc}$ can be used for further post-processing on the obtained damage classification map $\hat{\mathbf{Y}}^{clf}_{i} = \text{argmax}_{c}\mathrm{\mathbf{P}}_{i}^{clf}$ to improve the accuracy, e.g., object-based post-processing methods \cite{ZHENG2021Building}.

\subsection{Encoder Based on Visual State Space Model}\label{sec:encoder}

\par Mamba architecture causally processes the input data and thus can only capture information within the scanned part of the data. This attribute naturally aligns with natural language data, which poses significant challenges when adapting to non-causal data such as image \cite{liu2024vmamba}. Directly expanding image data along a certain dimension of space and inputting it into the S6 model will result in a lack of adequately modeled spatial contextual information. Recently, VMamba solved this problem well by proposing a 2D cross-scan mechanism. As shown in Fig. \ref{fig:attention_compare}, before inputting the tokens into the S6 model, the cross-scan mechanism rearranges the tokens in spatial dimensions, \emph{i.e.}, top-left to bottom-right, bottom-right to top-left, top-right to bottom-left, and bottom-left to top-right. Finally, the resulting features are then merged. In this way, any pixel can get spatial context information from different directions. Moreover, the computational complexity of the VMamba model under the cross-scan mechanism is still $O\left(N \right)$ compared to the self-attention operation in Transformer. 

\par Based on the VMamba architecture, the specific structure of the encoder network in our three proposed architectures is shown in Fig. \ref{fig:Mamba_encoder}. There are four stages, each of which first downsamples the input data, then fully models the spatial contextual information using a number of visual state space (VSS) blocks, and then outputs the features for that stage, \emph{i.e.}, $\mathbf{F}^{\mathrm{T}_{1}}_{i,j}$ and $\mathbf{F}^{\mathrm{T}_{2}}_{i,j}$. The structure of the VSS block is also depicted in Fig. \ref{fig:Mamba_encoder}. The input first passes through a linear embedding layer and the output is split into two information flows. One flow passes through a 3 $\times$3 depth-wise convolution (DWConv) layer, followed by a Silu activation function \cite{elfwing2018sigmoid} before entering the core SS2D module (i.e., the integration of the S6 model with the cross-scan mechanism). The output of the SS2D module passes through a layer normalization (LN) layer and is then summed up with the outputs of the other streams that have been activated by Silu. This combination produces the final output of the VSS block. Finally, the features extracted for pre-event and post-event images from the four stages $\left\{\mathbf{F}^{\mathrm{T}_{1}}_{i,j}\right\}^{4}_{j=1}$ and $\left\{\mathbf{F}^{\mathrm{T}_{2}}_{i,j}\right\}^{4}_{j=1}$ are then used in the subsequent decoders responsible for specific tasks.

\subsection{Task-Specific Decoders}\label{sec:decoder}

\begin{figure*}[!t]
  \centering
\includegraphics[width=6.95in]{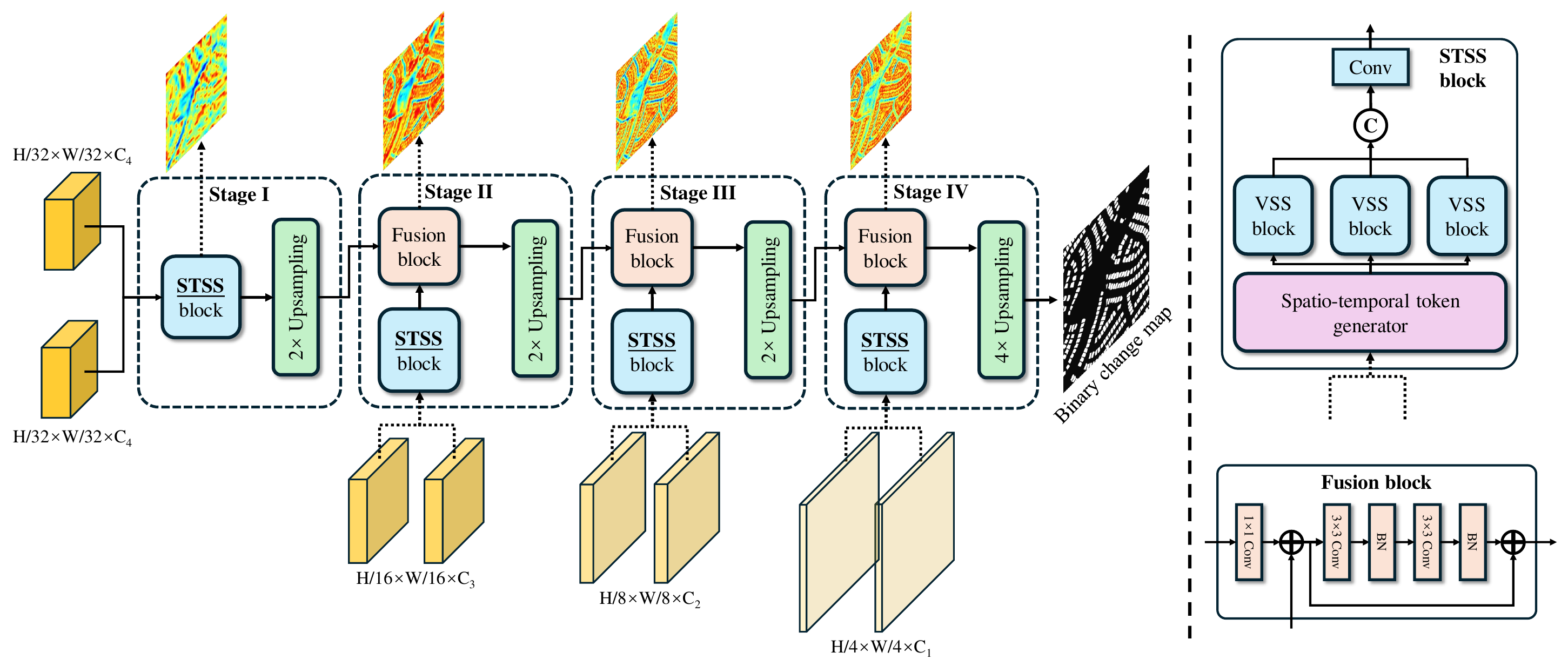}
  \caption{The specific structure of the change decoder used in MambaBCD, MambaSCD, and MambaBDA architectures based on the proposed spatio-temporal relationship modeling mechanisms in Fig. \ref{fig:spatio_temporal_modelling}.}
  \label{fig:Change_decoder}
\end{figure*}

\begin{figure}[!t]
  \centering
\includegraphics[width=3.4in]{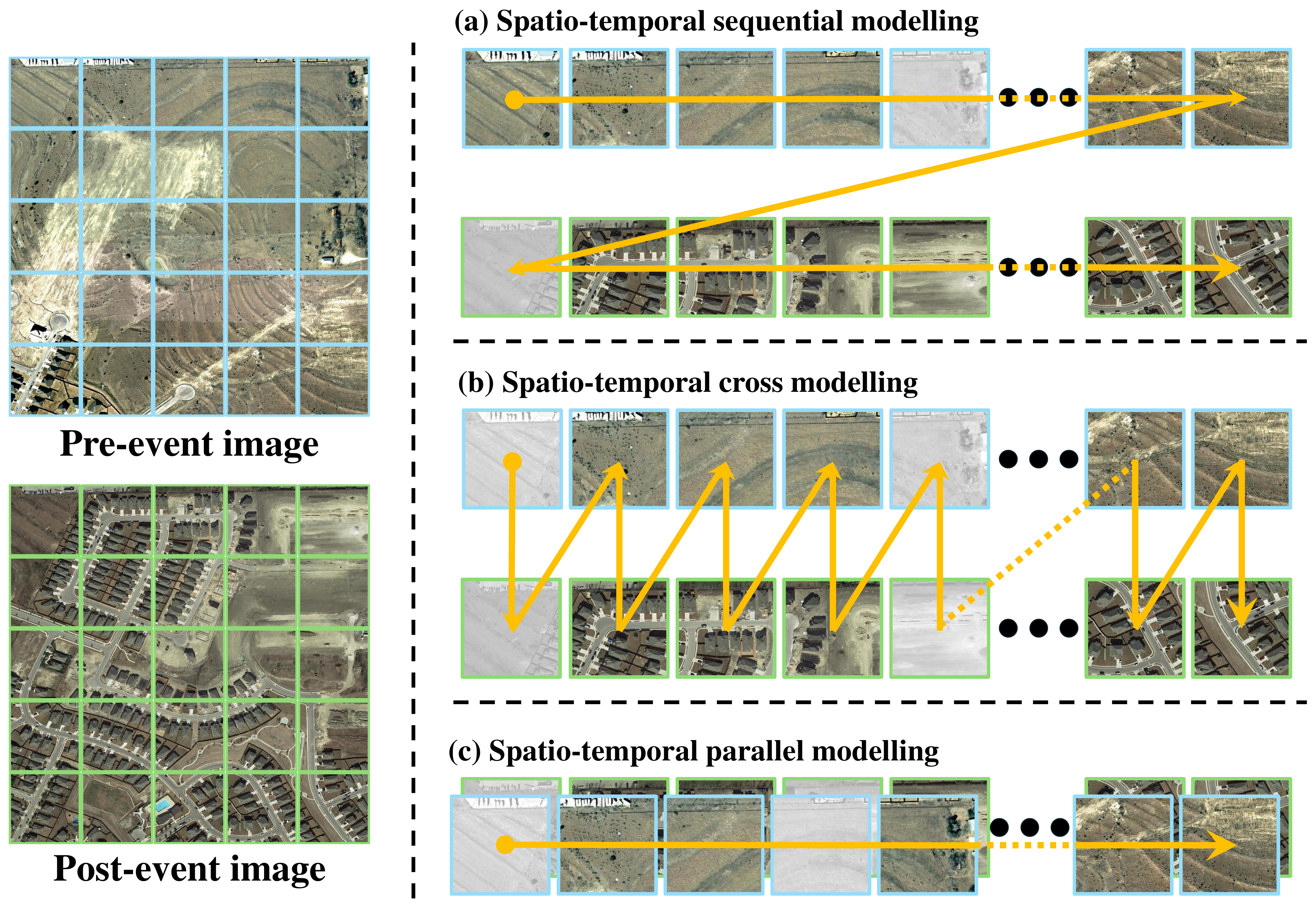}
  \caption{Illustration of three mechanisms for learning spatio-temporal relationships. }
  \label{fig:spatio_temporal_modelling}
\end{figure} 

\subsubsection{Spatio-Temporal Relationship Modeling Mechanism}
\par Although the employed encoder can extract robust and representative features, learning spatio-temporal relationships of multi-temporal images is also significant for CD tasks \cite{Mou2019, Zheng2022}. To this end, we propose three mechanisms for modeling spatio-temporal relationships that can be aligned with the attribute of the S6 model capable of adequately modeling the global contextual information of long sequential data. Fig. \ref{fig:spatio_temporal_modelling} illustrates the three modeling mechanisms. They are spatio-temporal sequential modeling, spatio-temporal cross modeling, and spatial-temporal parallel modeling. Sequential modeling is unfolding the tokens of two temporal phases of data and then arranging them in temporal order as $\mathbf{F}^{seq}_{i,j}=\left[\mathbf{F}^{\mathrm{T}_{1}}_{i,j}\left(1\right), \cdots, \mathbf{F}^{\mathrm{T}_{1}}_{i,j}\left(\frac{\mathrm{H}\mathrm{W}}{2^{1+j}}\right), \mathbf{F}^{\mathrm{T}_{2}}_{i,j}\left(1\right),\cdots,\mathbf{F}^{\mathrm{T}_{2}}_{i,j}\left(\frac{\mathrm{H}\mathrm{W}}{2^{1+j}}\right)\right]$. The cross-modeling mechanism, on the other hand, is modeling by cross-ordering the tokens of the two temporal phases as $\mathbf{F}^{crs}_{i,j}=\left[\mathbf{F}^{\mathrm{T}_{1}}_{i,j}\left(1\right), \mathbf{F}^{\mathrm{T}_{2}}_{i,j}\left(1\right), \cdots \mathbf{F}^{\mathrm{T}_{1}}_{i,j}\left(\frac{\mathrm{H}\mathrm{W}}{2^{1+j}}\right), \mathbf{F}^{\mathrm{T}_{2}}_{i,j}\left(\frac{\mathrm{H}\mathrm{W}}{2^{1+j}}\right)\right]$. Finally, parallel modeling is done by concatenating the tokens of the two temporal phases in the channel dimension and then performing joint modeling, that is, $\mathbf{F}^{pra}_{i,j}=\mathrm{concat}_{c}\left(\mathbf{F}^{\mathrm{T}_{1}}_{i,j}, \mathbf{F}^{\mathrm{T}_{2}}_{i,j}\right)$. Through these three mechanisms and Mamba architecture, the spatio-temporal relationships intrinsic in the multi-temporal features will be fully explored, helping the change decoder to obtain accurate change detection results.

\subsubsection{Change Decoder}

\par Based on the proposed three spatio-temporal learning mechanisms, the specific structure of the change decoder is then shown in Fig. \ref{fig:Change_decoder}. It fully learns the spatio-temporal relationships from the extracted multi-temporal features $\left\{\mathbf{F}^{\mathrm{T}_{1}}_{i,j}\right\}^{4}_{j=1}$ and $\left\{\mathbf{F}^{\mathrm{T}_{2}}_{i,j}\right\}^{4}_{j=1}$ in four stages and obtains accurate binary change maps. At the beginning of each stage, the spatio-temporal relationship of multi-temporal features is first modeled using a spatio-temporal state space (STSS) block. In the STSS block, a spatio-temporal tokens generator module will rearrange the input multi-temporal features, which are then fed into three VSS blocks. Each block is responsible for learning one of the spatio-temporal relationships in Fig. \ref{fig:spatio_temporal_modelling}. The output of the STSS block of the current stage is then integrated with the information from the feature map of the previous stage through a fusion module. In the fusion module, the number of channels of the low-level feature map is mapped to be consistent with the high-level feature map by a 1$\times$1 convolutional layer. Then, the high-level and low-level feature maps are summed. Finally, the resulting feature map is smoothed by a residual layer. After passing through an upsampling layer, the feature map is fed to the next stage. 

\subsubsection{Semantic Decoder}
\begin{figure*}[!t]
  \centering
\includegraphics[width=5.50in]{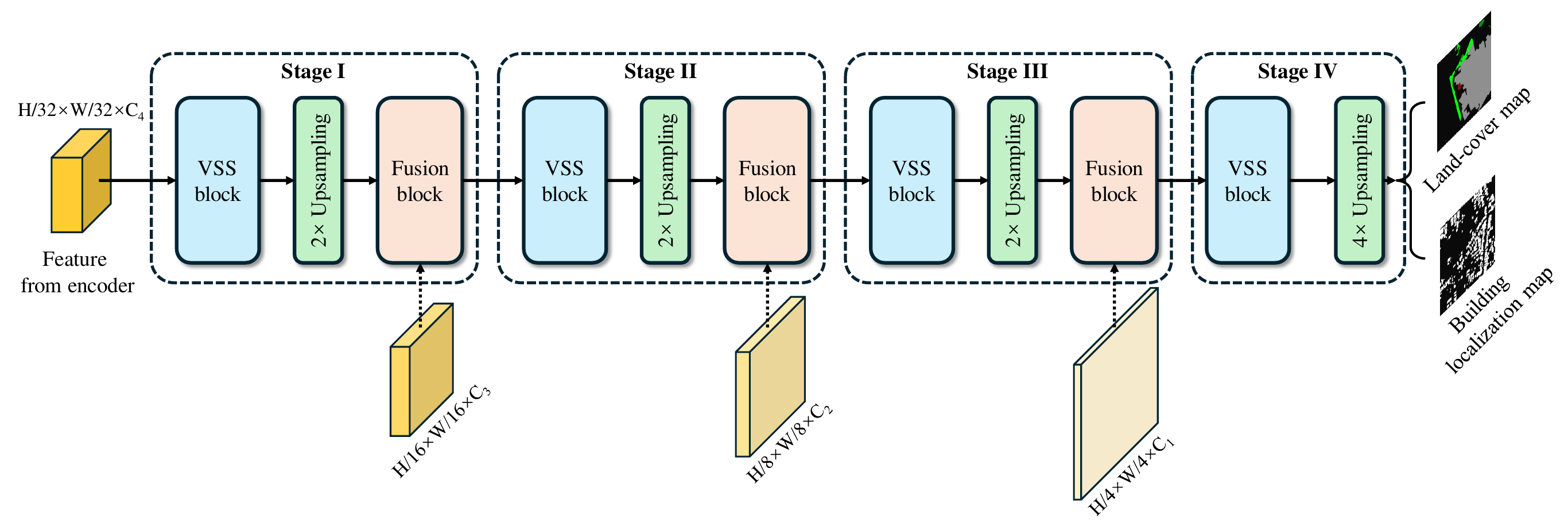}
  \caption{The specific structure of the semantic decoder in MambaSCD and MambaBDA architectures. }
  \label{fig:Semantic_decoder}
\end{figure*}

\par The specific structure of the semantic decoder is shown in Fig. \ref{fig:Semantic_decoder}. It is mainly responsible for gradually recovering the land-cover maps from the corresponding multi-level features extracted by the encoder. It also has four stages. At the beginning of each stage, the global spatial context information of the input data is first modeled using a VSS block. Then the feature map is up-sampled and integrated with information about the lower-level feature map with higher resolution through a fusion module. In the last stage, the features output by the VSS block are upsampled four times and used to predict the corresponding land cover map.

\subsection{Loss Function}
\par Since this paper focuses on exploring the potential of the Mamba architecture for CD tasks, we have optimized networks using commonly used loss functions in the field. Some tricks for improving accuracy such as focal loss \cite{lin2017focal}, deep supervision \cite{Zhang2020}, multi-scale training/testing \cite{chen2023land}, etc. are not adopted. Even so, we can find that the proposed architecture achieves very competitive and even SOTA performance on all three subtasks in the experimental section.

\subsubsection{BCD}
\par Since BCD can be regarded as a special semantic segmentation task \cite{Chen2023Exchange}, we directly optimize the network using the cross-entropy loss as
\begin{equation}
    \mathcal{L}^{bcd}_{ce} = -\frac{1}{N^{bcd}_{train}}  \sum_{i=1}^{N^{bcd}_{train}} \sum_{c=0}^{1}\mathrm{\mathbf{\tilde{Y}}}_{i}^{bcd}\left(c\right)log(\mathrm{\mathbf{P}}_{i}^{bcd}\left(c\right)),
\end{equation}
where $\mathrm{\mathbf{\tilde{Y}}}_{i}^{bcd}$ is the one-hot form of $\mathrm{\mathbf{Y}}_{i}^{bcd}$,  $\mathrm{\mathbf{P}}_{i}^{bcd}=\mathcal{F}_{\theta}^{bcd}\left(\mathrm{\mathbf{X}}^{\mathrm{T}_{1}}_{i}, \mathrm{\mathbf{X}}^{\mathrm{T}_{2}}_{i}\right)$ is the output of the binary change detector through the softmax activation.  

\par In addition, the Lovasz-softmax loss \cite{Berman_2018_CVPR} is introduced to alleviate the problem of imbalance in the number of samples between changed and unchanged pixels. The final loss function for the BCD task is formulated as 
\begin{equation}
    \mathcal{L}^{bcd}_{final} = \mathcal{L}^{bcd}_{ce} + \mathcal{L}^{bcd}_{lov}.
\end{equation}

\subsubsection{SCD}
\par For the SCD task, in addition to optimizing the BCD task, it is also necessary to optimize the land cover mapping task for both pre-event and post-event images, which can be optimized with cross-entropy loss as 

\begin{equation}
\left\{
    \begin{aligned}
      &\mathcal{L}^{\mathrm{T}_{1}}_{ce} = -\frac{1}{N^{scd}_{train}}\sum_{i=1}^{N^{scd}_{train}} \sum_{c=0}^{\mathbf{C}^{lcm}} \mathrm{\mathbf{\tilde{Y}}}_{i}^{\mathrm{T}_{1}}\left(c\right)log(\mathrm{\mathbf{P}}_{i}^{\mathrm{T}_{1}}\left(c\right))  \\
      &\mathcal{L}^{\mathrm{T}_{2}}_{ce} = -\frac{1}{N^{scd}_{train}} \sum_{i=1}^{N^{scd}_{train}} \sum_{c=0}^{\mathbf{C}^{lcm}} \mathrm{\mathbf{\tilde{Y}}}_{i}^{\mathrm{T}_{2}}\left(c\right)log(\mathrm{\mathbf{P}}_{i}^{\mathrm{T}_{2}}\left(c\right))
    \end{aligned},
\right.
\label{eq:2}
\end{equation}
where $\mathrm{\mathbf{P}}_{i}^{\mathrm{T}_{1}}$ and $\mathrm{\mathbf{P}}_{i}^{\mathrm{T}_{2}}$ are the probabilistic map of land-cover classification for the pre-event image and post-event image output by the semantic decoder of the MambaSCD architecture; $\mathcal{L}^{\mathrm{T}_{1}}_{ce}$ and $\mathcal{L}^{\mathrm{T}_{2}}_{ce}$ are the corresponding cross-entropy loss.

\par Similarly, the Lovasz-softmax loss function is used to solve the sample imbalance problem. Thus, the final loss function for the SCD task takes the form
\begin{equation}
    \mathcal{L}^{scd}_{final} = \mathcal{L}^{bcd}_{ce} + \mathcal{L}^{bcd}_{lov} + \frac{1}{2} \left(\mathcal{L}^{\mathrm{T}_{1}}_{ce} +  \mathcal{L}^{\mathrm{T}_{2}}_{ce} + \mathcal{L}^{{\mathrm{T}_{1}}}_{lov} 
+ \mathcal{L}^{{\mathrm{T}_{2}}}_{lov}\right). 
\end{equation}
\subsubsection{BDA}
\par BDA contains a building location task and a damage classification task. The cross-entropy loss for each task can be expressed as follows
\begin{equation}
\left\{
    \begin{aligned}
      &\mathcal{L}^{loc}_{ce} = -\frac{1}{N^{bda}_{train}}\sum_{i=1}^{N^{bda}_{train}} \sum_{c=0}^{1} \mathrm{\mathbf{\tilde{Y}}}_{i}^{loc}\left(c\right)log(\mathrm{\mathbf{P}}_{i}^{loc}\left(c\right))  \\
      &\mathcal{L}^{clf}_{ce} = -\frac{1}{N^{bda}_{train}} \sum_{i=1}^{N^{bda}_{train}} \sum_{c=0}^{\mathbf{C}^{dam}} \mathrm{\mathbf{\tilde{Y}}}_{i}^{clf}\left(c\right)log(\mathrm{\mathbf{P}}_{i}^{clf}\left(c\right))
    \end{aligned},
\right.
\label{eq:bda_ce}
\end{equation}
where $\mathrm{\mathbf{P}}_{i}^{loc}$ and $\mathrm{\mathbf{P}}_{i}^{clf}$ are the output of localization branch and classification branch of the MambaBDA architecture, respectively; $\mathcal{L}^{loc}_{ce}$ and $\mathcal{L}^{clf}_{ce}$ are the corresponding cross-entropy loss of building localization and damage classification tasks, respectively.

\par By utilizing the Lovasz-softmax loss function to solve the sample imbalance problem, the final loss function of BDA is formulated as 
\begin{equation}
    \mathcal{L}^{bda}_{final} = \mathcal{L}^{loc}_{ce} + \mathcal{L}^{loc}_{lov} + \mathcal{L}^{clf}_{ce} + \mathcal{L}^{clf}_{lov}. 
\end{equation}

\section{Experiments And Discussions}\label{sec:4}

\begin{table*}[!t]
  \renewcommand{\arraystretch}{1.2}
\caption{\centering{Information of the five benchmark datasets used for experiments.}}\label{tbl:dataset_info}
  \centering
  \begin{tabular}{c c c c c c}
  \toprule
    \hline	
 \textbf{Dataset}& \textbf{Study site} & \textbf{Number of image pairs} & \textbf{Size}  & \textbf{Resolution} & \textbf{Evaluation task}	\\
    \hline\hline
  SYSU & Hong Kong  & 20,000 & 256 $\times$ 256 & 0.5m &   BCD (category-agnostic)  \\
  LEVIR-CD+ & Texas of the U.S. & 985 & 1,024 $\times$ 1,024 & 0.5m &    BCD (single-category) \\
  WHU-CD  &  Christchurch, New
Zealand & 1 & 32,508 $\times$ 15,354 & 0.3m &    BCD (single-category)\\
  SECOND & Hangzhou, Chengdu, and Shanghai  & 4,662 & 512 $\times$ 512 & 0.5-3m &    SCD \\
  xBD & 15 countries  & 11,034 & 1,024 $\times$ 1,024 & \textless 0.8m &    BDA \\
   \hline
    \bottomrule
  \end{tabular}
\end{table*}

\subsection{Datasets}
\subsubsection{SYSU-CD \cite{Shi2022Deeply}}
 \par This dataset is a category-agnostic CD dataset, which introduces a comprehensive collection of 20,000 pairs of 0.5 meter/pixel resolution aerial images from Hong Kong, spanning 2007 to 2014. This dataset is distinguished by its focus on urban and coastal changes, featuring high-rise buildings and infrastructure developments. For effective deep learning application, the dataset is structured into training, validation, and test sets following a 6:2:2 ratio, with 20,000 patches of 256 $\times$ 256. This dataset encompasses a range of various change scenarios such as urban construction, suburban expansion, groundwork, vegetation changes, road expansion, and sea construction.
 
\subsubsection{LEVIR-CD+ \cite{Chen2020Spatial}} 
\par This advanced version of the LEVIR-CD dataset is a building CD dataset. It comprises 985 pairs of very high-resolution images at 0.5 meters/pixel, each with dimensions of 1024 $\times$ 1024 pixels. Spanning a time interval of 5 to 14 years, these multi-temporal images document significant building construction changes. It also encompasses a wide array of building types, including villa residences, tall apartments, small garages, and large warehouses, with a focus on both the emergence of new buildings and the decline of existing structures. 

\subsubsection{WHU-CD \cite{ji2018fully}}
\par The WHU-CD dataset, a subset of the larger WHU Building dataset, is tailored for the building CD task. It comprises two aerial datasets from Christchurch, New Zealand, captured in April 2012 and 2016, with a spatial resolution of 0.3 meters/pixel. This dataset is particularly focused on detecting changes in large and sparse building structures. The aerial images captured in 2012 cover an area of 20.5 $km^{2}$, featuring 12,796 buildings, while the image captured in 2016 shows an increase to 16,077 buildings within the same area, reflecting significant urban development over the four-year period. We follow the official protocol \cite{ji2018fully} to split the dataset into training and testing areas for experiments.

\subsubsection{SECOND \cite{Yang2022Asymmetric}} 
\par The SECOND dataset introduces a collection of 4,662 pairs of 512 $\times$ 512 aerial images at 0.5-3 meter/pixel resolution, annotated for SCD in cities such as Hangzhou, Chengdu, and Shanghai. It focuses on six primary land cover classes: non-vegetated ground surfaces, trees, low vegetation, water, buildings, and playgrounds to capture a wide range of geographical changes. The dataset uses land-cover map pairs and non-change masks to accurately represent change, facilitating the differentiation of changes from unchanged pixels within the same class.

\subsubsection{xBD \cite{Gupta_2019_CVPR_Workshops}}
\par The xBD dataset, pivotal for advancing BDA in disaster response, integrates pre- and post-disaster satellite imagery with over 850,000 building annotations across 45,362 $km^{2}$. The dataset adheres to stringent criteria, featuring high-resolution imagery (better than 0.8 meter/pixel), encapsulating multiple levels of damage (“no damage”, “minor damage”, “major damage”, and “destroyed”), and covering a diverse range of disaster types (earthquake, tsunami, storm, wildfire, volcano, and flood).

\par The information of these five datasets is summarized in Table \ref{tbl:dataset_info}. 

\subsection{Experimental Setup}
\subsubsection{Implementation Details} 
\begin{table}[!t]
  \renewcommand{\arraystretch}{1.2}
\caption{\centering{Number of layers and feature channels for encoders in ChangeMamba series.}}\label{tbl:encoder_type}
  \centering
  \begin{tabular}{c c c c}
  \toprule
    \hline	
 \textbf{Stage}	&	\textbf{Tiny}  & \textbf{Small}   & \textbf{Base} \\
    \hline\hline
  \multirow{2}{*}{I}  &   $L_{1}=2$ &  $L_{1}=2$  &  $L_{1}=2$ \\
   &   $C_{1}=96$  &   $C_{1}=96$   &  $C_{1}=128$  \\
     \hline
  \multirow{2}{*}{II}  &  $L_{2}=2$&  $L_{2}=2$  &  $L_{2}=2$   \\
  &  $C_{2}=192$  & $C_{2}=192$   &  $C_{2}=256$   \\
  \hline
   \multirow{2}{*}{III}  &  $L_{3}=4$  &  $L_{3}=15$ &  $L_{2}=15$  \\ 
   &  $C_{3}=384$  &  $C_{3}=384$   &  $C_{3}=512$   \\
    \hline
  \multirow{2}{*}{IV} &    $L_{4}=2$ &  $L_{4}=2$  &  $L_{4}=2$ \\
  &    $C_{4}=768$  & $C_{4}=768$   & $C_{4}=1024$  \\
    \hline
    \bottomrule
  \end{tabular}
\end{table}

\par All three of our proposed architectures are implemented in Pytorch. Depending on the size and depth of the encoder network, our proposed MambaBCD, MambaSCD, and MambaBDA architectures are available in Tiny, Small, and Base versions \cite{liu2024vmamba}. As shown in Table \ref{tbl:encoder_type}, the difference between them is the number of VSS blocks inside each stage and the number of channels in the feature map. For datasets other than the SYSU dataset, the multi-temporal image pairs and associated labels are cropped to 256$\times$256 pixels for input to the network, and then the data with the original size are inferred using the trained networks on the test set. During training, we optimize the network using the AdamW optimizer \cite{loshchilov2017decoupled} with a learning rate of 1$e^{-4}$ and a weight decay of 5$e^{-3}$. The batch size is set to 16. Except for the SYSU dataset, for which we set the number of training iterations to 20000, we set the number of training iterations to 50000 on the remaining four datasets. Random rotation, left-right, and top-bottom flip are used as training data augmentation methods. Our source code is open-sourced for community reproduction and subsequent research in \url{https://github.com/ChenHongruixuan/MambaCD}.

\begin{table*}[!t]
  \renewcommand{\arraystretch}{1.25}
\caption{\centering{Accuracy assessment for different BCD models on the SYSU dataset. In this and other tables, $\mathcal{C}$ indicates CNN-based models, $\mathcal{T}$ indicates Transformer-based models, and $\mathcal{M}$ indicates Mamba-based models. In this and other tables, we highlight the highest values in \textbf{\color{red}{red}}, and the second-highest results in \textbf{\color{blue}{blue}}}}\label{tbl:bcd_SYSU}
  \centering
  \begin{tabular}{c c c c c c c c}
  \toprule
    \hline	
 \textbf{Type}  &  \textbf{Method}	&	\textbf{Rec}	& \textbf{Pre} 	 &  \textbf{OA}  & \textbf{F1}   & \textbf{IoU}   & \textbf{KC} \\
    \hline\hline
\multirow{12}{*}{$\mathcal{C}$}  &   FC-EF \cite{CayeDaudt2018}  & 75.17 & 76.47& 88.69 & 75.81 & 61.04 & 68.43 \\
  &   FC-Siam-Diff \cite{CayeDaudt2018} & 75.30 & 76.28 & 88.65 & 75.79 & 61.01 & 68.38  \\
  &   FC-Siam-Conc \cite{CayeDaudt2018}  & 76.75 & 73.67 & 88.05 &  75.18 & 60.23 & 67.32 \\
  &   SiamCRNN-18 \cite{Chen2019a} & 76.83 & 84.80  & 91.29 & 80.62 & 67.54 &  75.02 \\
 &   SiamCRNN-34 \cite{Chen2019a}  & 76.85 & 85.13 & 91.37 & 80.78& 67.75& 75.23\\
 &   SiamCRNN-50 \cite{Chen2019a}  & 78.40 & 83.41 & 91.23 & 80.83 & 67.82 & 75.15 \\
&   SiamCRNN-101 \cite{Chen2019a}  & \textbf{\color{blue}{80.48}} & 80.40 & 90.77 & 80.44 & 67.28 & 74.40\\
  &   SNUNet \cite{Fang2022SNUNet} & 72.21 & 74.09 & 87.49 & 73.14 & 57.66 & 64.99 \\
  &   DSIFN \cite{Zhang2020} & \textbf{\color{red}{82.02}} & 75.83 & 89.59 & 78.80 & 65.02 & 71.92 \\
  &   HANet \cite{Han2023HANet} & 76.14 & 78.71 & 89.52 & 77.41 & 63.14 & 70.59 \\
  &   CGNet \cite{Han2023CGNet} & 74.37 & \textbf{\color{blue}{86.37}} & 91.19 & 79.92 & 66.55 & 74.31  \\
  &   SEIFNet \cite{Huang2024Spatiotemporal} & 78.29 & 78.61 & 89.86 & 78.45 & 64.54  & 71.82   \\
\hline
\multirow{13}{*}{$\mathcal{T}$} &   ChangeFormerV1 \cite{Bandara2022Transformer} & 75.82 & 79.65 & 89.73  & 77.69 & 63.52 & 71.02 \\
 &   ChangeFormerV2 \cite{Bandara2022Transformer}&	75.62 & 78.14 & 89.26 & 76.86 & 62.42 & 69.87 \\
 &   ChangeFormerV3 \cite{Bandara2022Transformer}&	75.24 & 79.46 & 89.57 & 77.29 & 62.99 & 70.53 \\
 &   ChangeFormerV4 \cite{Bandara2022Transformer}& 77.90	& 79.74 & 90.12 & 78.81 & 65.03 & 72.37 \\
   &   ChangeFormerV5 \cite{Bandara2022Transformer}& 74.00	& 77.26 & 88.73 & 75.60& 60.77 & 68.28 \\
   &   ChangeFormerV6 \cite{Bandara2022Transformer}& 	72.38 & 81.70 & 89.67 & 76.76 & 62.29 & 70.15 \\
 &   BIT-18 \cite{Chen2022Remote} & 76.42 & 84.85 & 91.22 & 80.41 & 67.24 & 74.78 \\
 &   BIT-34 \cite{Chen2022Remote} &74.63  & 82.40 & 90.26  & 78.32 & 64.37 & 72.06 \\
 &   BIT-50 \cite{Chen2022Remote} & 77.90 & 81.42  & 90.60 & 79.62  & 66.14 & 73.51\\
 &   BIT-101 \cite{Chen2022Remote} & 75.58 & 83.64 & 90.76 & 79.41 & 65.84 & 73.47\\
&  TransUNetCD \cite{LiTransUNetCD2022} & 77.73	& 82.59 & 90.88 & 80.09 & 66.79  & 74.18  \\
&  SwinSUNet \cite{Zhang2022SwinSUNet} &	79.75  & 83.50  & 91.51  & 81.58 & 68.89  & 76.06\\
&   CTDFormer \cite{Zhang2023Relation} &	75.53 & 80.80 & 90.00 & 78.08 & 64.04 & 71.61 \\
\hline
\multirow{3}{*}{$\mathcal{M}$} & MambaBCD-Tiny & 79.59 & 83.06 & 91.36 & 81.29 & 68.48	& 75.68 \\
& MambaBCD-Small & 78.25 & \textbf{\color{red}{87.99}} & \textbf{\color{red}{92.35}} & \textbf{\color{blue}{82.83}} & \textbf{\color{blue}{70.70}} & \textbf{\color{blue}{77.94}}	 \\
& MambaBCD-Base & 80.31 & 86.11 & \textbf{\color{blue}{92.30}} & \textbf{\color{red}{83.11}} & \textbf{\color{red}{71.10}} &	\textbf{\color{red}{78.13}} \\
    \hline
    \bottomrule
  \end{tabular}
\end{table*}

\begin{table*}[!t]
  \renewcommand{\arraystretch}{1.25}
\caption{\centering{Accuracy assessment for different binary CD models on the LEVIR-CD+ dataset.}}\label{tbl:acc_bcd}
  \centering
  \begin{tabular}{c c c c c c c c}
  \toprule
    \hline	
 \textbf{Type}  &  \textbf{Method}	&	\textbf{Rec}	& \textbf{Pre} 	 &  \textbf{OA}  & \textbf{F1}   & \textbf{IoU}   & \textbf{KC} \\
    \hline\hline
\multirow{12}{*}{$\mathcal{C}$} &   FC-EF \cite{CayeDaudt2018} & 71.77 & 69.12 & 97.54 & 70.42 & 54.34 & 69.14 \\
 &   FC-Siam-Diff \cite{CayeDaudt2018} & 74.02 & 81.49 & 98.26 & 77.57& 63.36& 76.67 \\
 &   FC-Siam-Conc \cite{CayeDaudt2018}  & 78.49 & 78.39 & 98.24 & 78.44& 64.53 & 77.52 \\
&   SiamCRNN-18 \cite{Chen2019a}  & 84.25 & 81.22 & 98.56& 82.71& 70.52& 81.96 \\
&   SiamCRNN-34 \cite{Chen2019a}  &83.88 & 82.28 & 98.61 & 83.08 & 71.05 & 82.35 \\
 &   SiamCRNN-50 \cite{Chen2019a}  & 81.61 & 85.39 & 98.68 & 83.46 & 71.61 & 82.77 \\
 &   SiamCRNN-101 \cite{Chen2019a}  & 80.96 & 85.56 & 98.67 & 83.20  & 71.23 & 82.50 \\
  &   DSIFN \cite{Zhang2020} & 84.36 & 83.78 & 98.70 & 84.07 & 72.52 &83.39  \\
 &   SNUNet \cite{Fang2022SNUNet} & 78.73 & 71.07 & 97.83 & 74.70 & 59.62 & 73.57  \\
 &   HANet \cite{Han2023HANet} &  75.53 & 79.70 &  98.22 & 77.56  &  63.34 &  76.63 \\
 &   CGNet \cite{Han2023CGNet} & 86.02 &  81.46 &  98.63 &  83.68 &  71.94  &  82.97  \\
  &   SEIFNet \cite{Huang2024Spatiotemporal} & 81.86 & 84.83  & 98.66 & 83.32 &  71.41 & 82.63   \\
				
\hline
\multirow{13}{*}{$\mathcal{T}$} &   ChangeFormerV1 \cite{Bandara2022Transformer}& 77.00	& 82.18  & 98.38  & 79.51  & 65.98& 78.66\\
 &   ChangeFormerV2 \cite{Bandara2022Transformer}&	81.32 & 79.10 & 98.36 & 80.20 & 66.94  & 79.35  \\
 &   ChangeFormerV3 \cite{Bandara2022Transformer}& 79.97	& 81.34 & 98.44 & 80.65 & 67.58 & 79.84 \\
 &   ChangeFormerV4 \cite{Bandara2022Transformer}&	76.68 & 75.07 & 98.01 & 75.87 & 61.12 & 74.83 \\
   &   ChangeFormerV5 \cite{Bandara2022Transformer}& 77.96	& 78.50 & 98.23 & 78.23 & 64.24 & 77.31 \\
   &   ChangeFormerV6 \cite{Bandara2022Transformer}& 78.57	& 67.66 & 97.60 & 72.71 & 57.12 & 71.46 \\
 &   BIT-18 \cite{Chen2022Remote} & 80.86  & 83.76 & 98.58 & 82.28 & 69.90 & 81.54 \\
 &   BIT-34 \cite{Chen2022Remote} & 80.96  & 85.87  & 98.68 & 83.34 & 71.44 & 82.66 \\
 &   BIT-50 \cite{Chen2022Remote} & 81.84 & 85.02 & 98.67 & 83.40 & 71.53 & 82.71 \\
 &   BIT-101 \cite{Chen2022Remote} & 81.20 & 83.91 & 98.60 & 82.53 & 70.26 & 81.80 \\
&  TransUNetCD \cite{LiTransUNetCD2022}&	84.18 & 83.08 & 98.66 & 83.63 & 71.86 & 82.93 \\
&   SwinSUNet \cite{Zhang2022SwinSUNet} &	85.85 & 85.34 & 98.92 & 85.60 & 74.82 & 84.98 \\
&  CTDFormer \cite{Zhang2023Relation} &	80.03  & 80.58 & 98.40 & 80.30 & 67.09  & 79.47  \\

\hline
\multirow{3}{*}{$\mathcal{M}$} & MambaBCD-Tiny & \textbf{\textcolor{blue}{87.26}} & 88.82 & \textbf{\textcolor{blue}{99.03}} & \textbf{\textcolor{blue}{88.04}} & \textbf{\textcolor{blue}{78.63}} &	\textbf{\textcolor{blue}{87.53}}  \\
& MambaBCD-Small & 86.49 & \textbf{\textcolor{blue}{89.17}} & 99.02 & 87.81 & 78.27 & 87.30	 \\

& MambaBCD-Base & \textbf{\textcolor{red}{87.57}} & \textbf{\textcolor{red}{89.24}} & \textbf{\textcolor{red}{99.06}} & \textbf{\textcolor{red}{88.39}} & \textbf{\textcolor{red}{79.20}} &	\textbf{\textcolor{red}{87.91}} \\
    \hline
    \bottomrule
  \end{tabular}
\end{table*}

\subsubsection{Evaluation Metrics}
\par For evaluating model performance across different CD subtasks, we employ a set of metrics tailored to the specific requirements of BCD, SCD and BDA. In evaluating the BCD performance of the model, we use the six commonly used metrics. They are recall rate (Rec), precision rate (Pre), overall accuracy (OA), F1 score (F1), intersection over union (IoU) and Kappa coefficient (KC). The higher the better for all six indicators. The specific formula of these metrics can be referred to \cite{Chen2023Exchange, Wu2022Unsupervised, CAO2023full}.

\par For SCD, we follow the existing setup in the literature \cite{Ding2024Joint} to use the four following metrics. They are OA, F1, mean IoU (mIoU), and separated Kappa coefficient (SeK). The higher the better for all four indicators. Their definitions can be referred to \cite{Ding2024Joint, Ding2022Semantic}.



\par For BDA, we follow the protocol of the xView2 Challenge \cite{Gupta_2019_CVPR_Workshops} to split BDA into two subtasks, i.e., building localization and damage classification tasks. F1 is used to evaluate these two subtasks. Firstly, F1 representing the model's accuracy in the building localization task is denoted as $F_{1}^{loc}$. Then, the model's performance across various damage levels are denoted as $F_{1}^{level_{i}}$. The harmonic mean of all the $F_{1}^{level_{i}}$ is calculated to represent as the performance of the model on the damage classification task, formulated as
\begin{equation}
    F_{1}^{clf}= \frac{1}{\frac{1}{\mathbf{C}^{dam}} \sum_{i=1}^{\mathbf{C}^{dam}} \frac{1}{F_{1}^{level_{i}}}},
\end{equation}
where the value of $\mathbf{C}^{dam}$ in the xBD dataset is 4, from 1 to 4 being “no damage”, “minor damage”, “major damage”, and “destroyed”, respectively. Finally, the overall performance of the model in building localization and damage classification tasks is calculated as $F_{1}^{overall}= 0.3 F_{1}^{loc} +  0.7F_{1}^{clf}$.

\subsection{Comparison Methods}
\par We select a number of representative methods based on CNN and Transformer architectures to compare across BCD, SCD, and BDA tasks.

\par For BCD, our comparison methods include a suite of CNN-based methods: FC-EF \cite{CayeDaudt2018}, FC-Siam-Diff \cite{CayeDaudt2018}, FC-Siam-Conc \cite{CayeDaudt2018}, SiamCRNN \cite{Chen2019a} (utilizing ResNet architectures ranging from ResNet-18 to ResNet-101 \cite{He2016}), SNUNet \cite{Fang2022SNUNet}, DSIFN \cite{Zhang2020}, HANet \cite{Han2023HANet}, CGNet \cite{Han2023CGNet}, and SEIFNet \cite{Huang2024Spatiotemporal}. For the Transformer-based methods, we compare against ChangeFormer (versions 1 to 6) \cite{Bandara2022Transformer}, BIT (with ResNet-18 to ResNet-101 as the backbone for feature extraction) \cite{Chen2022Remote}, TransUNetCD \cite{LiTransUNetCD2022}, SwinSUNet \cite{Zhang2022SwinSUNet}, and CTDFormer \cite{Zhang2023Relation}.

\par For the SCD arena, CNN-based competitors encompass HRSCD (variants S1 to S4) \cite{Rodrigo2019Multitask}, ChangeMask \cite{Zheng2022}, SSCD-1 \cite{Ding2022Semantic}, Bi-SRNet \cite{Ding2022Semantic}, and TED \cite{Ding2024Joint}. For Transformer-based methods, we evaluate against SMNet \cite{Niu2023SMNet} and ScanNet \cite{Ding2024Joint}.

\par Lastly, for the BDA task, we selected Siamese-UNet  \cite{Gupta_2019_CVPR_Workshops}, MTF \cite{weber2020building}, and ChangeOS \cite{ZHENG2021Building} (employing ResNet-18 to ResNet-101) as CNN-based benchmarks. DamFormer \cite{Chen2022Dual} stands as the solitary Transformer-based comparison method.

\par For the above methods, if the method does not have the open source code available, we directly use the accuracy reported in the original paper. Otherwise, we train and test the method on the corresponding dataset based on the hyperparameters recommended in the original paper, using the loss function and data augmentation methods consistent with our method. In addition, in the experimental tables, we will use $\mathcal{C}$ to denote CNN-based models, $\mathcal{T}$ to denote Transformer-based models, and $\mathcal{M}$ to denote Mamba-based models.

\subsection{Detection Results and Benchmark Comparison in Three Tasks}\label{sec:benchmark_comparison}

\begin{table*}[!t]
  \renewcommand{\arraystretch}{1.25}
\caption{\centering{Accuracy assessment for different binary CD models on the WHU-CD dataset.}}\label{tbl:bcd_WHU}
  \centering
  \begin{tabular}{c c c c c c c c}
  \toprule
    \hline	
 \textbf{Type}  &  \textbf{Method}	&	\textbf{Rec}	& \textbf{Pre} 	 &  \textbf{OA}  & \textbf{F1}   & \textbf{IoU}   & \textbf{KC} \\
    \hline\hline
\multirow{12}{*}{$\mathcal{C}$}   &   FC-EF \cite{CayeDaudt2018} 
 & 86.33 & 83.50 & 98.87 & 84.89 & 73.74 & 84.30  \\
&   FC-Siam-Diff \cite{CayeDaudt2018}  & 84.69 & 90.86 & 99.13 & 87.67 & 78.04 & 87.22  \\
&   FC-Siam-Conc  \cite{CayeDaudt2018} & 87.72 & 84.02 & 98.94 & 85.83 &75.18  & 85.28   \\
 &   SiamCRNN-18 \cite{Chen2019a}  & 90.48 & 91.56& 99.34 & 91.02& 83.51 &90.68 \\
 &   SiamCRNN-34 \cite{Chen2019a}  & 89.10 & 93.88 & 99.39 & 91.42 & 84.20 & 91.11 \\
 &   SiamCRNN-50 \cite{Chen2019a}  & 91.45 & 86.70 & 99.30 & 90.57 & 82.76 & 90.20 \\
 &   SiamCRNN-101 \cite{Chen2019a}  & 90.45 & 87.79 & 99.19 & 89.10 & 80.34 & 88.68  \\
  &   DSIFN \cite{Zhang2020}  & 83.45 & \textbf{\textcolor{red}{97.46}} & 99.31 & 89.91& 81.67 & 89.56 \\
 &   SNUNet \cite{Fang2022SNUNet} & 87.36 & 88.04  & 99.10 & 87.70 & 78.09 & 87.23 \\
&   HANet \cite{Han2023HANet}  & 88.30 & 88.01 & 99.16 & 88.16 & 78.82  & 87.72 \\
&   CGNet \cite{Han2023CGNet} & 90.79 & 94.47 & 99.48 & 92.59 & 86.21 & 92.33 \\
 &  SEIFNet \cite{Huang2024Spatiotemporal} & 90.66 & 91.93  & 99.36 & 91.29 & 83.98 & 90.96   \\
\hline
\multirow{13}{*}{$\mathcal{T}$} &   ChangeFormerV1 \cite{Bandara2022Transformer}&	84.30 & 90.80 & 99.11 & 87.43 & 77.67 & 86.97  \\
&   ChangeFormerV2 \cite{Bandara2022Transformer}&	83.41 & 88.77 & 99.00 & 86.00 & 75.45 & 85.49  \\
 &   ChangeFormerV3 \cite{Bandara2022Transformer}&	85.55 & 88.25  & 99.05 & 86.88 & 76.80  & 86.39  \\
 &   ChangeFormerV4 \cite{Bandara2022Transformer}& 84.85 & 90.09 & 99.10 & 87.39 & 77.61 & 86.93  \\
  &   ChangeFormerV5 \cite{Bandara2022Transformer}& 84.87 & 90.20 & 99.11 & 87.45 & 77.70 & 86.99 \\
 &   ChangeFormerV6 \cite{Bandara2022Transformer}& 81.90	& 85.49 & 98.83 & 83.66 & 71.91 & 83.05 \\
  &   BIT-18 \cite{Chen2022Remote} & 90.36 & 90.30 & 99.29 & 90.33 & 82.37 & 89.96 \\
 &   BIT-34 \cite{Chen2022Remote} & 90.10 & 89.14  & 99.23 & 89.62 & 81.19 & 89.22\\
 &   BIT-50 \cite{Chen2022Remote} & 90.33  & 89.70 & 99.26 & 90.01 & 81.84  & 89.63  \\
 &   BIT-101 \cite{Chen2022Remote} & 90.24 & 89.83 & 99.27 & 90.04 & 81.88 & 89.66 \\
&  TransUNetCD \cite{LiTransUNetCD2022}& 90.50	& 85.48 & 99.09 & 87.79 & 78.44  & 87.44 \\
&   SwinSUNet \cite{Zhang2022SwinSUNet} &	92.03 & 94.08 & 99.50 & 93.04 & 87.00 & 92.78 \\
&   CTDFormer \cite{Zhang2023Relation} & 85.37 & 92.23 & 99.20 & 88.67 & 79.65 & 88.26 \\

\hline
\multirow{3}{*}{$\mathcal{M}$} & MambaBCD-Tiny & 91.94 & 94.76 & 99.52 & 93.33 &87.49  &93.08 	 \\
& MambaBCD-Small & \textbf{\textcolor{red}{92.29}} & 95.90 & \textbf{\textcolor{blue}{99.57}} & \textbf{\textcolor{blue}{94.06}}  & \textbf{\textcolor{blue}{88.79}}  & \textbf{\textcolor{blue}{93.84}}  	 \\
& MambaBCD-Base & \textbf{\textcolor{blue}{92.23}} & \textbf{\textcolor{blue}{96.18}} & \textbf{\textcolor{red}{99.58}} & \textbf{\textcolor{red}{94.19}} & \textbf{\textcolor{red}{89.02}} &	\textbf{\textcolor{red}{93.98}} \\
    \hline
    \bottomrule
  \end{tabular}
\end{table*}

\begin{table}[!t]
  \renewcommand{\arraystretch}{1.2}
\caption{\centering{Comparison of different binary CD models in computational cost.}}\label{tbl:bcd_param_flops}
  \centering
  \begin{tabular}{c c c c}
  \toprule
    \hline	
 \textbf{Type}  &  \textbf{Method}	&	\textbf{Params (M)}	& \textbf{GFLOPs}	 	 \\
    \hline\hline
\multirow{12}{*}{$\mathcal{C}$}   &   FC-EF \cite{CayeDaudt2018} &  1.35 &  14.13  \\
&   FC-Siam-Diff \cite{CayeDaudt2018} & 1.35 & 18.66 \\
&   FC-Siam-Conc  \cite{CayeDaudt2018} & 1.54  &  21.07 \\
 &   SiamCRNN-18 \cite{Chen2019a} &	18.85 &  86.64  \\ 
 &   SiamCRNN-34 \cite{Chen2019a} & 28.96	&  113.28     \\ 
 &   SiamCRNN-50 \cite{Chen2019a}  &	44.45 &  185.30    \\ 
 &   SiamCRNN-101 \cite{Chen2019a}  & 63.44	& 224.30   \\ 
 &   DSIFN \cite{Zhang2020}  &   35.73 &  329.03 \\
 &   SNUNet \cite{Fang2022SNUNet} &  10.21 &  176.36 \\
&   HANet \cite{Han2023HANet}  &  2.61 &  70.68 \\
&   CGNet \cite{Han2023CGNet} &   33.68 &  329.58    \\
&  SEIFNet \cite{Huang2024Spatiotemporal} & 39.08 & 167.75 \\

\hline
\multirow{13}{*}{$\mathcal{T}$} &   ChangeFormerV1 \cite{Bandara2022Transformer}&	29.84 &   46.62    \\  
&   ChangeFormerV2 \cite{Bandara2022Transformer}&	24.30 & 44.54 \\
 &   ChangeFormerV3 \cite{Bandara2022Transformer}&	24.30 &  33.68     \\ 
 &   ChangeFormerV4 \cite{Bandara2022Transformer}& 33.61	&  852.53   \\
   &   ChangeFormerV5 \cite{Bandara2022Transformer}& 55.27 & 841.08 \\
   &   ChangeFormerV6 \cite{Bandara2022Transformer}& 41.03 & 811.15 \\

 &   BIT-18 \cite{Chen2022Remote}  &	11.50 &  106.14    \\ 
 &   BIT-34 \cite{Chen2022Remote}  & 21.61	&  190.83     \\ 
 &   BIT-50 \cite{Chen2022Remote}  &	 24.28 & 224.61  \\ 
 &   BIT-101 \cite{Chen2022Remote}  & 43.27	&  380.62
 \\ 
 &  TransUNetCD \cite{LiTransUNetCD2022} & 28.37	& 244.54     \\ 
&   SwinSUNet \cite{Zhang2022SwinSUNet} &	39.28 &  43.50     \\ 
&  CTDFormer \cite{Zhang2023Relation} &	3.85 & 303.77 \\ 

\hline
\multirow{3}{*}{$\mathcal{M}$} & MambaBCD-Tiny & 17.13 	&  45.74     \\ 	 
& MambaBCD-Small  &	49.94 &  114.82  \\ 
& MambaBCD-Base &  84.70 &	 179.32  \\ 
    \hline
    \bottomrule
  \end{tabular}
\end{table}

\subsubsection{BCD}
\par
\begin{figure}[!t]
  \centering
\includegraphics[width=3.2in]{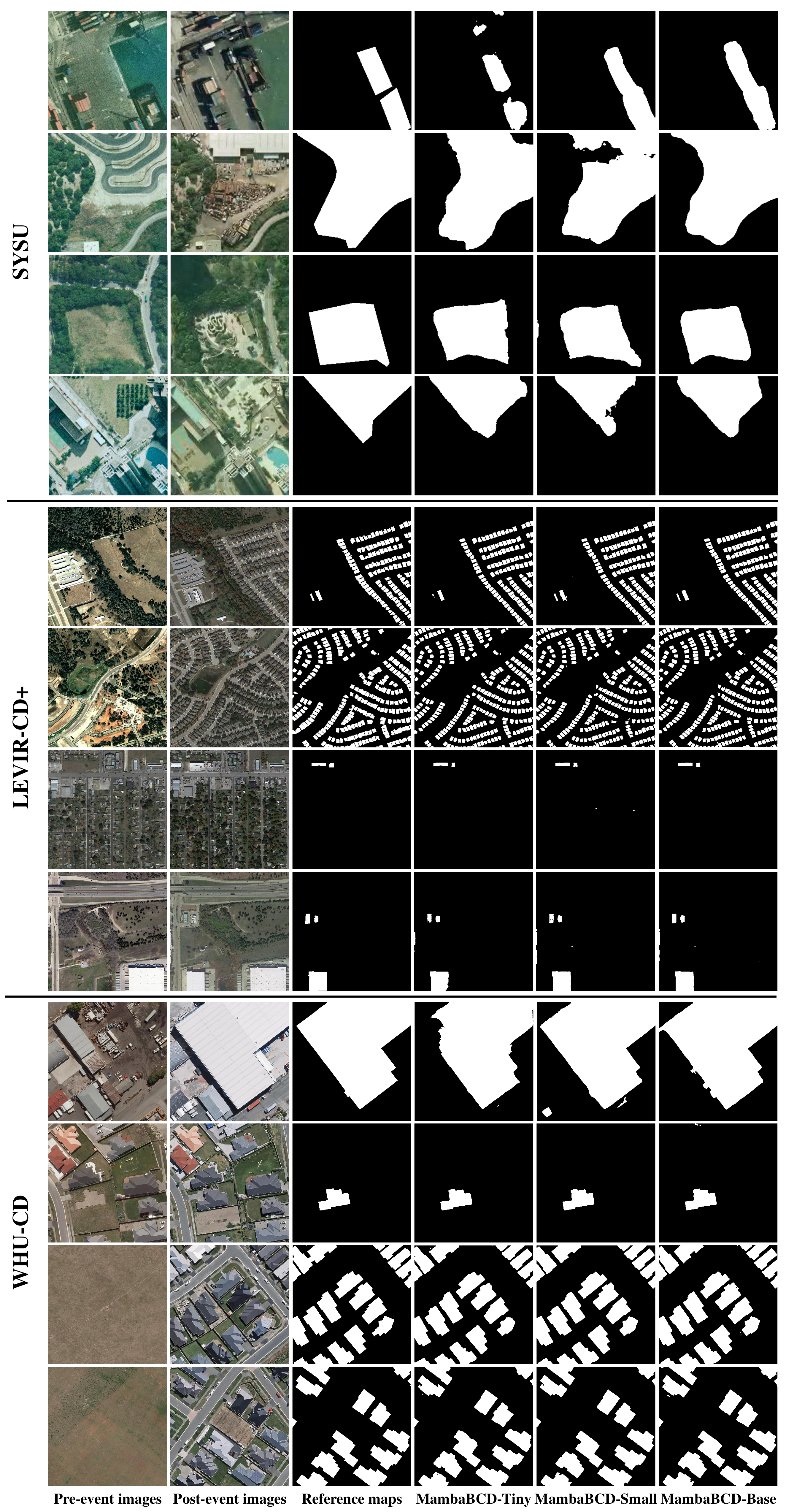}
  \caption{Some binary change maps obtained by our methods on the test set of the three BCD dataset. }
  \label{fig:ALL_BCD}
\end{figure}

 Tables \ref{tbl:bcd_SYSU} to \ref{tbl:bcd_WHU} list the performance in BCD of the proposed MambaBCD architecture and comparison methods on the three benchmark datasets. It can be seen that our approaches significantly outperform both the CNN-based approaches and the Transformer-based architectures. The proposed MambaBCD-Base and MambaBCD-Small architectures achieve the highest and second-highest OA, F1, IoU, and KC for both category-agnostic CD (SYSU dataset) and single-category CD tasks (LEVIR-CD+ and WHU-CD datasets), fully demonstrating the potential of the Mamba architecture for the BCD task. Fig. \ref{fig:ALL_BCD} then shows some binary change maps predicted by our three architectures on the test sets of the three datasets. It can be seen that the proposed methods accurately detect changes with varying types, scales, and numbers contained in these image pairs, very close to the change reference maps.

\par Table \ref{tbl:bcd_param_flops} lists the network parameters and GFLOPs\footnote{https://github.com/Lyken17/pytorch-OpCounter} of these BCD architectures. Benefiting from the S6 model, the MambaBCD architecture can have only linear computational complexity while modeling global context information. It can be seen that even the largest model MambaBCD-Base has GFLOPs of 179.32, which is lower than many CNN and Transformer architectures, like DSIFN, CFNet, and ChangeFormerV4, comparable to some CNN SOTA methods such as SNUNet and SiamCRNN-50. However, the accuracy advantage is obvious. In particular, the MambaBCD-Tiny architecture has a clear accuracy advantage over other architectures with a similar number of parameters and computational consumption. For example, compared to BIT-18, MambaBCD-Tiny has 56.9$\%$ lower GFLOPs but 0.88$\%$, 1.33$\%$ and 3.00$\%$ higher F1 scores on the three datasets.

\begin{table}[!t]
  \renewcommand{\arraystretch}{1.25}
\caption{\centering{Accuracy assessment for different SCD models on the SECOND dataset. Note that unlike \cite{Ding2022Semantic, Zheng2022, Ding2024Joint} and others that divide the training set for testing, we train on the training set and then test on the test set.}}\label{tbl:scd_second}
  \centering
  \begin{tabular}{c c c c c c}
  \toprule
    \hline	
 \textbf{Type}  &  \textbf{Method}	&	 \textbf{OA}  & \textbf{F1}   & \textbf{mIoU}   & \textbf{$\mathbf{SeK_{37}}$} \\
    \hline\hline
    
\multirow{8}{*}{$\mathcal{C}$}   &   HRSCD-S1 \cite{Rodrigo2019Multitask} &  62.56  & 33.04 &  47.57  & 6.62     \\
&     HRSCD-S2 \cite{Rodrigo2019Multitask}&  82.93  & 39.79  & 63.80 & 8.46   \\
&     HRSCD-S3 \cite{Rodrigo2019Multitask}&  81.17 & 41.78 & 64.96 & 10.86     \\
&   HRSCD-S4 \cite{Rodrigo2019Multitask}&  83.88 &   47.68  &  70.58 &  16.81  \\

&   ChangeMask \cite{Zheng2022} & 85.53 & 50.54 & 71.39 & 18.36  \\
&   SSCD-1 \cite{Ding2022Semantic} & 85.62  & 52.80  & 72.14  & 20.15 \\
&   BiSRNet \cite{Ding2022Semantic} & 85.80 & 53.59 & 72.69   & 21.18  \\
&   TED \cite{Ding2024Joint}  &  85.50 & 52.61 & 72.19 & 20.29 \\
\hline
\multirow{2}{*}{$\mathcal{T}$} &   SMNet \cite{Niu2023SMNet}  & 84.29 &  50.48 & 71.62  & 18.98  \\
& ScanNet \cite{Ding2024Joint}&  86.13 & 55.21 & 72.48 & 21.57 \\
\hline
\multirow{3}{*}{$\mathcal{M}$} & MamabaSCD-Tiny & \textbf{\textcolor{red}{90.44}} & 54.51 & 73.33 & 22.08 	 \\

& MamabaSCD-Small  & 86.48 &  \textbf{\textcolor{red}{55.72}} & \textbf{\textcolor{blue}{73.41}}  & \textbf{\textcolor{blue}{22.83}}	 \\

& MamabaSCD-Base  &  \textbf{\textcolor{blue}{90.36}}  &  \textbf{\textcolor{blue}{55.32}} &   \textbf{\textcolor{red}{73.68}} &  \textbf{\textcolor{red}{22.92}}  \\
    \hline
    \bottomrule
  \end{tabular}
\end{table}

\subsubsection{SCD}
\par Table \ref{tbl:scd_second} compares the SCD performance of the proposed MamabaSCD architectures and representative SCD approaches. Compared to the BCD task, the SCD task requires the change detector capable of effectively modeling not only the spatio-temporal relationships of multi-temporal images, but also learning representative and discriminative representations for accurate land-cover mapping. From Table \ref{tbl:scd_second}, we can see that the MambaSCD architecture can satisfy both requirements well. The MambaSCD-Small architecture surpasses the Transformer-based SOTA approach, ScanNet, on all four metrics of the SCD task. Note that our methods have achieved the current accuracy without employing advanced loss functions specifically proposed for the SCD like the ones in ScanNet \cite{Ding2024Joint}. 

\begin{figure}[!t]
    \centering
  \subfloat[]{
    \includegraphics[width=1.6in]{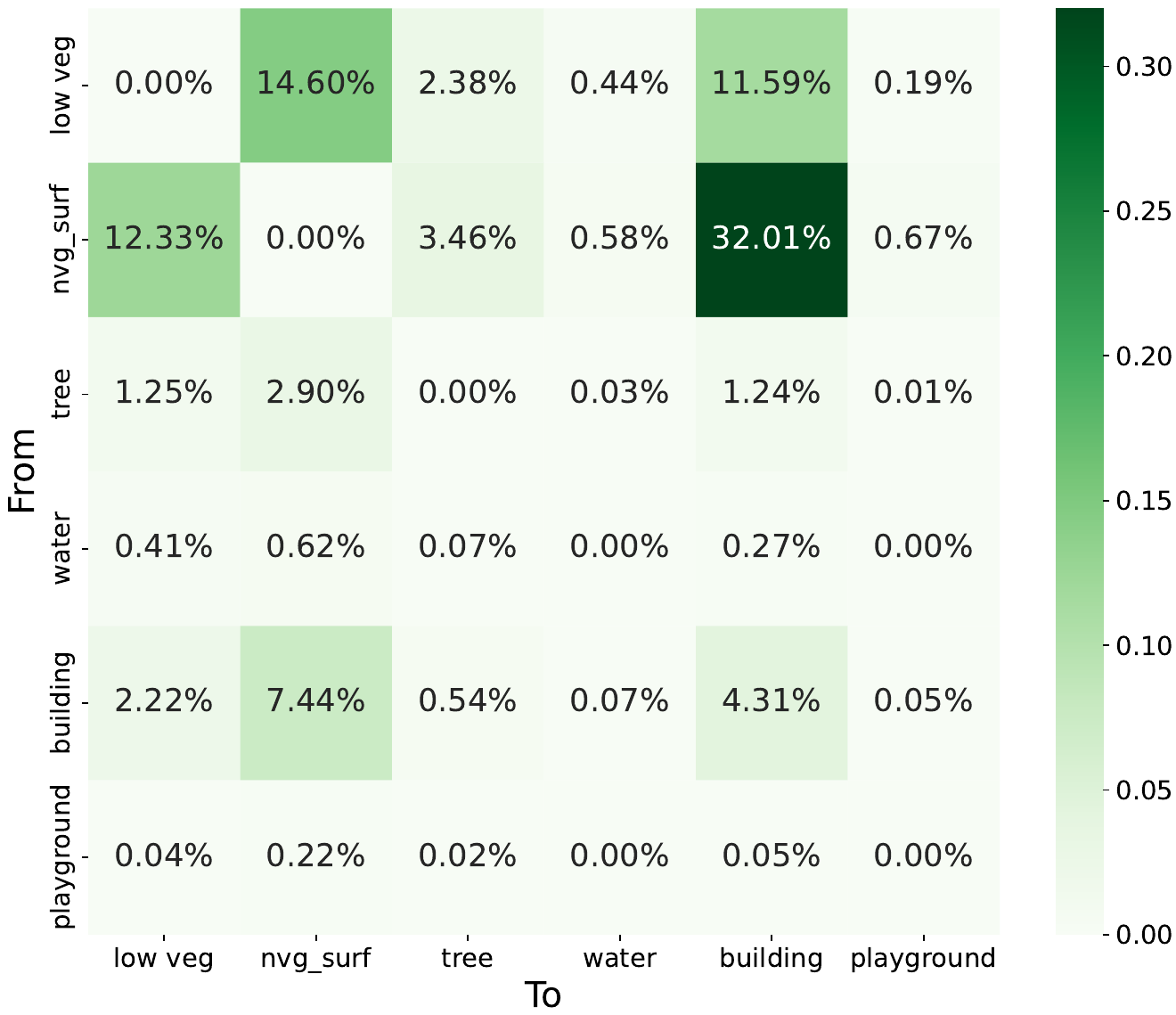}
  \label{fig_second_case}}  
  \subfloat[]{
    \includegraphics[width=1.6in]{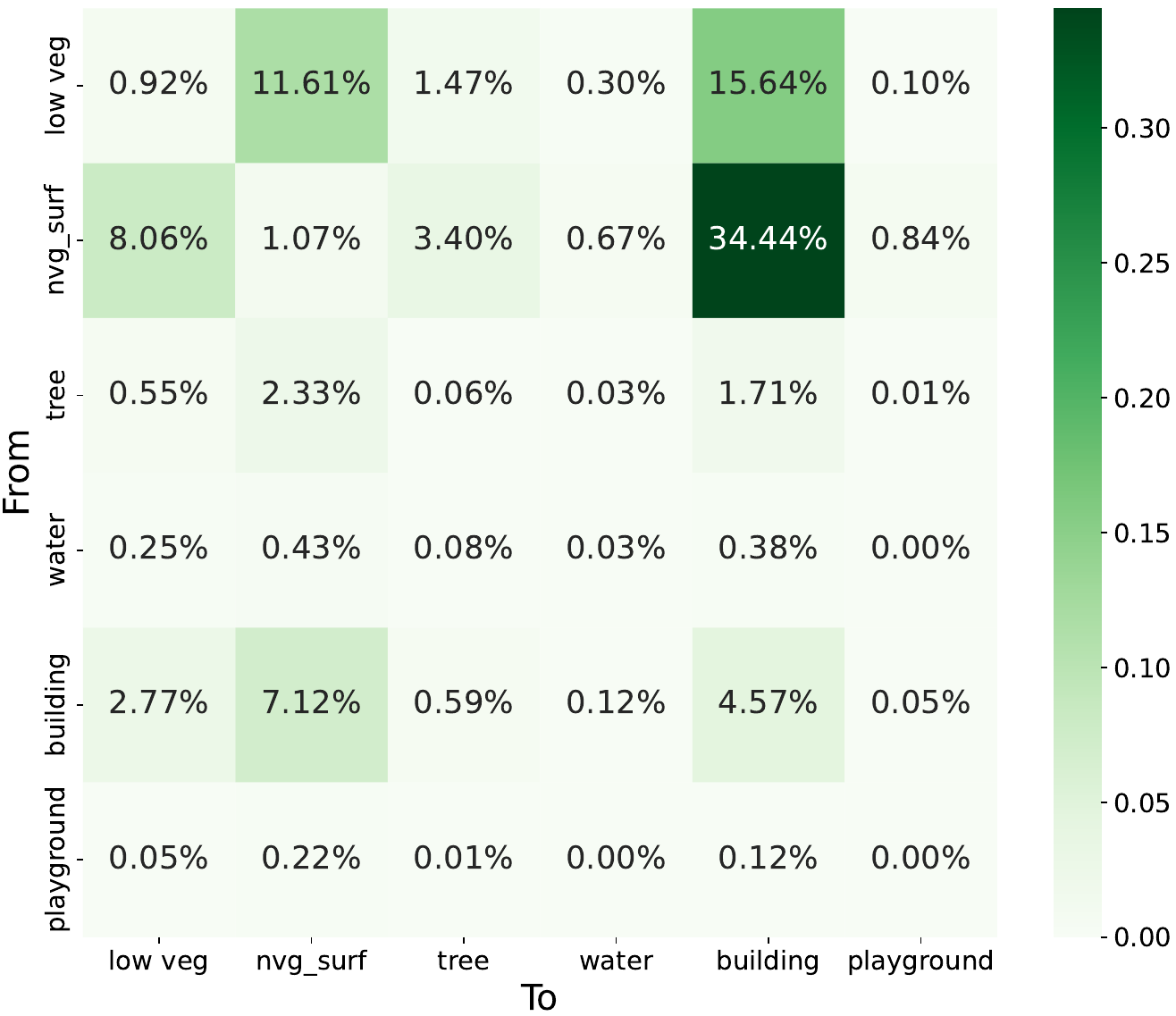}
  \label{fig_second_case}} 
    \hfil
  \subfloat[]{
    \includegraphics[width=1.6in]{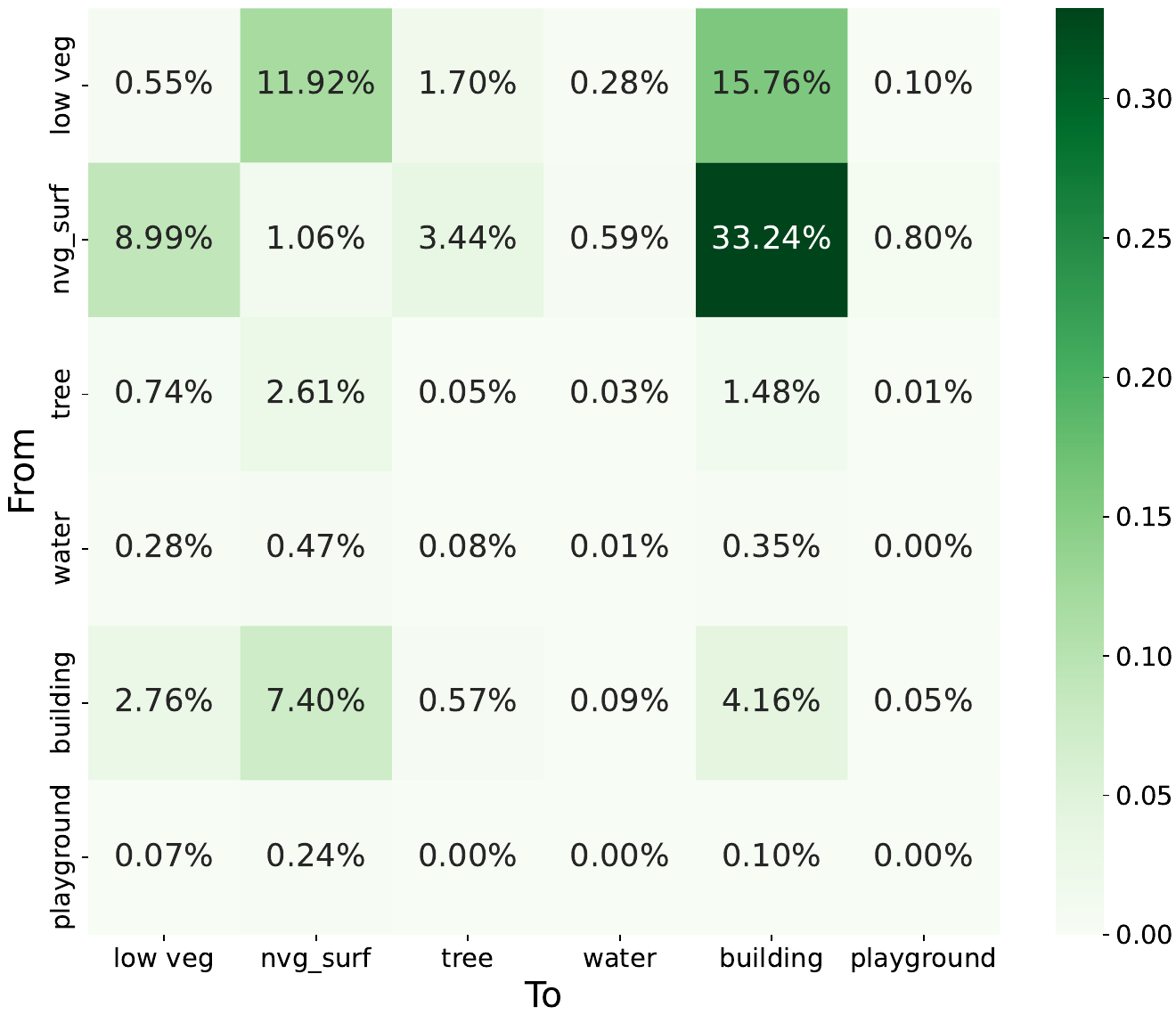}
  \label{fig_second_case}}  
     \subfloat[]{
    \includegraphics[width=1.6in]{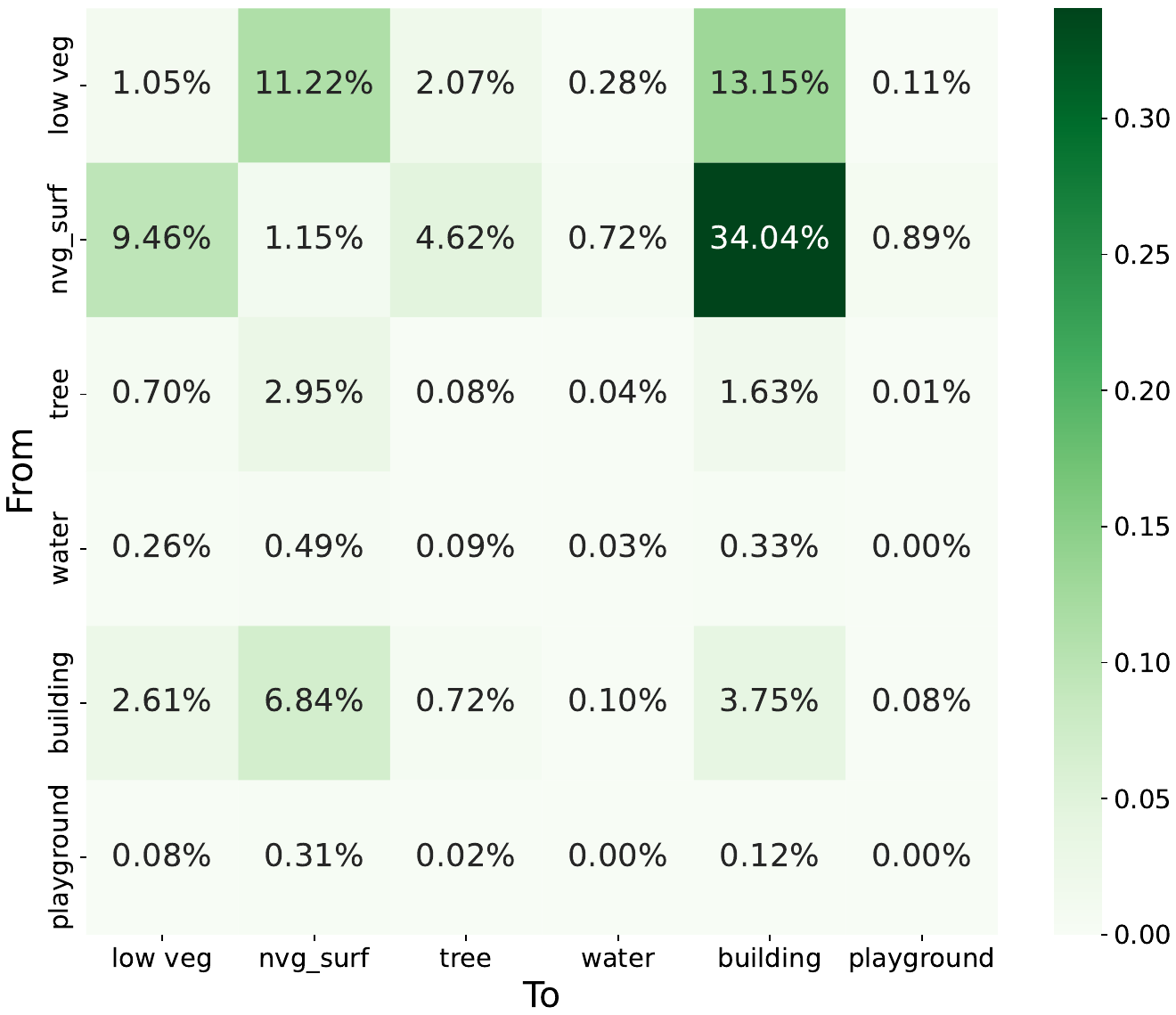}
  \label{fig_second_case}}  
  
   \caption{Semantic change matrices providing “from-to” transition information derived from the SCD results. (a) Reference. (b) MambaSCD-Tiny. (c) MambaSCD-Small. (d) MambaSCD-Base. }

  \label{fig:confusion_matrix_SCD}
\end{figure}

\par Based on the SCD results, the detailed “from-to" semantic change information can be obtained, which is presented in Fig. \ref{fig:confusion_matrix_SCD}. Firstly, we can find that the “from-to” transition matrices obtained by our methods are very close to the ground truth, demonstrating the effectiveness of our MambaSCD architecture again. Then, based on the contents of the matrices, we can find that the three semantic changes with the highest percentage between the two temporal phases are “non-vegetated ground surface $\rightarrow$ building”, “low vegetation $\rightarrow$ building”, “low vegetation $\rightarrow$ non-vegetated ground surface”. These results indicate that the areas covered by the SECOND dataset was in the process of urbanization during this period.

\begin{table}[!t]
  \renewcommand{\arraystretch}{1.2}
\caption{\centering{Comparison of different SCD models in computational cost.}}\label{tbl:param_scd}
  \centering
  \begin{tabular}{c c c c }
  \toprule
    \hline	
 \textbf{Type}  &  \textbf{Method}	&	\textbf{Params}	& \textbf{GFLOPs} \\
    \hline\hline
\multirow{8}{*}{$\mathcal{C}$}   &   HRSCD-S1 \cite{Rodrigo2019Multitask} &  3.36  & 8.02 \\
&     HRSCD-S2 \cite{Rodrigo2019Multitask}&  6.39 & 14.29  \\
&     HRSCD-S3 \cite{Rodrigo2019Multitask} & 12.77  & 42.67 \\
&   HRSCD-S4 \cite{Rodrigo2019Multitask} & 13.71 & 43.69 \\
&   ChangeMask \cite{Zheng2022}  &  2.97  & 37.16 \\
&   SSCD-1 \cite{Ding2022Semantic}  & 23.31  & 189.57 \\
&   Bi-SRNet \cite{Ding2022Semantic}  &  23.39  & 189.91 \\
&   TED \cite{Ding2024Joint} &  24.19 & 204.29 \\
\hline
\multirow{2}{*}{$\mathcal{T}$} &   SMNet \cite{Niu2023SMNet}  & 42.16  &  75.79 \\
& ScanNet \cite{Ding2024Joint} & 27.90  & 264.95 \\
\hline
\multirow{3}{*}{$\mathcal{M}$} & MamabaSCD-Tiny &  21.51 & 73.42 \\
& MamabaSCD-Small &  54.28 &  146.70 \\
& MamabaSCD-Base  & 89.99& 211.55 \\
    \hline
    \bottomrule
  \end{tabular}
\end{table}

\par Table \ref{tbl:param_scd} further lists the number of parameters and GFLOPs of these methods. Compared to the Transformer-based ScanNet method, the GFLOPs of our three architectures are lower. Compared to SMNet, another Transformer-based method, the proposed MamabaSCD-Tiny can achieve higher SeK values with a lower number of parameters and GFLOPs. These results demonstrate the effectiveness of the proposed methods and the potential of the Mamba architecture on the SCD task.

\begin{figure}[!t]
  \centering
\includegraphics[width=3.4in]{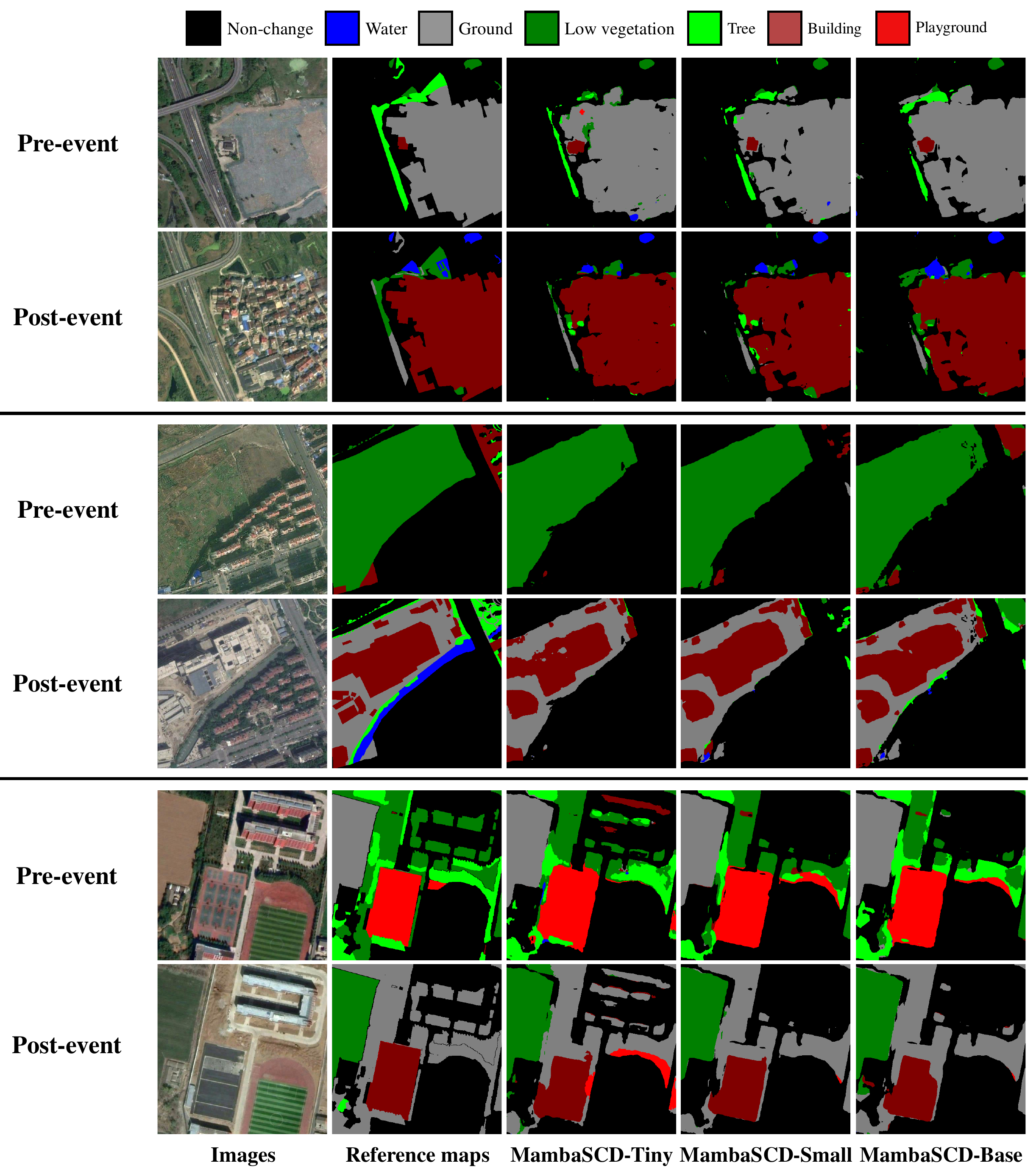}
  \caption{Some semantic change maps obtained by the proposed MamabaSCD architectures on the test set of the SECOND dataset. }
  \label{fig:SECOND_SCD}
\end{figure}

\par Fig. \ref{fig:SECOND_SCD} shows some semantic change maps on the test set of the SECOND dataset. Due to land-cover category combinations, SCD often involves a considerable number of categories. For ease of visualization, we follow the visualization way in \cite{Zheng2022, Tian2022Large, Yang2022Asymmetric}, that is, only plotting the land-cover categories of the changed regions in both pre- and post-event images. In this way, the “from-to” semantic change information can be effectively reflected. From Fig. \ref{fig:SECOND_SCD}, we can observe that the MambaSCD architecture is not only capable of distinguishing changed features at different scales but also accurately identifying the specific semantic categories of these changes.

\subsubsection{BDA}

\begin{table*}[ht]
  \renewcommand{\arraystretch}{1.25}
  \caption{\centering{Accuracy assessment for different building damage assessment models on the xBD dataset. The suffix PPS means post-processing.}}
  \label{tbl:bda_xbd}
  \centering
  \begin{tabular}{c c c c c c c c c}
    \toprule
    \hline	
   \multirow{2}{*}{\textbf{Type} }& \multirow{2}{*}{\textbf{Method}}
    & \multirow{2}{*}{\textbf{$F_{1}^{loc}$}}  & \multirow{2}{*}{\textbf{$F_{1}^{clf}$}}  & \multirow{2}{*}{\textbf{$F_{1}^{overall}$}} &\multicolumn{4}{c}{\textbf{Damage $F_{1}$ per class}}   \\
    \cline{6-9} 
    & &  & & & \textbf{No damage} &  \textbf{Minor damage} &  \textbf{Major damage} & \textbf{Destroyed}  \\
    \hline\hline
   \multirow{10}{*}{$\mathcal{C}$}   & Siamese-UNet \cite{Gupta_2019_CVPR_Workshops} &	85.92&	65.58&71.68&		86.74&	50.02&	64.43&	71.68\\ 
     & MTF \cite{weber2020building} &	83.60&	70.02& 74.10&	90.60 &	49.30&	72.20&	83.70 \\ 
     & ChangeOS-18 \cite{ZHENG2021Building}  &	 84.62 & 69.87  &74.30 &88.61  &52.10  &70.36  & 79.65 \\ 
     & ChangeOS-34 \cite{ZHENG2021Building} &85.16 & 70.28 &	74.74  & 88.63 &52.38 &71.16 & 80.08  \\ 
     & ChangeOS-50 \cite{ZHENG2021Building} & 85.41 &70.88  &	75.24 &88.98 &53.33& 71.24 &80.60 \\
     & ChangeOS-101 \cite{ZHENG2021Building}&	85.69&	71.14& 75.50&		89.11&	53.11&	72.44&	80.79\\ 
     & ChangeOS-18-PPS \cite{ZHENG2021Building}  & 84.62 & 73.89  & 77.11  & 92.38  & 57.41  & 72.54  & 82.62 \\ 
     & ChangeOS-34-PPS \cite{ZHENG2021Building} & 85.16 & 74.25 & 77.52  & 92.19 & 58.07 & 72.84 & 82.79  \\ 
     & ChangeOS-50-PPS \cite{ZHENG2021Building} & 85.41 & 75.64  &	78.57 & 92.66 & 60.14 & 74.18 & 83.45 \\ 
     & ChangeOS-101-PPS \cite{ZHENG2021Building}& 85.69 & 75.44 & 78.52 & 92.81 &	59.38 & 74.65 &	83.29 \\ 
     \hline
   $\mathcal{T}$  & DamFormer \cite{Chen2022Dual}  &	86.86 &	 72.81 &	77.02 &	89.86&	56.78 &	72.56 &	80.51\\ 
    \hline    
   \multirow{3}{*}{$\mathcal{M}$}  & MambaBDA-Tiny &  	 \textbf{\textcolor{blue}{87.20}} & 78.30   &  80.97 & 95.84  & 60.96  & \textbf{\textcolor{red}{77.43}}  &  88.23 \\ 
     & MambaBDA-Small &  86.61  & \textbf{\textcolor{blue}{78.80}}  &  \textbf{\textcolor{blue}{81.14}}  & \textbf{\textcolor{red}{95.99}} & \textbf{\textcolor{red}{62.82}} & 76.26 & \textbf{\textcolor{blue}{88.37}} \\ 
      & MambaBDA-Base &	\textbf{\textcolor{red}{87.38}} &	\textbf{\textcolor{red}{78.84}} &	\textbf{\textcolor{red}{81.41}} &	\textbf{\textcolor{blue}{95.94}} &	\textbf{\textcolor{blue}{62.74}} &	\textbf{\textcolor{blue}{76.46}} & \textbf{\textcolor{red}{88.58}} \\ 
    \hline
    \bottomrule
\end{tabular}
\end{table*}

\par Table \ref{tbl:bda_xbd} reports the accuracy of the MambaBDA architecture and some Transformer and CNN-based architectures on the xBD dataset. The performance of the MambaBDA architecture significantly outperforms these representative BDA approaches. Compared to the current Transformer-based SOTA method, DamFormer \cite{Chen2022Dual}, our three MambaBDA architectures have a 2.64$\%$, 4.12$\%$, and 4.39$\%$ higher $F_{1}^{overall}$, respectively. Moreover, it can be noticed that the accuracy improvement compared to DamFormer mainly comes from the more difficult damage classification task rather than the building localization task, which demonstrates that the proposed MambaBDA architecture can adequately learn the spatio-temporal relationships between the multi-temporal images so as to effectively distinguish different building damage levels.

\begin{figure*}[!t]
    \centering
  \subfloat[]{
    \includegraphics[width=2.1in]{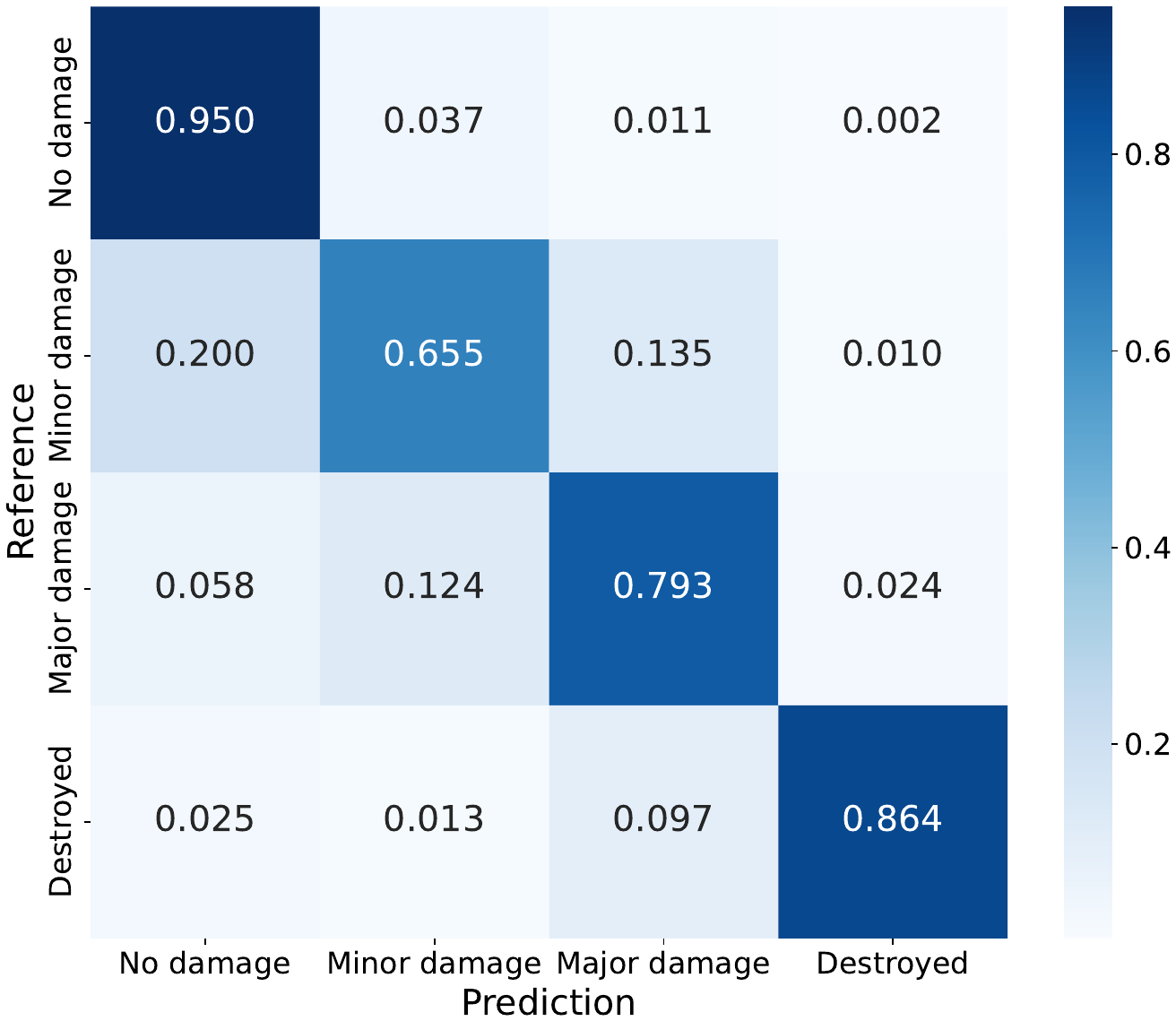}
  \label{fig_second_case}}  
  \subfloat[]{
    \includegraphics[width=2.1in]{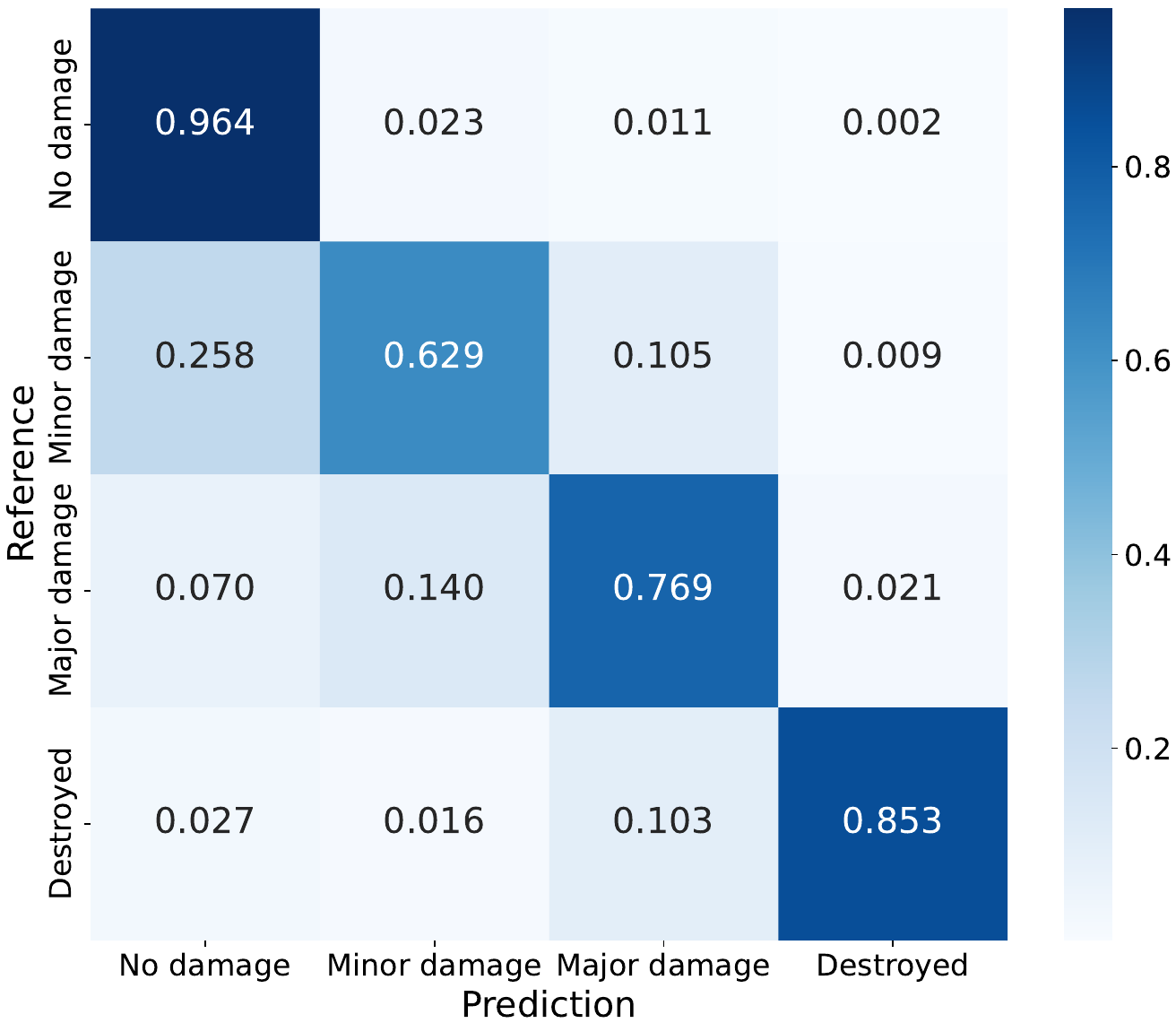}
  \label{fig_second_case}}  
  \subfloat[]{
    \includegraphics[width=2.1in]{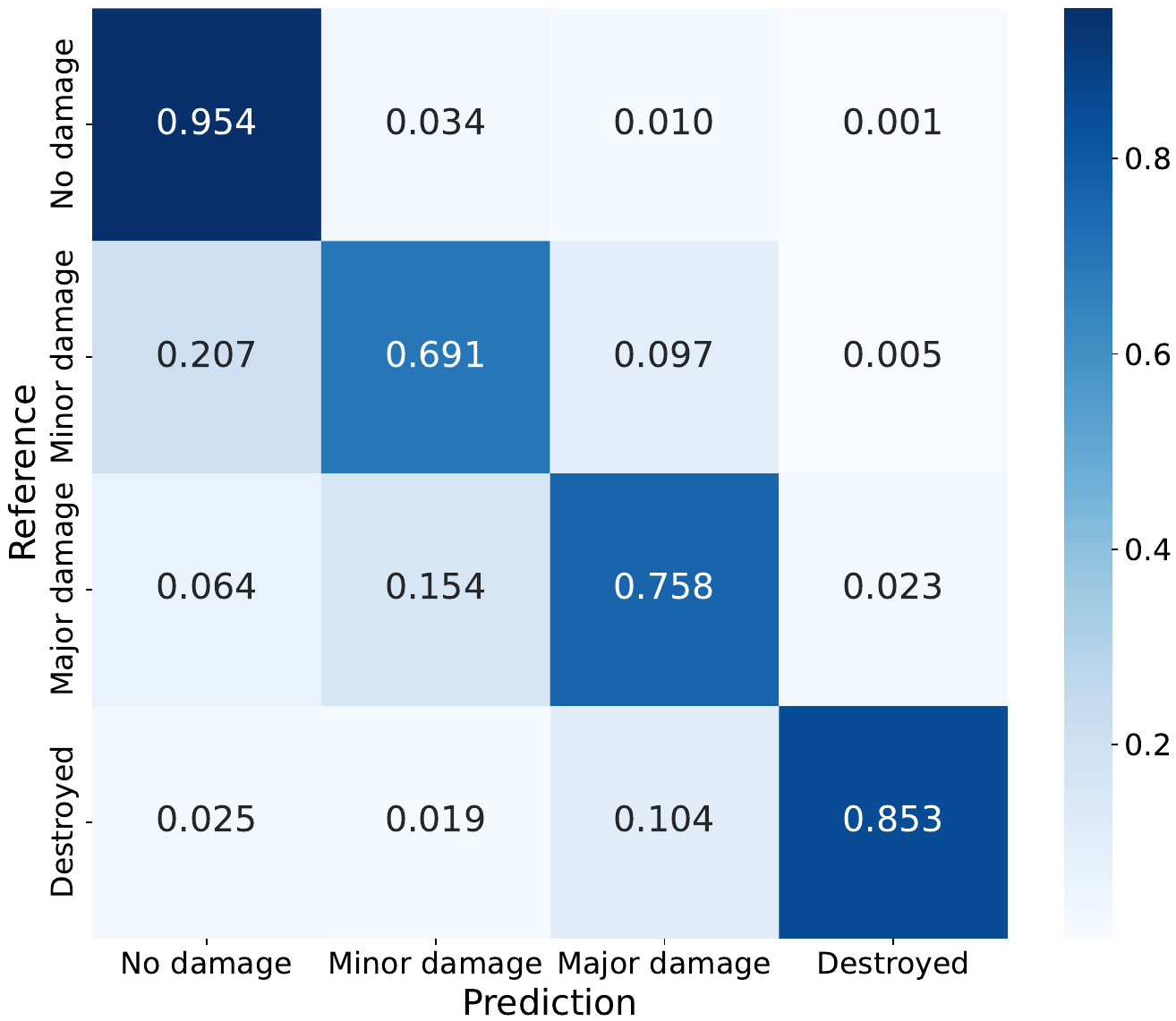}
  \label{fig_second_case}}  
  \hfil
     \subfloat[]{
    \includegraphics[width=2.1in]{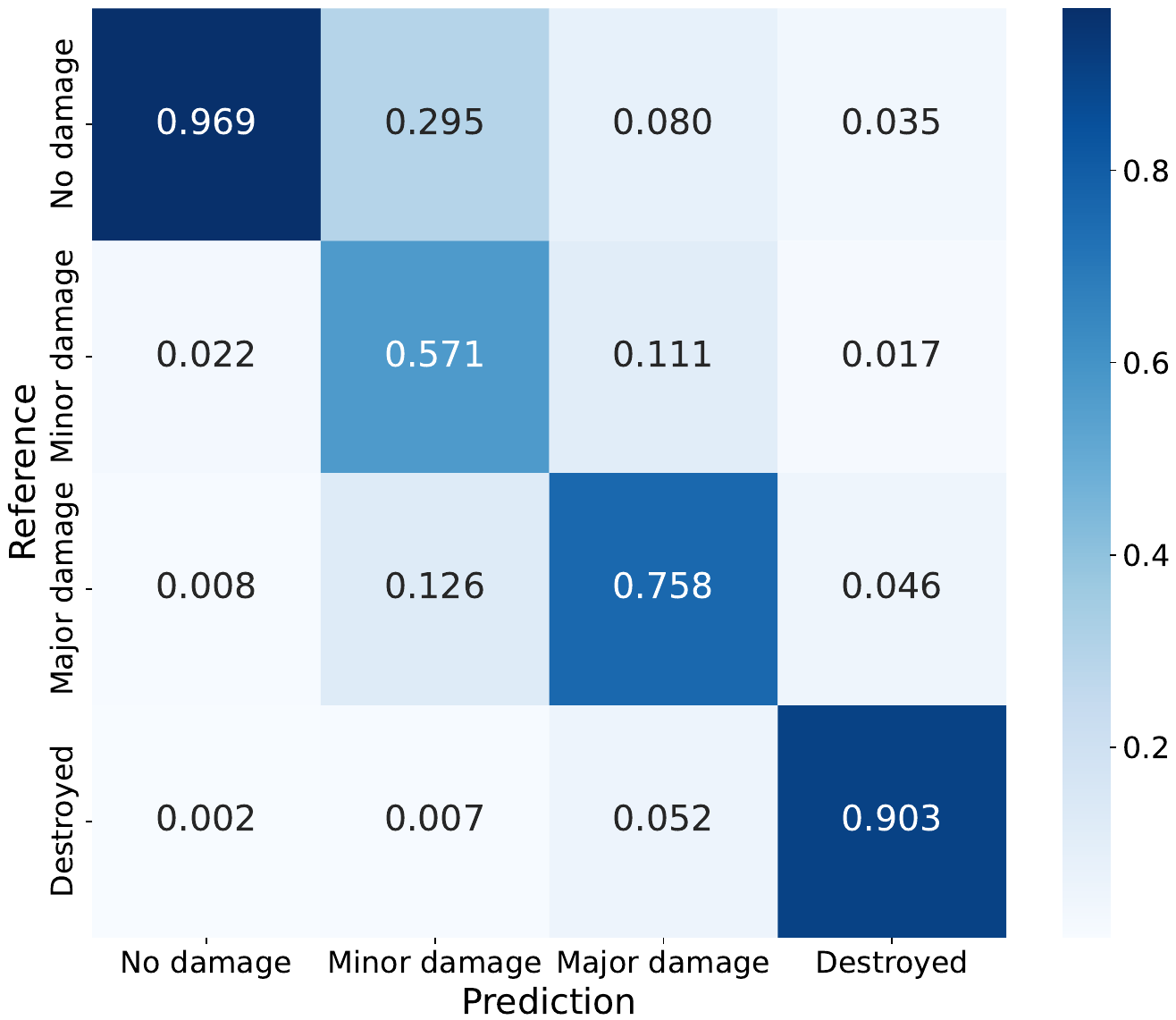}
  \label{fig_second_case}}  
  \subfloat[]{
    \includegraphics[width=2.1in]{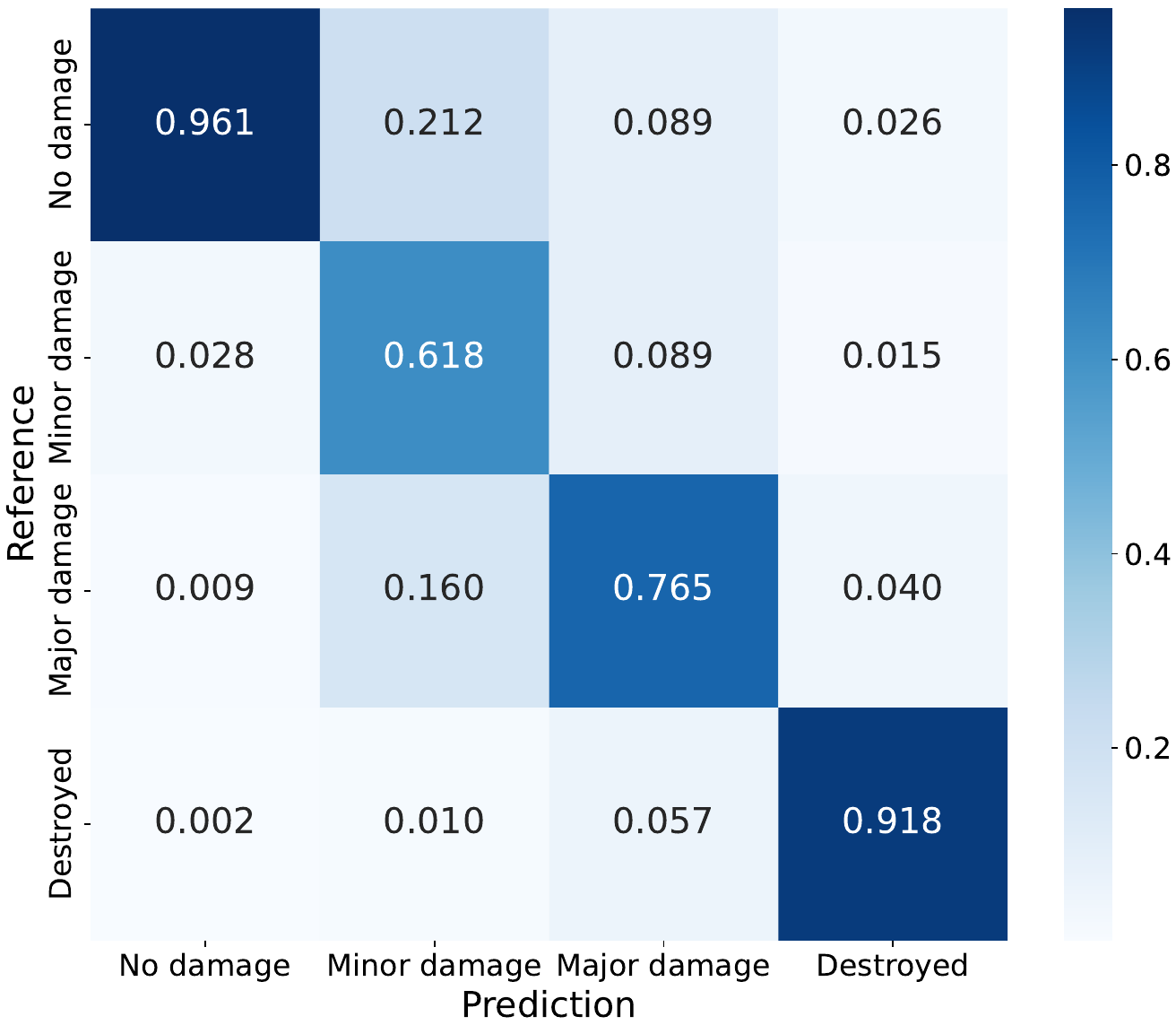}
  \label{fig_second_case}}  
  \subfloat[]{
    \includegraphics[width=2.1in]{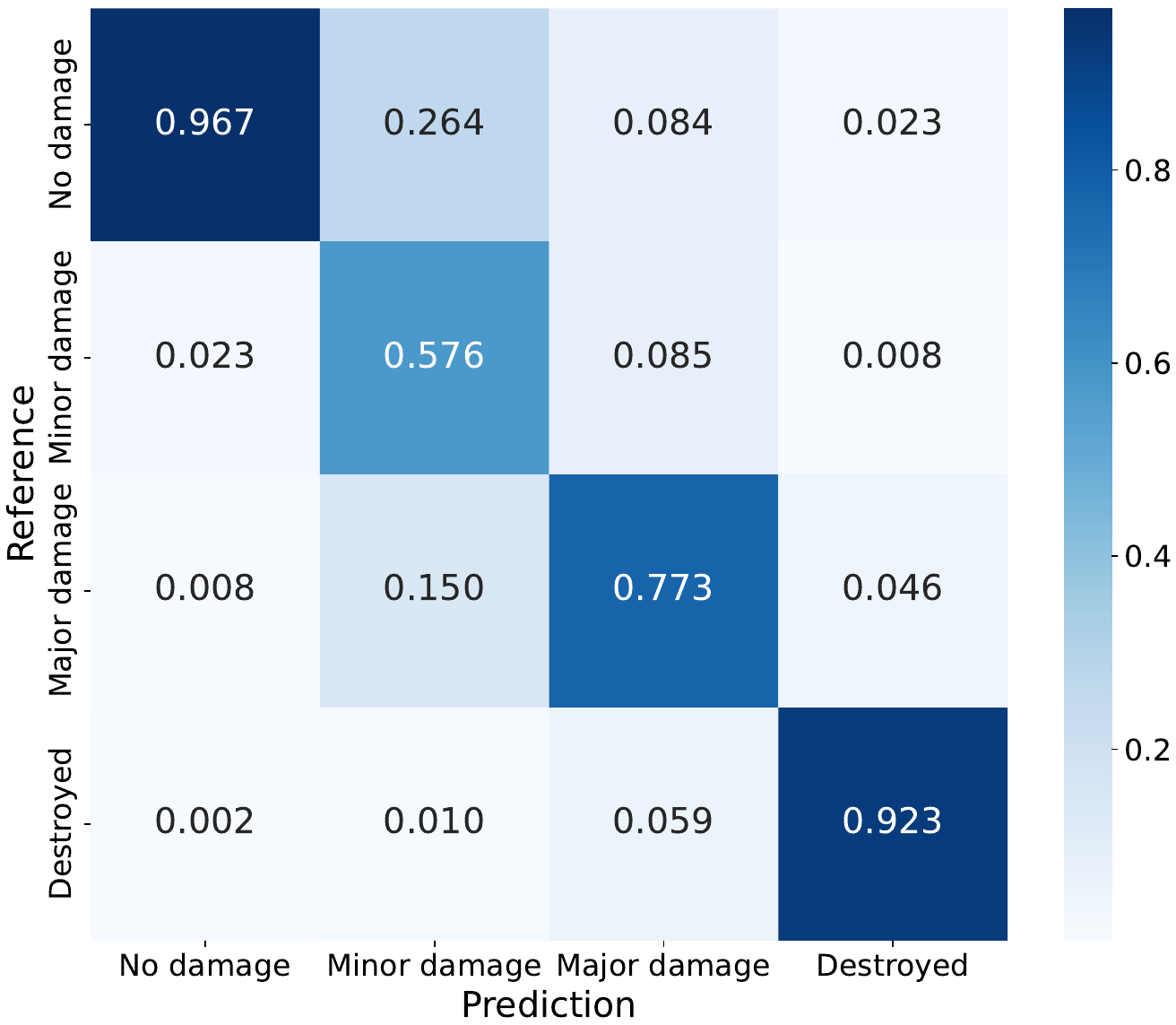}
  \label{fig_second_case}}  
  
  \caption{Confusion matrices of MambaBDA-Tiny, MambaBDA-Small, and MambaBDA-Base. (a)-(c) are matrices normalized by rows. (d)-(f) are matrices normalized by columns. }
  
  \label{fig:confusion_matrix_BDA}
\end{figure*}

\par In order to further show the misclassification between the different damage levels in more detail, we show their confusion matrices in Fig. \ref{fig:confusion_matrix_BDA}. As can be seen from the confusion matrices, the categories “no damage” and “destroyed” are classified with high accuracy and are not easily misclassified into the remaining two categories. On the contrary, the categories of “minor damage” and “major damage” are easily misclassified as other categories, especially the “minor damage” category. A large portion of the “minor damage” category is misclassified as “no damage” and “major damage”. There are mainly the following reasons. First, the optical images in the xBD dataset were acquired with a relatively small viewing angle of the sensor. In many cases, “minor damage” is nuance damages to the structure of the building. Little relevant information can be obtained from the nearly orthorectified optical image for such cases. Furthermore, there is a semantic ambiguity in the definition of the categories of minor damage and major damage. We argue that it is difficult to address this problem by starting from the model architecture perspective. It can be considered to introduce auxiliary information, such as height information or multi-view imagery \cite{Song_2024_WACV, CAO2021deep}, etc. to help the model make a more accurate classification. 

\begin{table}[!t]
  \renewcommand{\arraystretch}{1.2}
\caption{\centering{Comparison of different BDA models in computational cost.}}\label{tbl:param_scd}
  \centering
  \begin{tabular}{c c c c }
  \toprule
    \hline	
 \textbf{Type}  &  \textbf{Method}	&	\textbf{Params}	& \textbf{GFLOPs} \\
    \hline\hline
\multirow{7}{*}{$\mathcal{C}$}
&     Siamese-UNet \cite{Gupta_2019_CVPR_Workshops}&   51.46 & 138.88   \\
& MTF \cite{weber2020building} &  44.40 &   268.76  \\
&     ChangeOS-18 \cite{ZHENG2021Building} &  25.25  &  91.55  \\
&   ChangeOS-34 \cite{ZHENG2021Building} &   35.36 &  110.89  \\
&   ChangeOS-50 \cite{ZHENG2021Building}  &   39.05 &  118.41 \\
&   ChangeOS-101 \cite{ZHENG2021Building}  &  58.05 &  157.26  \\

\hline
$\mathcal{T}$ &   DamFormer \cite{Chen2022Dual} & 32.52  & 169.34 \\
\hline
\multirow{3}{*}{$\mathcal{M}$} & MamabaBDA-Tiny & 19.74  & 59.57 	 \\
& MamabaBDA-Small & 52.11 & 130.80 \\
& MamabaBDA-Base  & 87.76&   195.43   \\
    \hline
    \bottomrule
  \end{tabular}
\end{table}

\par Table \ref{tbl:param_scd} lists the number of parameters and GFLOPs of these BDA approaches. For MambaBDA-Small, our architecture yields more accurate results on building damage assessment compared to Siamese-UNet, ChangeOS-101, and MTF, with an improvement in $F^{overall}_{1}$ from 5.64 $\%$ to 9.46 $\%$ for a similar number of parameters and lower GFLOPs.

\begin{figure}[!t]
  \centering
\includegraphics[width=3.4in]{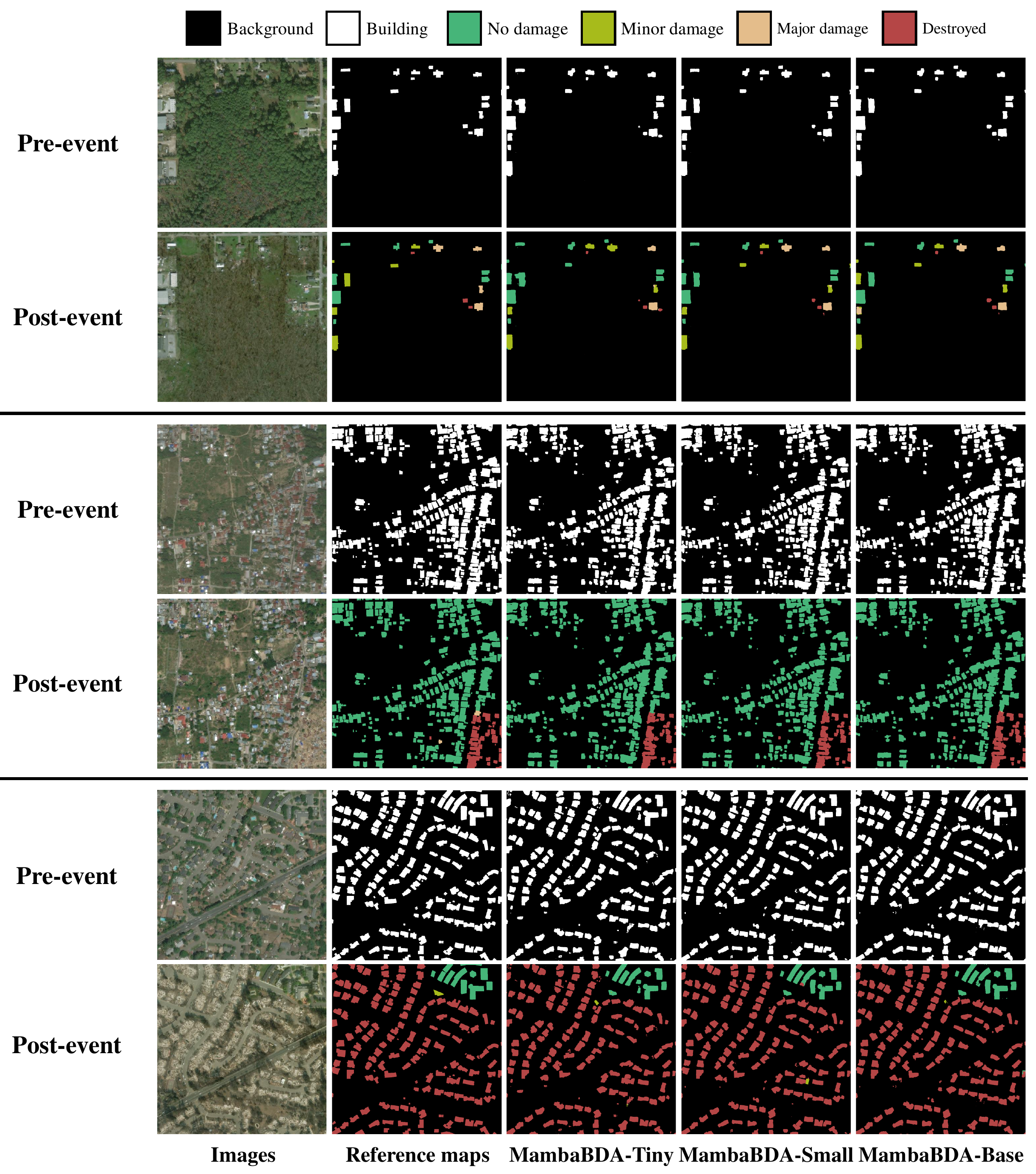}
  \caption{Some building damage assessment results obtained by our methods on the test set of the xBD dataset. From top to bottom, the disaster events are typhoons, tsunamis, and wildfires. }
  \label{fig:xBD_BDA}
\end{figure}

\par Fig. \ref{fig:xBD_BDA} shows the visualization results of the building damage caused by different disaster events obtained by MambaBDA on the xBD dataset. MambaBDA can accurately localize and then differentiate the damage levels of buildings despite the varying type of building and the disaster event, implying the potential of the MambaBDA architecture for practical disaster response applications.

\subsection{Different Spatio-Temporal Modeling Methods}
\begin{table*}[!t]
  \renewcommand{\arraystretch}{1.25}
\caption{\centering{Comparison of different approaches for modeling spatio-temporal relationships on three tasks. } }\label{tbl:st_model_com}
  \centering
  \begin{tabular}{c c c c c c c c c  c c c c}
  \toprule
    \hline	
 \multirow{2}{*}{\textbf{Method}}&	 \multicolumn{4}{c}{\textbf{SYSU}} & \multicolumn{4}{c}{\textbf{SECOND}}  & \multicolumn{3}{c}{\textbf{xBD}}\\
    \cline{2-12} 
&  \textbf{OA}  & \textbf{F1}   & \textbf{IoU}   & \textbf{KC}    & \textbf{OA}  & \textbf{F1}   & \textbf{mIoU}   & \textbf{SeK}  &  \textbf{$F_{1}^{loc}$}  & \textbf{$F_{1}^{clf}$} & \textbf{$F_{1}^{overall}$} \\
    \hline\hline
  Baseline &  91.71 & 81.11 & 68.22 & 75.84 & 86.11 & 54.20 & 72.42 & 21.15 &  84.85 &  74.74  & 77.77 \\
  STT \cite{Zheng2022} & 92.09  & \textbf{\textcolor{blue}{82.22}} &  \textbf{\textcolor{blue}{69.80}}  & 77.16 & 86.35 & 54.70 &  72.86  & 21.83 &  \textbf{\textcolor{red}{86.74}} & 75.92  & 79.17 \\
  
  RNN \cite{Chen2019a} &  \textbf{\textcolor{blue}{92.97}} & 82.02  & 69.52  & 76.94 &  86.46 &  54.76  & 72.89  & 21.90  & 85.74&  76.48  & 79.26     \\
  Transformer \cite{Chen2022Remote} &   92.14&  82.17 & 69.74 & \textbf{\textcolor{blue}{77.17}}  &   \textbf{\textcolor{red}{86.66}} & \textbf{\textcolor{blue}{55.50}} & \textbf{\textcolor{blue}{73.01}}  & \textbf{\textcolor{blue}{22.31}}  &  86.09  & \textbf{\textcolor{blue}{77.45}}  & \textbf{\textcolor{blue}{80.04}}    \\
  
  Ours & \textbf{\textcolor{red}{92.35}} & \textbf{\textcolor{red}{82.83}} & \textbf{\textcolor{red}{70.70}} & \textbf{\textcolor{red}{77.94}} &   \textbf{\textcolor{blue}{86.48}} &  \textbf{\textcolor{red}{55.72}} & \textbf{\textcolor{red}{73.41}}  & \textbf{\textcolor{red}{22.83}} & \textbf{\textcolor{blue}{86.61}}  & \textbf{\textcolor{red}{78.80}}  & \textbf{\textcolor{red}{81.14}} \\
    \hline
    \bottomrule
  \end{tabular}
\end{table*}

\par Modeling spatio-temporal relationships is significant for CD tasks \cite{Mou2019, Chen2019a, Zheng2022}. To demonstrate the effectiveness of our proposed modeling mechanisms, we compare them with several commonly used spatial-temporal modeling methods in CD. They are concatenation operations with feature pyramid network (FPN) \cite{Zheng2022} (baseline), temporal-symmetric transformation based on 3D convolutional layers (TST) \cite{Zheng2022}, RNN-based method \cite{Chen2019a}, and Tranformer-based method \cite{Chen2022Remote}. Table \ref{tbl:st_model_com} lists the performance of these modeling methods on the three CD subtasks. On the simpler BCD tasks, Transformer does not yield significant advantages in performance compared to RNN. However, on the more difficult SCD and BDA tasks, the advantages of the Transformer architecture are illustrated. Compared with these current representative spatio-temporal relationship modeling approaches, our proposed modeling approach based on the Mamba architecture, which inherits the advantages of RNN and Transformer, shows superior performance on the three CD tasks of BCD, SCD, and BDA. Compared to the Transformer-based spatio-temporal modeling approach, our method improves the F1 value by 0.77$\%$ on BCD, SeK by 0.52$\%$ on the SCD task and $F_{1}^{overall}$ by 1.10$\%$ on the BDA task.

\begin{table}[!t]
  \renewcommand{\arraystretch}{1.25}
\caption{\centering{Ablation study of the proposed spatio-temporal modelling mechanisms on the SYSU dataset.}}\label{tbl:three_st_model}
  \centering
  \begin{tabular}{c c c c c c c}
  \toprule
    \hline	
 \multicolumn{3}{c}{\textbf{Modelling Mechanism}}	&	\multirow{2}{*}{\textbf{OA}}  & \multirow{2}{*}{\textbf{F1}}   & \multirow{2}{*}{\textbf{IoU}}    & \multirow{2}{*}{\textbf{KC}} \\
 \cline{1-3}
Sequential & Cross &  Parallel &  &  &  & \\ 
\hline\hline
 & &  &  91.71 & 81.11 & 68.22 & 75.84 \\ 
 $\checkmark$ &   &  & 92.16 & 82.20 & 69.78  & 77.20  \\ 
 & $\checkmark$ &  &  \textbf{\textcolor{blue}{92.20}} & \textbf{\textcolor{blue}{82.24}} & \textbf{\textcolor{blue}{69.84}} & \textbf{\textcolor{blue}{77.28}}  \\ 
 & & $\checkmark$ &  92.12 & 82.17 & 69.73  & 77.14  \\ 
$\checkmark$ & $\checkmark$ & $\checkmark$ & \textbf{\textcolor{red}{92.35}} &  \textbf{\textcolor{red}{82.83}}  & \textbf{\textcolor{red}{70.70}}  & \textbf{\textcolor{red}{77.94}} \\ 
    \hline
    \bottomrule
  \end{tabular}
\end{table}

\par In Table \ref{tbl:three_st_model}, we list the detection performance obtained using one of the three modeling mechanisms alone. Firstly, using any of the modeling mechanisms can significantly improve detection performance compared to the baseline. Then, it can be seen that the performance gap between the three modeling mechanisms is not significant, with spatio-temporal cross modeling being slightly better than the other two. Finally, the detection performance of the network can be further improved by using the three modeling mechanisms simultaneously. 

\begin{figure}[!t]
  \centering
\includegraphics[width=3.1in]{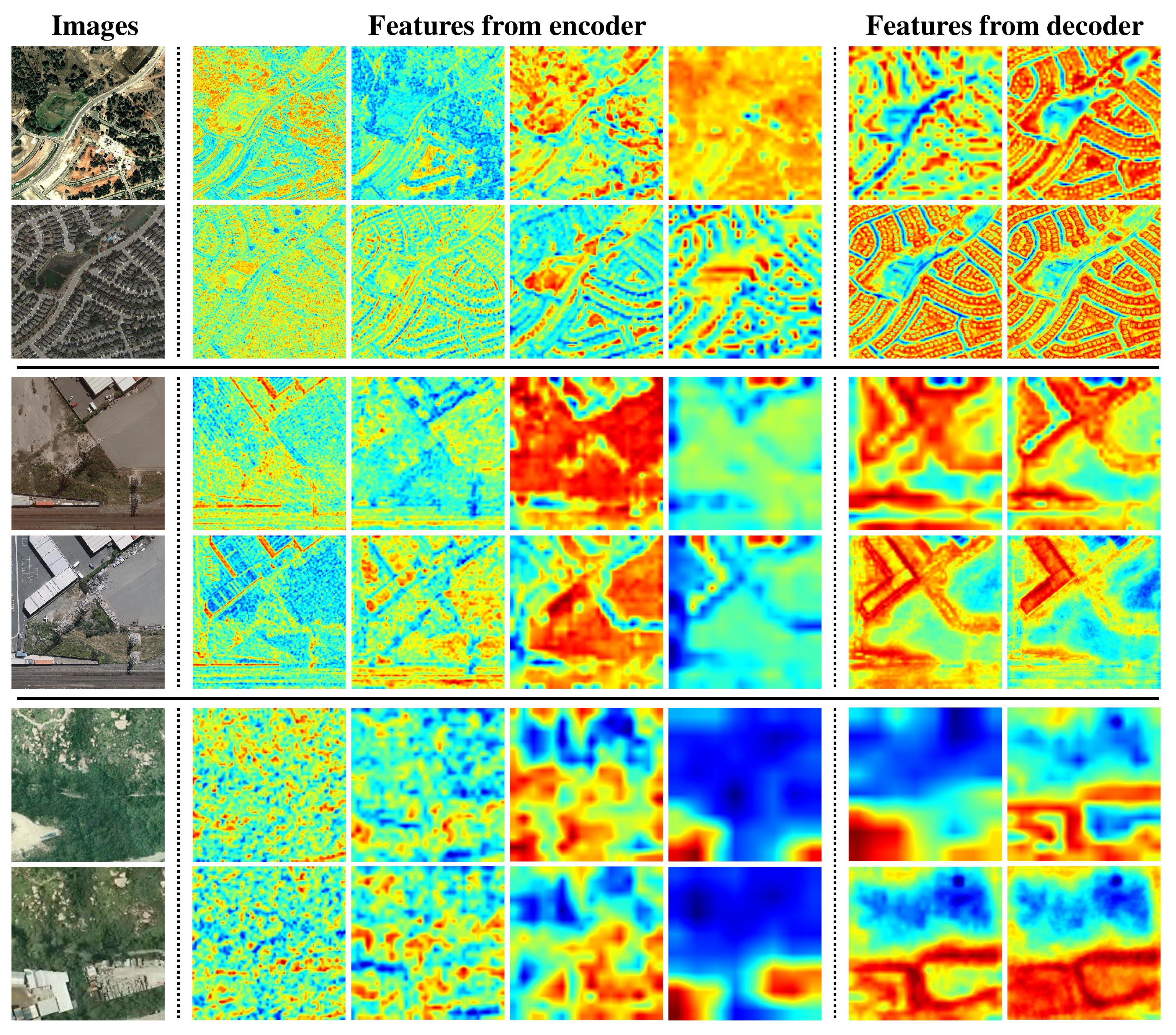}
  \caption{The visualization of features from different layers of MambaBCD. }
  \label{fig:visualization_features}
\end{figure}

\par We also visualize the features extracted by the MambaBCD architecture in Fig. \ref{fig:visualization_features}. In the features extracted by the encoder, the features in the region of high response values are not necessarily regions of interest, i.e., changed regions. Whereas in the decoder, after the spatio-temporal relationships are modeled by the proposed spatio-temporal modeling mechanisms, changed regions gradually exhibit high response values.

\subsection{Comparison to Other Backbone Networks}

\begin{table*}[!t]
  \renewcommand{\arraystretch}{1.3}
\caption{\centering{Comparison of different commonly used backbones in BCD on the SYSU dataset. }}\label{tbl:backbone_network}
  \centering
  \begin{tabular}{c c c c c c c}
  \toprule
    \hline	
 \textbf{Encoder}	&	\textbf{OA}  & \textbf{F1}   & \textbf{IoU}   & \textbf{KC} 	& \textbf{Param (M)} & \textbf{GFLOPs} \\
    \hline\hline
  ResNet-101 \cite{He2016} &   91.10  &81.10  &68.22  & 75.28 & 46.90  &  131.36 \\
  EfficientNet-B5 \cite{Tan2019EfficientNet} &  91.60  &   \textbf{\color{blue}{82.26}}  &  \textbf{\color{blue}{69.86}} &  76.84 & 31.63  & 76.88   \\
  MixFormer-v3 \cite{Xie2021SegFormer} & 87.70   & 75.00 & 60.00  & 66.87  &    47.03 &  82.98  \\
  Swin-Small \cite{Liu_2021_ICCV} &   \textbf{\color{blue}{92.09}} & 82.22 & 69.80   & \textbf{\color{blue}{77.16}}  &   37.75 & 97.99 \\
  VMamba-Small \cite{liu2024vmamba} & \textbf{\color{red}{92.35}} & \textbf{\color{red}{82.83}} & \textbf{\color{red}{70.70}} & \textbf{\color{red}{77.94}} & 49.94 & 114.82 \\
    \hline
    \bottomrule
  \end{tabular}
\end{table*}

\par we also simply compare the Mamba with some representative backbone networks in CD tasks on the SYSU dataset in Table \ref{tbl:backbone_network}. It can be seen that Mamba outperforms the two Transformer backbone networks, Swin-Transformer and MixFormer. This is because, in order to reduce the computational overhead of self-attention operation, these two Transformer backbone networks either adopt a local attention mechanism (Swin-Transformer) \cite{Liu_2021_ICCV} or a strided convolutional layer (MixFormer) \cite{Xie2021SegFormer} to reduce the size of the feature maps, which will negatively affect the learning of global contextual information. The adopted VMamba architecture, on the other hand, can reduce its computational consumption to a linear relationship with the number of tokens, thus eliminating the need for these methods, which cause information loss, and allowing for the fully learning of global context information. As for the remote sensing change detection task, due to the multi-scale phenomenon of land-cover features and the differences in spatial resolution of different sensors, the sizes of changed objects often vary greatly. Therefore, the features extracted by Mamba through fully learning the global contextual information are particularly important for the subsequent detection of these changed objects with different sizes.

\subsection{GPU Memory Footprint}
\begin{figure}[!t]
  \centering
\includegraphics[width=3.0in]{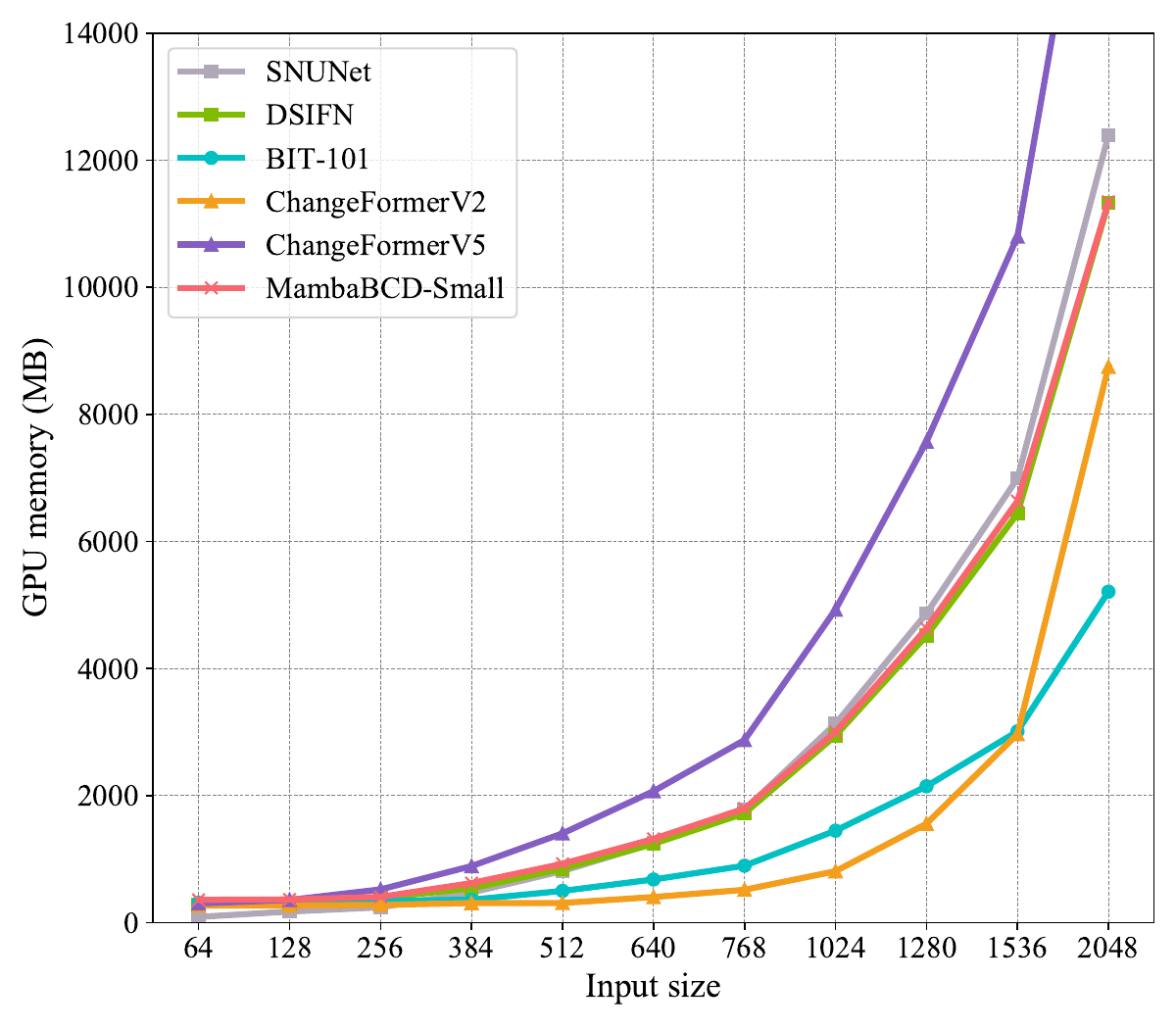}
  \caption{GPU memory footprint of different models with varying input sizes.}
  \label{fig:GPU_memory}
\end{figure}

\par In Section \ref{sec:benchmark_comparison}, we list the number of parameters and GFLOPs of different models. In this subsection, we further compare the GPU memory footprint of some models. Fig. \ref{fig:GPU_memory} compares the GPU memory footprint of some CNN-based and Transformer-based methods with our MambaBCD architecture for varying input sizes. We can see that the trend of GPU memory footprint with input size for MambaBCD-Small is similar to some of the advanced CNN architectures (DSIFN and SNUNet). However, our MambaBCD is able to learn global spatial contextual information as well as adequately model spatio-temporal relationships, thereby exhibiting better detection performance. The GPU memory footprint of MambaBCD-Small is significantly lower compared to ChangeFormerV5, which employs the Transformer architecture as its backbone. Note that ChangeFormerV5 itself has employed strided convolutional layers to reduce the computational overhead of self-attention operations. Also, we can find that the lightweight architecture ChangeFormerV2 has surpassed the Mamba architecture and the two CNN architectures, DSIFN and SNUNet, in terms of the increasing trend of GPU memory footprint when the input size is increased from 1536 to 2048. These results imply that Transformer to be applicable to large-scale scenarios of remote sensing images often comes at the cost of information loss. In comparison, Mamba architecture can avoid this problem.

\subsection{Robustness Against Degraded Input Data}

\begin{figure*}[!t]
   \centering
  \subfloat[]{
    \includegraphics[width=2.1in]{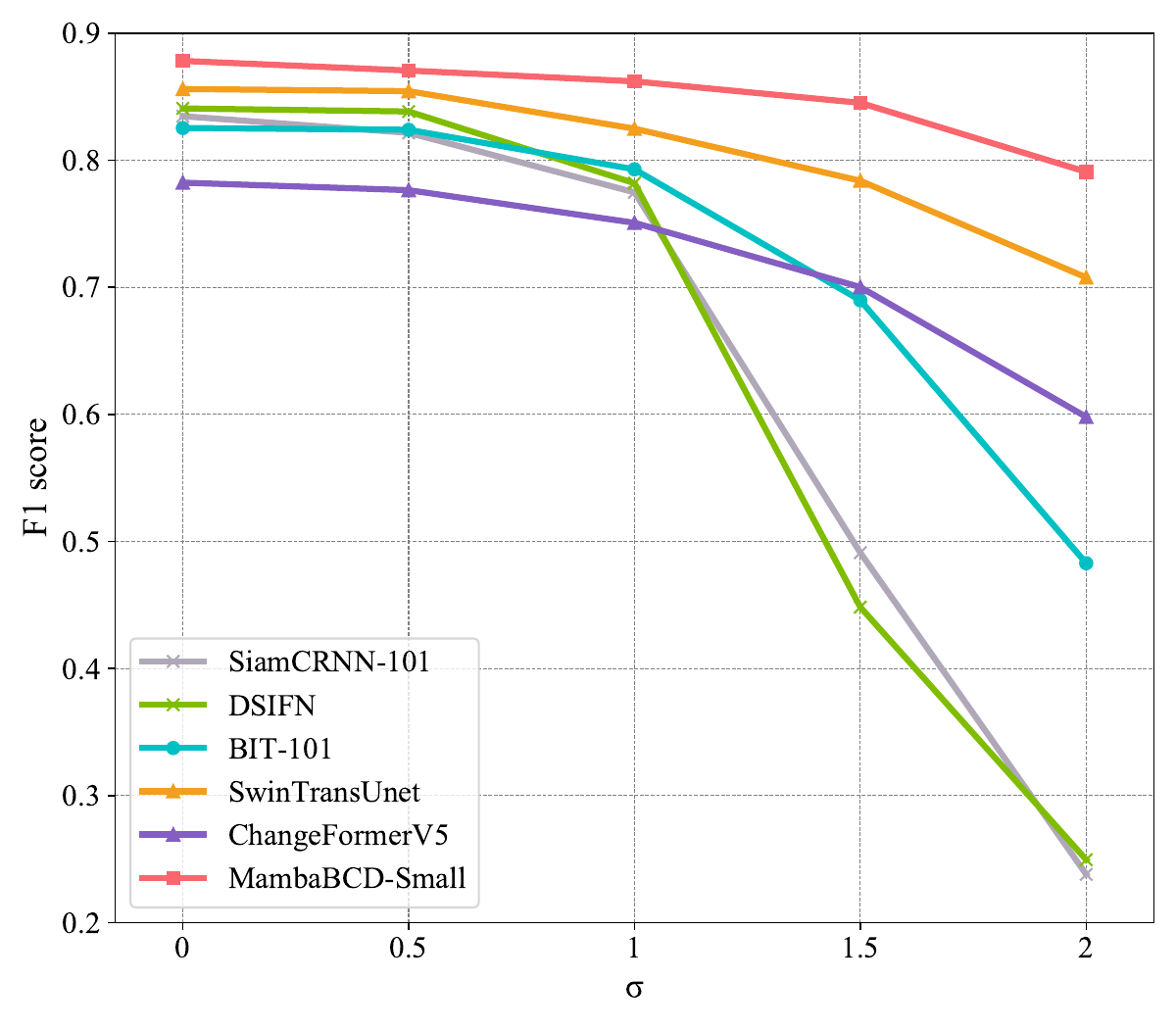}
  \label{fig_second_case}}  
      \hfil
  \subfloat[]{
    \includegraphics[width=2.1in]{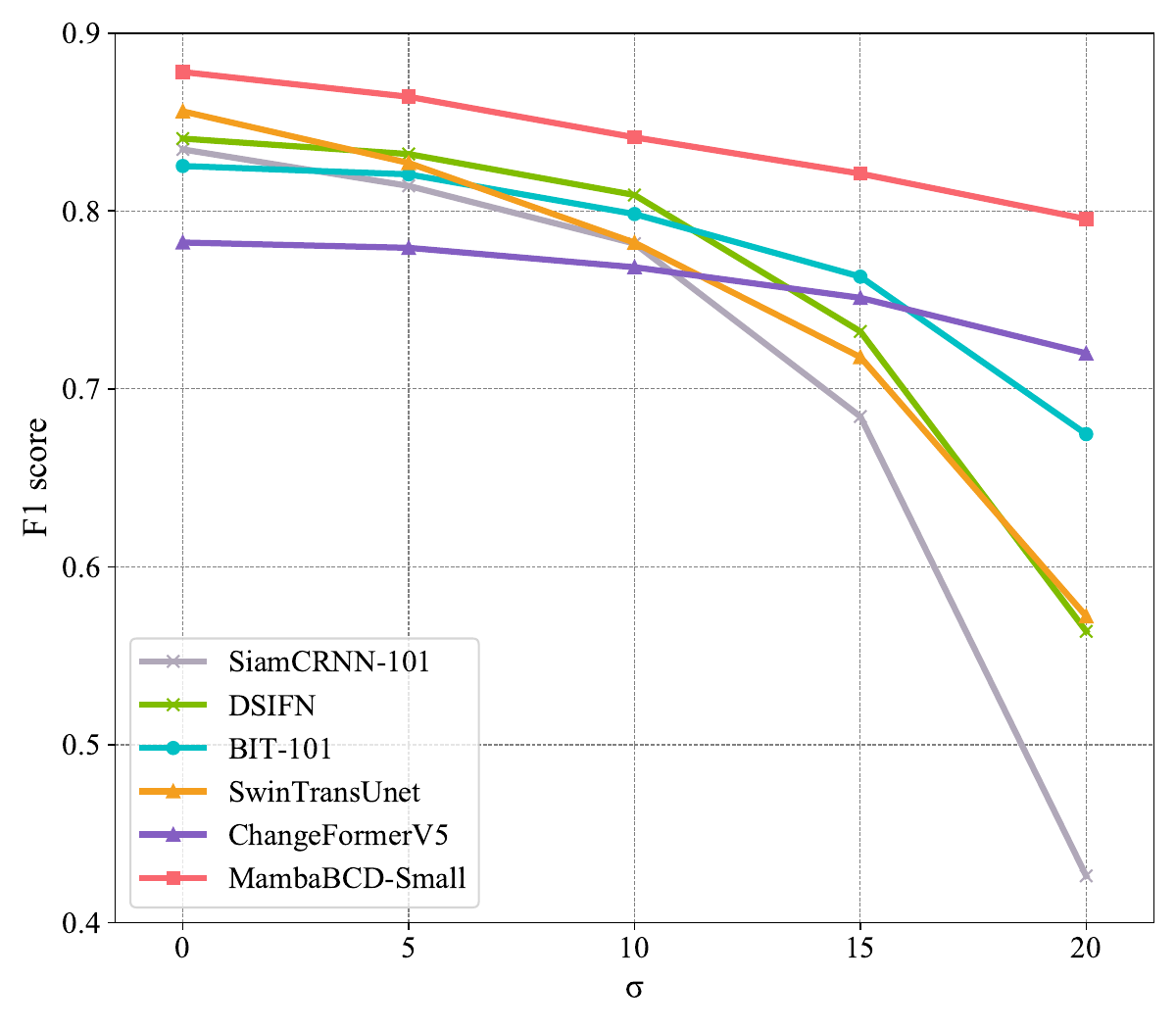}
  \label{fig_second_case}} 
    \hfil
  \subfloat[]{
    \includegraphics[width=2.1in]{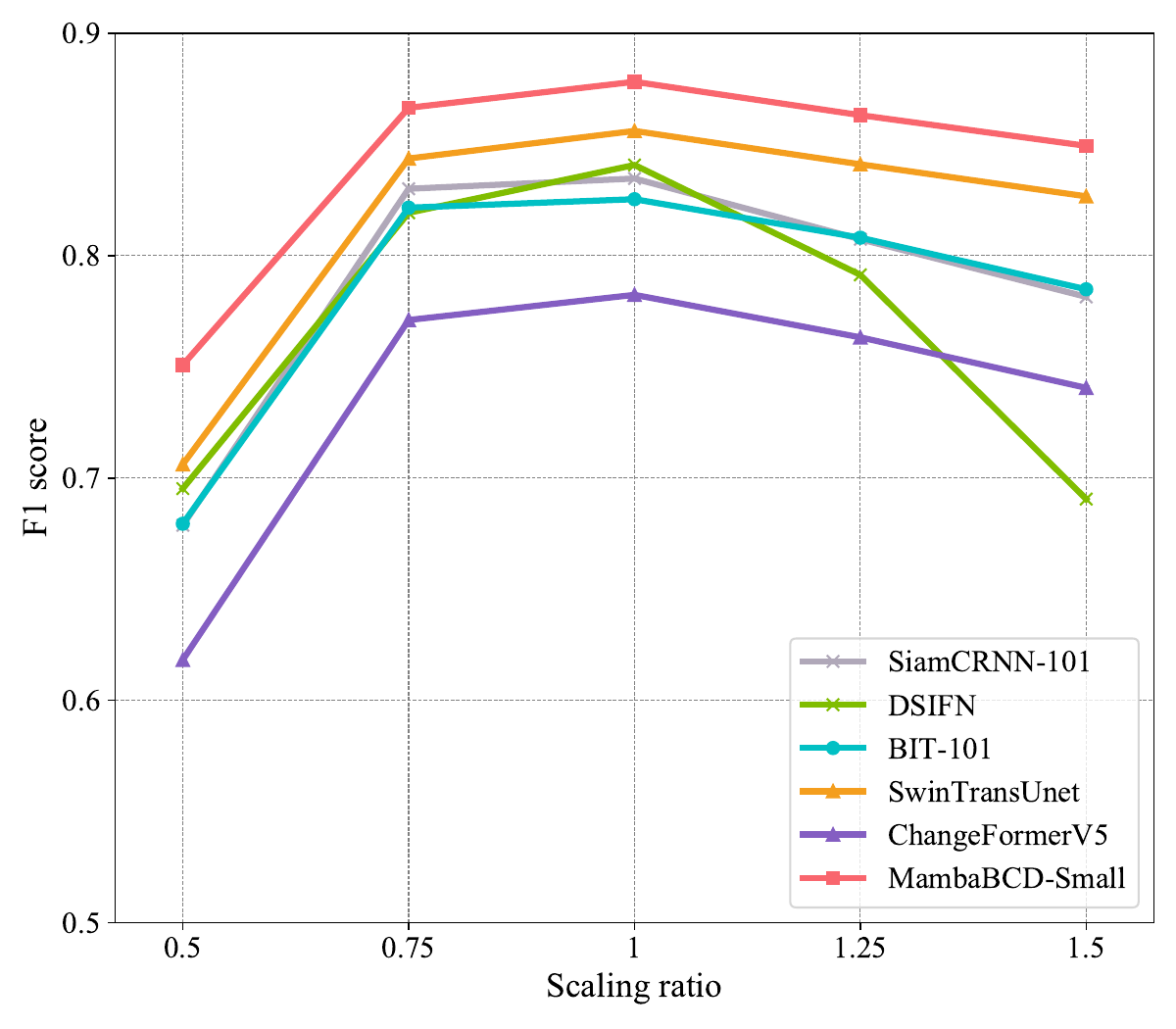}
  \label{fig_second_case}}  
  
   \caption{Robustness of different models against degraded data on the LEVIR-CD+ dataset. (a) Gaussian blur. (b) Gaussian noise. (c) Scaling. }
  \label{fig:robustness}
\end{figure*}

\par Finally, we test the robustness of our method to degraded input data. In the inference phase, the data input to the model is subjected to different levels of Gaussian blurring, Gaussian noise, and scaling of the data by different proportions, respectively. Fig. \ref{fig:robustness} shows the degradation of the F1 score of different methods with degraded data on the LEVIR-CD+ dataset. First, we can see that the CNN-based methods are very poorly robust to Gaussian blurred data. F1 scores of SiamCRNN-101 and DSIFN drop by nearly 60$\%$ when increasing the standard deviation of the Gaussian blur from 0 to 2. Whereas the two methods using Transformer as the backbone, SwinTransUNet and ChangeFormerV5 are relatively robust to Gaussian blurred data, with F1 scores decreasing by 14.83 $\%$ and 18.43 $\%$, respectively. In contrast, the F1 score for our MambaBCD architecture drops by only 8.73$\%$. 

\par Fig. \ref{fig:robustness}-(b) shows the effect of adding different levels of Gaussian noise to the input data. With the exception of ChangeFormerV5 and our MambaBCD, the remaining methods have significant F1 decreases ranging from 40.83$\%$ to 15.07$\%$. Our MambaBCD architecture, on the other hand, shows robustness to Gaussian noise, with only an 8.26$\%$ drop in the F1 score for Gaussian noise with a standard deviation of 20.

\par Fig. \ref{fig:robustness}-(c) compares the robustness of models trained at a single scale to test data at different scales. The F1 value of MambaBCD decreases by only 2.88$\%$ when the scale increases from 1 to 1.5. When the scaling ratio decreases from 1 to 0.75, the F1 value of MambaBCD decreases only by 1.18$\%$. When the scaling factor is reduced to 0.5, all methods, including MambaBCD, show a larger decrease in the F1 score. However, compared to the decline of 14.99 $\%$ of SwinTransUNet and the decline of 16.41 $\%$ of ChangeFormerV5, the decline of the MambaBCD decline is smaller, at 12.03 $\%$. All the above results again demonstrate the superiority of our proposed Mamba-based CD architecture.

\section{Conclusion}\label{sec:5}
\par In this paper, we explore the potential of the emerging Mamba architecture for remote sensing image change detection tasks. Three architectures, MambaBCD, MambaSCD, and MambaBDA are developed for binary change detection, semantic change detection, and building damage assessment tasks, respectively. The encoders of all three architectures are cutting-edge VMamba architecture, which can adequately learn the global contextual information of the input image with linear complexity. Then, for the change decoder, in order to adequately learn spatio-temporal features, we propose three ways of modeling spatio-temporal relationships that can adequately exploiting the attributes and advantages of Mamba architecture. Experiments on five benchmark datasets have fully revealed the potential of Mamba for multi-temporal remote sensing image data processing tasks. Compared to the CNN-based and Transformer-based approaches, the proposed three architectures can achieve quite competitive performance on their respective tasks. Compared to some commonly used backbone networks, the Mamba architecture can extract features that are more suitable for downstream CD tasks. The combination of our proposed three spatio-temporal relationship modeling mechanisms and the Mamba architecture provides a new perspective on modeling spatio-temporal relationships, showing much better detection results on all three subtasks. 

\par Our future work includes, but is not limited to, developing Mamba architectures more suited to the characteristics of remote sensing data, applying them to multimodal and time-series remote sensing tasks, and exploring the potential of Mamba for remote sensing foundation models.

\ifCLASSOPTIONcaptionsoff
  \newpage
\fi

\bibliographystyle{IEEEtran}
\bibliography{ChangeMamba.bib}

\begin{thebibliography}{10}
\providecommand{\url}[1]{#1}
\csname url@samestyle\endcsname
\providecommand{\newblock}{\relax}
\providecommand{\bibinfo}[2]{#2}
\providecommand{\BIBentrySTDinterwordspacing}{\spaceskip=0pt\relax}
\providecommand{\BIBentryALTinterwordstretchfactor}{4}
\providecommand{\BIBentryALTinterwordspacing}{\spaceskip=\fontdimen2\font plus
\BIBentryALTinterwordstretchfactor\fontdimen3\font minus \fontdimen4\font\relax}
\providecommand{\BIBforeignlanguage}[2]{{%
\expandafter\ifx\csname l@#1\endcsname\relax
\typeout{** WARNING: IEEEtran.bst: No hyphenation pattern has been}%
\typeout{** loaded for the language `#1'. Using the pattern for}%
\typeout{** the default language instead.}%
\else
\language=\csname l@#1\endcsname
\fi
#2}}
\providecommand{\BIBdecl}{\relax}
\BIBdecl

\bibitem{Lu2004}
D.~Lu, P.~Mausel, E.~Brond{\'{i}}zio, and E.~Moran, ``{Change detection techniques},'' \emph{Int. J. Remote Sens.}, vol.~25, no.~12, pp. 2365--2407, 2004.

\bibitem{Coppin2004}
P.~Coppin, I.~Jonckheere, K.~Nackaerts, B.~Muys, and E.~Lambin, ``{Digital change detection methods in ecosystem monitoring: A review},'' \emph{Int. J. Remote Sens.}, vol.~25, no.~9, pp. 1565--1596, 2004.

\bibitem{ZHENG2021Building}
Z.~Zheng, Y.~Zhong, J.~Wang, A.~Ma, and L.~Zhang, ``Building damage assessment for rapid disaster response with a deep object-based semantic change detection framework: From natural disasters to man-made disasters,'' \emph{Remote Sens. Environ.}, vol. 265, p. 112636, 2021.

\bibitem{guo2021deep}
H.~Guo, Q.~Shi, A.~Marinoni, B.~Du, and L.~Zhang, ``Deep building footprint update network: A semi-supervised method for updating existing building footprint from bi-temporal remote sensing images,'' \emph{Remote Sens. Environ.}, vol. 264, p. 112589, 2021.

\bibitem{chen2022unsupervised}
H.~Chen, N.~Yokoya, C.~Wu, and B.~Du, ``{Unsupervised Multimodal Change Detection Based on Structural Relationship Graph Representation Learning},'' \emph{IEEE Trans. Geosci. Remote Sens.}, pp. 1--18, 2022.

\bibitem{chen2023land}
H.~Chen, C.~Lan, J.~Song, C.~Broni-Bediako, J.~Xia, and N.~Yokoya, ``Land-cover change detection using paired openstreetmap data and optical high-resolution imagery via object-guided transformer,'' \emph{arXiv preprint arXiv:2310.02674}, 2023.

\bibitem{Xiao2024TTST}
Y.~Xiao, Q.~Yuan, K.~Jiang, J.~He, C.-W. Lin, and L.~Zhang, ``Ttst: A top-k token selective transformer for remote sensing image super-resolution,'' \emph{IEEE Trans. Image Process.}, vol.~33, pp. 738--752, 2024.

\bibitem{Wu2014}
C.~Wu, B.~Du, and L.~Zhang, ``{Slow feature analysis for change detection in multispectral imagery},'' \emph{IEEE Trans. Geosci. Remote Sens.}, vol.~52, no.~5, pp. 2858--2874, 2014.

\bibitem{Nielsen2007}
A.~A. Nielsen, ``{The regularized iteratively reweighted MAD method for change detection in multi- and hyperspectral data},'' \emph{IEEE Trans. Image Process.}, vol.~16, no.~2, pp. 463--478, 2007.

\bibitem{Sharma2007}
L.~Bruzzone and {Diego Fern{\`{a}}ndez Prieto}, ``{Automatic Analysis of the Difference Image for Unsupervised Change Detection},'' \emph{IEEE Trans. Geosci. Remote Sens.}, vol.~38, no.~3, pp. 1171--1182, 2000.

\bibitem{Hussain2013}
M.~Hussain, D.~Chen, A.~Cheng, H.~Wei, and D.~Stanley, ``{Change detection from remotely sensed images: From pixel-based to object-based approaches},'' \emph{ISPRS J. Photogramm. Remote Sens.}, vol.~80, pp. 91--106, 2013.

\bibitem{Sun2021a}
Y.~Sun, L.~Lei, D.~Guan, and G.~Kuang, ``{Iterative Robust Graph for Unsupervised Change Detection of Heterogeneous Remote Sensing Images},'' \emph{IEEE Trans. Image Process.}, vol.~30, pp. 6277--6291, 2021.

\bibitem{Chen2023Fourier}
H.~Chen, N.~Yokoya, and M.~Chini, ``Fourier domain structural relationship analysis for unsupervised multimodal change detection,'' \emph{ISPRS J. Photogramm. Remote Sens.}, vol. 198, pp. 99--114, 2023.

\bibitem{Wu2022Unsupervised}
C.~Wu, H.~Chen, B.~Du, and L.~Zhang, ``Unsupervised change detection in multitemporal vhr images based on deep kernel pca convolutional mapping network,'' \emph{IEEE Trans. Cybern}, vol.~52, no.~11, pp. 12\,084--12\,098, 2022.

\bibitem{Xiao2024EDiffSR}
Y.~Xiao, Q.~Yuan, K.~Jiang, J.~He, X.~Jin, and L.~Zhang, ``Ediffsr: An efficient diffusion probabilistic model for remote sensing image super-resolution,'' \emph{IEEE Trans. Geosci. Remote Sens.}, vol.~62, pp. 1--14, 2024.

\bibitem{Gong2017Superpixel}
M.~Gong, T.~Zhan, P.~Zhang, and Q.~Miao, ``Superpixel-based difference representation learning for change detection in multispectral remote sensing images,'' \emph{IEEE Trans. Geosci. Remote Sens.}, vol.~55, no.~5, pp. 2658--2673, 2017.

\bibitem{Chen2019a}
H.~Chen, C.~Wu, B.~Du, L.~Zhang, and L.~Wang, ``{Change Detection in Multisource VHR Images via Deep Siamese Convolutional Multiple-Layers Recurrent Neural Network},'' \emph{IEEE Trans. Geosci. Remote Sens.}, vol.~58, no.~4, pp. 2848--2864, 2020.

\bibitem{Chen2019Deep}
H.~Chen, C.~Wu, B.~Du, and L.~Zhang, ``{Deep Siamese Multi-scale Convolutional Network for Change Detection in Multi-Temporal VHR Images},'' in \emph{2019 10th International Workshop on the Analysis of Multitemporal Remote Sensing Images (MultiTemp)}, 2019, pp. 1--4.

\bibitem{Shi2022Deeply}
Q.~Shi, M.~Liu, S.~Li, X.~Liu, F.~Wang, and L.~Zhang, ``A deeply supervised attention metric-based network and an open aerial image dataset for remote sensing change detection,'' \emph{IEEE Trans. Geosci. Remote Sens.}, vol.~60, pp. 1--16, 2022.

\bibitem{Yang2022Asymmetric}
K.~Yang, G.-S. Xia, Z.~Liu, B.~Du, W.~Yang, M.~Pelillo, and L.~Zhang, ``Asymmetric siamese networks for semantic change detection in aerial images,'' \emph{IEEE Trans. Geosci. Remote Sens.}, vol.~60, pp. 1--18, 2022.

\bibitem{Song_2024_WACV}
J.~Song, H.~Chen, and N.~Yokoya, ``Syntheworld: A large-scale synthetic dataset for land cover mapping and building change detection,'' in \emph{Proceedings of the IEEE/CVF Winter Conference on Applications of Computer Vision (WACV)}, January 2024, pp. 8287--8296.

\bibitem{CayeDaudt2018}
R.~{Caye Daudt}, B.~{Le Saux}, and A.~Boulch, ``{Fully convolutional siamese networks for change detection},'' in \emph{Proceedings of the International Conference on Image Processing (ICIP)}, 2018, pp. 4063--4067.

\bibitem{Zhang2020}
C.~Zhang, P.~Yue, D.~Tapete, L.~Jiang, B.~Shangguan, L.~Huang, and G.~Liu, ``{A deeply supervised image fusion network for change detection in high resolution bi-temporal remote sensing images},'' \emph{ISPRS J. Photogramm. Remote Sens.}, vol. 166, no. June, pp. 183--200, 2020.

\bibitem{Zheng2022}
Z.~Zheng, Y.~Zhong, S.~Tian, A.~Ma, and L.~Zhang, ``{ChangeMask: Deep multi-task encoder-transformer-decoder architecture for semantic change detection},'' \emph{ISPRS J. Photogramm. Remote Sens.}, vol. 183, no. March 2021, pp. 228--239, 2022.

\bibitem{CAO2023full}
Y.~Cao and X.~Huang, ``A full-level fused cross-task transfer learning method for building change detection using noise-robust pretrained networks on crowdsourced labels,'' \emph{Remote Sens. Environ.}, vol. 284, p. 113371, 2023.

\bibitem{Lv2023Hierarchical}
Z.~Lv, J.~Liu, W.~Sun, T.~Lei, J.~A. Benediktsson, and X.~Jia, ``Hierarchical attention feature fusion-based network for land cover change detection with homogeneous and heterogeneous remote sensing images,'' \emph{IEEE Trans. Geosci. Remote Sens.}, vol.~61, pp. 1--15, 2023.

\bibitem{Liu2022Building}
T.~Liu, M.~Gong, D.~Lu, Q.~Zhang, H.~Zheng, F.~Jiang, and M.~Zhang, ``Building change detection for vhr remote sensing images via local–global pyramid network and cross-task transfer learning strategy,'' \emph{IEEE Trans. Geosci. Remote Sens.}, vol.~60, pp. 1--17, 2022.

\bibitem{Chen2022Remote}
H.~Chen, Z.~Qi, and Z.~Shi, ``Remote sensing image change detection with transformers,'' \emph{IEEE Trans. Geosci. Remote Sens.}, vol.~60, pp. 1--14, 2022.

\bibitem{dosovitskiy2020image}
A.~Dosovitskiy, L.~Beyer, A.~Kolesnikov, D.~Weissenborn, X.~Zhai, T.~Unterthiner, M.~Dehghani, M.~Minderer, G.~Heigold, S.~Gelly \emph{et~al.}, ``An image is worth 16x16 words: Transformers for image recognition at scale,'' \emph{arXiv preprint arXiv:2010.11929}, 2020.

\bibitem{Bandara2022Transformer}
W.~G.~C. Bandara and V.~M. Patel, ``A transformer-based siamese network for change detection,'' in \emph{IEEE International Geoscience and Remote Sensing Symposium (IGARSS)}, 2022, pp. 207--210.

\bibitem{Chen2022Dual}
H.~Chen, E.~Nemni, S.~Vallecorsa, X.~Li, C.~Wu, and L.~Bromley, ``Dual-tasks siamese transformer framework for building damage assessment,'' in \emph{International Geoscience and Remote Sensing Symposium (IGARSS)}, 2022, pp. 1600--1603.

\bibitem{Zhang2022SwinSUNet}
C.~Zhang, L.~Wang, S.~Cheng, and Y.~Li, ``Swinsunet: Pure transformer network for remote sensing image change detection,'' \emph{IEEE Trans. Geosci. Remote Sens.}, vol.~60, pp. 1--13, 2022.

\bibitem{Zhang2023Relation}
K.~Zhang, X.~Zhao, F.~Zhang, L.~Ding, J.~Sun, and L.~Bruzzone, ``Relation changes matter: Cross-temporal difference transformer for change detection in remote sensing images,'' \emph{IEEE Trans. Geosci. Remote Sens.}, vol.~61, pp. 1--15, 2023.

\bibitem{LiTransUNetCD2022}
Q.~Li, R.~Zhong, X.~Du, and Y.~Du, ``Transunetcd: A hybrid transformer network for change detection in optical remote-sensing images,'' \emph{IEEE Trans. Geosci. Remote Sens.}, vol.~60, pp. 1--19, 2022.

\bibitem{Liu_2021_ICCV}
Z.~Liu, Y.~Lin, Y.~Cao, H.~Hu, Y.~Wei, Z.~Zhang, S.~Lin, and B.~Guo, ``Swin transformer: Hierarchical vision transformer using shifted windows,'' in \emph{Proceedings of the IEEE/CVF International Conference on Computer Vision (ICCV)}, October 2021, pp. 10\,012--10\,022.

\bibitem{Xie2021SegFormer}
E.~Xie, W.~Wang, Z.~Yu, A.~Anandkumar, J.~M. Alvarez, and P.~Luo, ``Segformer: Simple and efficient design for semantic segmentation with transformers,'' in \emph{Advances in Neural Information Processing Systems (NIPS)}, vol.~34, 2021, pp. 12\,077--12\,090.

\bibitem{gu2021efficiently}
A.~Gu, K.~Goel, and C.~R{\'e}, ``Efficiently modeling long sequences with structured state spaces,'' \emph{arXiv preprint arXiv:2111.00396}, 2021.

\bibitem{gu2023mamba}
A.~Gu and T.~Dao, ``Mamba: Linear-time sequence modeling with selective state spaces,'' \emph{arXiv preprint arXiv:2312.00752}, 2023.

\bibitem{liu2024vmamba}
Y.~Liu, Y.~Tian, Y.~Zhao, H.~Yu, L.~Xie, Y.~Wang, Q.~Ye, and Y.~Liu, ``Vmamba: Visual state space model,'' \emph{arXiv preprint arXiv:2401.10166}, 2024.

\bibitem{zhu2024vision}
L.~Zhu, B.~Liao, Q.~Zhang, X.~Wang, W.~Liu, and X.~Wang, ``Vision mamba: Efficient visual representation learning with bidirectional state space model,'' \emph{arXiv preprint arXiv:2401.09417}, 2024.

\bibitem{Fang2022SNUNet}
S.~Fang, K.~Li, J.~Shao, and Z.~Li, ``Snunet-cd: A densely connected siamese network for change detection of vhr images,'' \emph{IEEE Geosci. Remote Sens. Lett.}, vol.~19, pp. 1--5, 2022.

\bibitem{Han2023HANet}
C.~Han, C.~Wu, H.~Guo, M.~Hu, and H.~Chen, ``Hanet: A hierarchical attention network for change detection with bitemporal very-high-resolution remote sensing images,'' \emph{IEEE J. Sel. Topics Appl. Earth Observ. Remote Sens.}, vol.~16, pp. 3867--3878, 2023.

\bibitem{Han2023CGNet}
C.~Han, C.~Wu, H.~Guo, M.~Hu, J.~Li, and H.~Chen, ``Change guiding network: Incorporating change prior to guide change detection in remote sensing imagery,'' \emph{IEEE J. Sel. Topics Appl. Earth Observ. Remote Sens.}, vol.~16, pp. 8395--8407, 2023.

\bibitem{Zhao2023Exchanging}
S.~Zhao, X.~Zhang, P.~Xiao, and G.~He, ``Exchanging dual-encoder–decoder: A new strategy for change detection with semantic guidance and spatial localization,'' \emph{IEEE Trans. Geosci. Remote Sens.}, vol.~61, pp. 1--16, 2023.

\bibitem{Chen2023Exchange}
H.~Chen, J.~Song, C.~Wu, B.~Du, and N.~Yokoya, ``Exchange means change: An unsupervised single-temporal change detection framework based on intra- and inter-image patch exchange,'' \emph{ISPRS J. Photogramm. Remote Sens.}, vol. 206, pp. 87--105, 2023.

\bibitem{Rodrigo2019Multitask}
R.~{Caye Daudt}, B.~{Le Saux}, A.~Boulch, and Y.~Gousseau, ``Multitask learning for large-scale semantic change detection,'' \emph{Comput. Vis. Image Underst.}, vol. 187, p. 102783, 2019.

\bibitem{Mou2019}
L.~Mou, L.~Bruzzone, and X.~X. Zhu, ``{Learning spectral-spatialoral features via a recurrent convolutional neural network for change detection in multispectral imagery},'' \emph{IEEE Trans. Geosci. Remote Sens.}, vol.~57, no.~2, pp. 924--935, 2019.

\bibitem{Ding2022Semantic}
L.~Ding, H.~Guo, S.~Liu, L.~Mou, J.~Zhang, and L.~Bruzzone, ``Bi-temporal semantic reasoning for the semantic change detection in hr remote sensing images,'' \emph{IEEE Trans. Geosci. Remote Sens.}, vol.~60, pp. 1--14, 2022.

\bibitem{peng2021scdnet}
D.~Peng, L.~Bruzzone, Y.~Zhang, H.~Guan, and P.~He, ``Scdnet: A novel convolutional network for semantic change detection in high resolution optical remote sensing imagery,'' \emph{Int. J. Appl. Earth Obs. Geoinf.}, vol. 103, p. 102465, 2021.

\bibitem{tian2023temporal}
S.~Tian, X.~Tan, A.~Ma, Z.~Zheng, L.~Zhang, and Y.~Zhong, ``Temporal-agnostic change region proposal for semantic change detection,'' \emph{ISPRS J. Photogramm. Remote Sens.}, vol. 204, pp. 306--320, 2023.

\bibitem{Saha2019}
S.~Saha, F.~Bovolo, and L.~Bruzzone, ``{Unsupervised deep change vector analysis for multiple-change detection in VHR Images},'' \emph{IEEE Trans. Geosci. Remote Sens.}, vol.~57, no.~6, pp. 3677--3693, 2019.

\bibitem{Gupta_2019_CVPR_Workshops}
R.~Gupta, B.~Goodman, N.~Patel, R.~Hosfelt, S.~Sajeev, E.~Heim, J.~Doshi, K.~Lucas, H.~Choset, and M.~Gaston, ``Creating xbd: A dataset for assessing building damage from satellite imagery,'' in \emph{Proceedings of the IEEE/CVF Conference on Computer Vision and Pattern Recognition (CVPR) Workshops}, June 2019.

\bibitem{weber2020building}
E.~Weber and H.~Kan{\'e}, ``Building disaster damage assessment in satellite imagery with multi-temporal fusion,'' \emph{arXiv preprint arXiv:2004.05525}, 2020.

\bibitem{He2017Mask}
K.~He, G.~Gkioxari, P.~Dollár, and R.~Girshick, ``Mask r-cnn,'' in \emph{IEEE International Conference on Computer Vision (ICCV)}, 2017, pp. 2980--2988.

\bibitem{Niu2023SMNet}
Y.~Niu, H.~Guo, J.~Lu, L.~Ding, and D.~Yu, ``Smnet: Symmetric multi-task network for semantic change detection in remote sensing images based on cnn and transformer,'' \emph{Remote Sens.}, vol.~15, no.~4, 2023.

\bibitem{Ding2024Joint}
L.~Ding, J.~Zhang, H.~Guo, K.~Zhang, B.~Liu, and L.~Bruzzone, ``Joint spatio-temporal modeling for semantic change detection in remote sensing images,'' \emph{IEEE Trans. Geosci. Remote Sens.}, vol.~62, pp. 1--14, 2024.

\bibitem{smith2022simplified}
J.~T. Smith, A.~Warrington, and S.~W. Linderman, ``Simplified state space layers for sequence modeling,'' \emph{arXiv preprint arXiv:2208.04933}, 2022.

\bibitem{fu2022hungry}
D.~Y. Fu, T.~Dao, K.~K. Saab, A.~W. Thomas, A.~Rudra, and C.~R{\'e}, ``Hungry hungry hippos: Towards language modeling with state space models,'' \emph{arXiv preprint arXiv:2212.14052}, 2022.

\bibitem{he2024pan}
X.~He, K.~Cao, K.~Yan, R.~Li, C.~Xie, J.~Zhang, and M.~Zhou, ``Pan-mamba: Effective pan-sharpening with state space model,'' \emph{arXiv preprint arXiv:2402.12192}, 2024.

\bibitem{chen2024rsmamba}
K.~Chen, B.~Chen, C.~Liu, W.~Li, Z.~Zou, and Z.~Shi, ``Rsmamba: Remote sensing image classification with state space model,'' \emph{arXiv preprint arXiv:2403.19654}, 2024.

\bibitem{Vaswani2017Attention}
A.~Vaswani, N.~Shazeer, N.~Parmar, J.~Uszkoreit, L.~Jones, A.~N. Gomez, L.~u. Kaiser, and I.~Polosukhin, ``Attention is all you need,'' in \emph{Advances in Neural Information Processing Systems (NIPS)}, vol.~30, 2017.

\bibitem{elfwing2018sigmoid}
S.~Elfwing, E.~Uchibe, and K.~Doya, ``Sigmoid-weighted linear units for neural network function approximation in reinforcement learning,'' \emph{Neural Netw.}, vol. 107, pp. 3--11, 2018.

\bibitem{lin2017focal}
T.-Y. Lin, P.~Goyal, R.~Girshick, K.~He, and P.~Doll{\'a}r, ``Focal loss for dense object detection,'' in \emph{Proceedings of the IEEE international Conference on Computer Vision (ICCV)}, 2017, pp. 2980--2988.

\bibitem{Berman_2018_CVPR}
M.~Berman, A.~R. Triki, and M.~B. Blaschko, ``The lovász-softmax loss: A tractable surrogate for the optimization of the intersection-over-union measure in neural networks,'' in \emph{Proceedings of the IEEE Conference on Computer Vision and Pattern Recognition (CVPR)}, June 2018.

\bibitem{Chen2020Spatial}
H.~Chen and Z.~Shi, ``A spatial-temporal attention-based method and a new dataset for remote sensing image change detection,'' \emph{Remote Sens.}, vol.~12, no.~10, 2020.

\bibitem{ji2018fully}
S.~Ji, S.~Wei, and M.~Lu, ``Fully convolutional networks for multisource building extraction from an open aerial and satellite imagery data set,'' \emph{IEEE Trans. Geosci. Remote Sens.}, vol.~57, no.~1, pp. 574--586, 2018.

\bibitem{loshchilov2017decoupled}
I.~Loshchilov and F.~Hutter, ``Decoupled weight decay regularization,'' \emph{arXiv preprint arXiv:1711.05101}, 2017.

\bibitem{Huang2024Spatiotemporal}
Y.~Huang, X.~Li, Z.~Du, and H.~Shen, ``Spatiotemporal enhancement and interlevel fusion network for remote sensing images change detection,'' \emph{IEEE Trans. Geosci. Remote Sens.}, vol.~62, pp. 1--14, 2024.

\bibitem{He2016}
K.~He, X.~Zhang, S.~Ren, and J.~Sun, ``{Deep residual learning for image recognition},'' \emph{Proceedings of the IEEE Computer Society Conference on Computer Vision and Pattern Recognition (CVPR)}, pp. 770--778, 2016.

\bibitem{Tian2022Large}
S.~Tian, Y.~Zhong, Z.~Zheng, A.~Ma, X.~Tan, and L.~Zhang, ``Large-scale deep learning based binary and semantic change detection in ultra high resolution remote sensing imagery: From benchmark datasets to urban application,'' \emph{ISPRS J. Photogramm. Remote Sens.}, vol. 193, pp. 164--186, 2022.

\bibitem{CAO2021deep}
Y.~Cao and X.~Huang, ``A deep learning method for building height estimation using high-resolution multi-view imagery over urban areas: A case study of 42 chinese cities,'' \emph{Remote Sens. Environ.}, vol. 264, p. 112590, 2021.

\bibitem{Tan2019EfficientNet}
M.~Tan and Q.~Le, ``Efficientnet: Rethinking model scaling for convolutional neural networks,'' in \emph{Proceedings of the International Conference on Machine Learning (ICML)}, vol.~97, 2019, pp. 6105--6114.

\end{thebibliography}

\end{document}